\documentclass[twocolumn,prd,aps,longbibliography,superscriptaddress,preprintnumbers,tightenlines,
showpacs,nofootinbib,eqsecnum,amsfonts,amsmath]{revtex4}
\pdfoutput=1

\usepackage{color}
\usepackage{calc}
\usepackage{graphicx}
\usepackage{tensor}
\usepackage{bm}
\usepackage{url}
\usepackage{times}
\usepackage{multirow}
\usepackage[varg]{txfonts}
\usepackage{float}
\usepackage{dcolumn}
\usepackage[nolist,nohyperlinks]{acronym}
\usepackage{xspace}
\usepackage[english]{babel}
\usepackage[abs]{overpic}
\usepackage{pict2e}
\usepackage[caption=false]{subfig} 
\allowdisplaybreaks[1]
\usepackage[utf8]{inputenc}

\DeclareMathAlphabet{\mathcalstd}{OMS}{cmsy}{m}{n}
\DeclareMathAlphabet{\mathpzc}{OT1}{pzc}{m}{it}

\newcommand{\UIB}{Departament de F\'isica, Universitat de les Illes Balears, IAC3 -- IEEC, Crta. Valldemossa km 7.5, E-07122 Palma, Spain}

\begin{document}

\preprint{LIGO-P1800267}


\title{Simple procedures to reduce eccentricity of binary black hole simulations}

\author{Antoni Ramos-Buades}
\affiliation{\UIB}
\author{Sascha Husa}
\affiliation{\UIB}
\author{Geraint Pratten}
\affiliation{\UIB}

\begin{abstract}
We present simple procedures to construct quasicircular initial data for numerical evolutions of binary black hole spacetimes.
        Our method consists of using post-Newtonian (PN) theory in three ways: first to provide an initial guess for the initial momenta at $3.5$PN order that implies
         low residual eccentricity, second to measure the resulting eccentricity, and third to calculate corrections to the momenta or initial separation that further reduce the eccentricity.
         Regarding the initial guess, we compare numerical evolutions in post-Newtonian theory to the postcircular and post-postcircular analytical approximations  to
         quasicircular data. We discuss a robust fitting procedure to measure eccentricity from numerical simulations using the orbital frequency $\Omega$, and    
          derive from the quasi-Keplerian parametrization at 1PN order the correction factors for the tangential and radial momentum components required to 
          reduce the measured eccentricity to zero.
          We first test our procedure integrating PN equations of motion at $3.5$PN where low eccentric initial data are easily obtained, and then apply our method to sets of binary black hole numerical relativity simulations with different mass ratios ($q=m_2/m_1=1,2,...,8$), spin configurations, and separations. Our set of simulations contains non-spinning, spin-aligned and precessing simulations. We observe that the iterative procedure produces low eccentric simulations with eccentricities of the order $\mathcal{O}\left(10^{-4}\right)$ with only one iteration. The simplicity of the procedure allows one to obtain low eccentric numerical relativity simulations easily and save computational resources. Moreover, the analytical PN formulas derived in this paper will be useful to generate eccentric hybrid waveforms.
\end{abstract}


\today

\maketitle

\acrodef{PN}{post-Newtonian}
\acrodef{EOB}{effective-one-body}
\acrodef{NR}{numerical relativity}
\acrodef{GW}{gravitational-wave}
\acrodef{BBH}{binary black hole}
\acrodef{BH}{black hole}
\acrodef{BNS}{binary neutron star}
\acrodef{NSBH}{neutron star-black hole}
\acrodef{SNR}{signal-to-noise ratio}
\acrodef{aLIGO}{Advanced LIGO}
\acrodef{AdV}{Advanced Virgo}

\newcommand{\PN}[0]{\ac{PN}\xspace}
\newcommand{\EOB}[0]{\ac{EOB}\xspace}
\newcommand{\NR}[0]{\ac{NR}\xspace}
\newcommand{\BBH}[0]{\ac{BBH}\xspace}
\newcommand{\BH}[0]{\ac{BH}\xspace}
\newcommand{\BNS}[0]{\ac{BNS}\xspace}
\newcommand{\NSBH}[0]{\ac{NSBH}\xspace}
\newcommand{\GW}[0]{\ac{GW}\xspace}
\newcommand{\SNR}[0]{\ac{SNR}\xspace}
\newcommand{\aLIGO}[0]{\ac{aLIGO}\xspace}
\newcommand{\AdV}[0]{\ac{AdV}\xspace}

\section{Introduction}\label{sec:introduction}

The  first detection of a gravitational wave (GW) signal \cite{PhysRevLett.116.061102} in 2015 by the LIGO detectors \cite{TheLIGOScientific:2014jea}, as well as the subsequent detections \cite{PhysRevLett.116.241103,Abbott:2017vtc,Abbott:2017gyy,Abbott:2017oio,TheLIGOScientific:2017qsa}, has been found to be consistent with models of the waveform emitted from
 the merger of compact objects under the assumption of quasicircularity of the binary's orbit prior to the merger. These models have been used to infer the parameters of the sources from the measured data; see e.g., the detailed discussion of parameter estimation results for the first detection \cite{TheLIGOScientific:2016wfe}. 
Indeed, efforts to model the gravitational wave signals from compact binary coalescence have to a large degree neglected eccentricity, as motivated by the efficient circularization 
of binaries as a  consequence of the emission of gravitational waves \cite{PhysRev.131.435,PhysRev.136.B1224}.

Only a decade before the first detection of gravitational waves, breakthroughs in numerical relativity (NR) \cite{Pretorius:2005gq,Campanelli:2005dd,Baker:2005vv} have made it possible to compute the evolution of binary black holes until the merger in general relativity (GR) and to extract the gravitational waves emitted from such systems.  Numerical simulations
of compact binaries are now performed routinely \cite{PhysRevLett.111.241104,Healy:2017psd,Jani:2016wkt,Bruegmann:2006at}, and models synthesized from numerical parameter studies and perturbative results are routinely used to analyze the data from the LIGO and Virgo detectors \cite{Husa:2015iqa,Khan:2015jqa,Damour:2008te,Bohe:2016gbl,Blackman:2017pcm}.

Initial data for numerical relativity simulations of black hole binaries are typically constructed in a five-step procedure, which can be roughly summarized as follows:
\begin{itemize}
\item[ \textbf{1)} ] One chooses the separation and the spin components.
\item[\textbf{2)}  ] One chooses the momenta or velocities of the black holes (BHs) such as to result in a low eccentricity. This step is usually guided by post-Newtonian (PN) approximations \cite{Blanchet2014}.
\item[\textbf{3)}  ] The constraint equations of general relativity are solved numerically for the chosen parameters, often using the approximation of conformal flatness.
\item[\textbf{4)}  ] The data are evolved numerically until the eccentricity can be estimated reliably from the corresponding oscillations in the separation, or orbital and gravitational wave frequency, as well as in other quantities. Residual eccentricity can lead to parameter biases when using the resulting waveforms for parameter estimation in gravitational wave analysis  and can complicate the construction of quasicircular waveform models from the numerical data.
In GR there is, however, no unique definition of eccentricity, and a specific quantity usually referred to as ``eccentricity estimator''  needs to be chosen, which reduces to the Newtonian concept of eccentricity in the Newtonian limit. 
Determining eccentricity from the orbital frequency $\Omega$, one would, for example, typically choose the eccentricity estimator 
\begin{equation}
e_\Omega = \frac{\Omega(t)-\Omega(e=0)}{2 \Omega(e=0)},
\label{eq:e_est_intro}
\end{equation}
which measures the time dependent oscillations in the orbital frequency relative to the case with vanishing eccentricity. The factor of 2 normalizes the  quantity $e_\Omega$ to be consistent eccentricity in radial oscillations (without the corresponding factor of 2).
\item[\textbf{5)}  ] A correction to the initial parameters is applied, and steps 2-5 (or 1-5) are applied until the eccentricity is deemed low enough for applications, taking into account the computational cost of short evolutions required to measure the eccentricity and the human effort to carry out or automatize the procedure. 

\end{itemize}

In this paper we first discuss steps two, four, and five, and then a version of changing the momenta in step two where we also correct for the initial coordinate separation, thus changing step one. In order to guess initial conditions, to determine the eccentricity of a numerical simulation, and to guess improved initial momenta, PN approximations at different orders in the PN expansion parameter $v/c$ are used. A key problem in relating post-Newtonian quantities to numerically constructed spacetimes are the different coordinate systems that are employed. For our numerical evolutions, we use the moving puncture approach \cite{Campanelli:2005dd,Hannam:2006vv,Bruegmann:2006at,Hannam:2008sg} with conformally flat Bowen-York initial data \cite{PhysRevD.21.2047}. The coordinates used to construct the initial data for the numerical relativity simulations are close to the Arnowitt-Deser-Misner transverse-traceless (ADMTT) coordinates
\cite{SchaeferADMTT} typically employed in the Hamiltonian approach to the post-Newtonian expansion. However,  the standard puncture data we employ are consistent with the PN description in the ADMTT gauge only up to order  $(v/c)^3$, see \cite{Tichy:2002ec,Yunes:2006iw,Yunes:2005nn}.  In \cite{Tichy:2010qa} it is argued that as a consequence, only low order PN expressions should be used in constructing low-eccentricity initial data. However, since high-order PN expressions are routinely used in modeling the gravitational wave signal from compact binaries, and therefore are readily available, in this work we take the point of view that it is simplest  to just use the highest PN order available to guess the initial momenta in step two. In addition, we show that while a low PN order expression of the radial initial momentum is enough to build low eccentric initial data, the tangential momentum benefits from the knowledge of high PN orders, and the higher the PN order the closer to the low eccentric value. 

The simplest post-Newtonian description of quasicircular (QC) initial parameters is to set the radial momentum to zero, which is inconsistent with an actual inspiral (at least in the absence of precession).  A straightforward way to improve the post-Newtonian description is to numerically solve the PN/effective-one-body (EOB) \cite{Buonanno:1998gg} dynamics from a larger separation down to the desired starting separation for a numerical  relativity simulation,  and to use the momenta read from this numerical calculation as input parameters to numerically solve the constraints  \cite{Husa:2007rh}. This procedure benefits from the fact that radiation reaction circularizes the orbit during the long inspiral, and for a sufficiently long inspiral, the eccentricity present in the PN data can be neglected. This will not lead to negligible eccentricity of the NR evolution due to the finite order used for the PN expansion, and the difference in the PN and NR coordinate systems as discussed above.
A second method \cite{Healy:2017zqj} specifies the values of the initial momenta at a given separation using analytical expressions at 3PN derived from a Hamiltonian formalism, which approximately take into account the radial momentum. In this work we follow the second approach, since it simplifies the construction of precessing initial data with chosen directions of the spins at a given separation. When numerically integrating the PN equations from a larger distance, constructing low eccentricity momenta with fixed spin directions would require an iteration of numerical integrations of the PN equations, which complicates setting up a grid of NR simulations to cover (portions of) the precessing parameter space. 

In Sec. \ref{sec:PNID} we discuss and compare these different approaches in more detail, as well as provide analytical formulas for the momenta in terms of initial separation, mass ratio and spins, including spin precession, updating the expressions presented in \cite{Healy:2017zqj} to $3.5$PN order. We also implement the post-postcircular (PPC) approximation  \cite{Damour:2007xr,Damour:2007yf} commonly used in the EOB theory and provide a recipe to compute it. This approximation consists of correcting analytically for the tangential momenta by iterating over the postcircular (PC) approximation. 

In Sec. \ref{sec:eccRedtheory} we develop the post-Newtonian methods to deal with steps four and five: we first discuss our procedure to determine the eccentricity of numerical data using the eccentricity estimator defined in Eq. \eqref{eq:e_est_intro}. Then, from the 1PN quasi-Keplerian parametrization \cite{1985AIHS...43..107D} we compute explicit  expressions for the correction factors for the tangential and radial momentum to achieve approximately vanishing eccentricity. Because of the deviations between the post-Newtonian equations and the full Einstein equations in the chosen gauge, as well as the noise that is present in numerical relativity simulations, this procedure may have to be iterated, although in many cases we find that a single step is sufficient for our purposes. Finally, we compute a similar formula that instead corrects the radial momentum and separation, thus directly compensating for the difference between the PN and NR coordinate systems.

We test our procedures in Sec. \ref{sec:ecc_red}, first applying them to post-Newtonian data, and check that the PPC approximation is indeed an excellent approximation to carry out full numerical solutions of the post-Newtonian inspiral. One practical application of such low-eccentricity post-Newtonian data is the construction of hybrid waveforms, where residual eccentricity in the post-Newtonian part leads to undesired oscillations \cite{GeraintPhenomX}. Finally, we apply our procedures to several precessing and nonprecessing numerical relativity simulations.

To date the most accurate procedures used to construct low eccentricity inspirals in numerical relativity are two iterative methods \cite{Pfeiffer:2007yz,Purrer:2012wy}. The method consists of running first a simulation with QC parameters, modifying the tangential and radial velocities of the simulation, and rerunning the simulation with the updated values. The iterative method in \cite{Pfeiffer:2007yz} is highly successful and can reduce eccentricities to below $10^{-5}$ in two iterations. Nevertheless, as discussed in \cite{Purrer:2012wy} its application to moving puncture simulations shows some difficulties. The iterative method presented in \cite{Purrer:2012wy} is designed for moving puncture simulation, but it is computationally expensive, and we have found it significantly more cumbersome than the method presented here .

We summarize and discuss our results in Sec.~\ref{sec:summary}.

Throughout this text we are working in geometric units $G=c=1$. To simplify expressions we will also set the total mass of the system $M=1$, and we 
 define the mass ratio $q=m_2/m_1$ with the choice $m_2 > m_1$, so that $q>1$. We also introduce the symmetric mass ratio $\eta=q/(1+q)^2$, and we will denote the black hole's dimensionless spin vectors by $\vec{\chi}_i= \vec{S}_i/m^2_i$, for $i=1,2$. 
\section{Post-Newtonian initial data} \label{sec:PNID}
We prepare initial data for our simulations within PN theory in the ADMTT gauge. We describe the particles in the center of mass (CM) frame, so that the motion of the two point particles can be described by the motion of one effective particle. We choose our $z$ axis in the direction of the initial orbital angular momentum, and initially locate the particles on the x axis  with $y=z=0$, and then locate $p_\phi=L_z$ with the standard definition of spherical polar coordinates. We define the tangential momentum as
\begin{equation}
p_t = \frac{p_\phi}{r}.
\label{eq11.1}
\end{equation}
Using the standard relation between Cartesian and polar coordinates one can write $(p_x, p_y)$ in terms of $(p_\phi, p_t)$ as
\begin{equation}
p_x=\frac{x p_r-y p_t}{\sqrt{x^2+y^2}}=p_r \text{cos} \phi - p_t \text{sin} \phi ,
\label{eq11}
\end{equation}
\begin{equation}
p_y=\frac{x p_t+ y p_r}{\sqrt{x^2+y^2}}=p_r \text{sin} \phi + p_t \text{cos} \phi .
\label{eq12}
\end{equation}

To compute the initial parameters we use the ADMTT Hamiltonian in the CM frame which is currently completely known up to $3.5$PN order,
\begin{equation}
H= H_{NS}+ H_{SO}+ H_{SS}+ H_{SSS},
\label{eq13.1}
\end{equation}
where $H_{NS}$ is the nonspinning part of the Hamiltonian, 
\begin{equation}
H_{NS}= H_{Newt}+H_{1PN}+ H_{2PN}+ H_{3PN}.
\label{eq13.2}
\end{equation}
The Hamiltonians in Eq. \eqref{eq13.2} can be found in \cite{Buonanno:2005xu}. The spin-orbit Hamiltonian is 
\begin{equation}
H_{SO}= H_{SO,1.5PN}+H_{SO,2.5PN}+ H_{SO,3.5PN}.
\label{eq13.3}
\end{equation}
The expressions for $H_{SO,1.5PN}$, $H_{SO,2.5PN}$ and $H_{SO,3.5PN}$ can be found in \cite{Buonanno:2005xu}, \cite{Damour:2007nc} and \cite{Hartung:2011te}, respectively. The spin-spin interaction Hamiltonian is
\begin{equation}
H_{SS}= H_{S^2,2PN}+H_{S_1S_2,2PN}+H_{S^2,3PN}+H_{S_1S_2,3PN},
\label{eq13.4}
\end{equation}
where explicit formulas for $H_{S^2,2PN}$ and $H_{S_1S_2,2PN}$ can be found in \cite{Buonanno:2005xu} while for $H_{S^2,3PN}$, $H_{S_1S_2,3PN}$ in \cite{Steinhoff:2008ji} and \cite{Steinhoff:2007mb}, respectively. Finally,  $H_{SSS}$ is given in \cite{Levi:2014gsa}.

\subsection{Postcircular approximation} \label{sec:pc}
Using the Hamiltonian of \eqref{eq13.1} one can compute the circular conditions for the orbit of the binary  in the  absence of radiation reaction:
\begin{equation}
p_r = 0, \quad \quad \quad 	\left( \frac{\partial H}{\partial r}\right)_{p_r=0}=0.
\label{26}
\end{equation}
Equation \eqref{26} gives a set of conditions to solve in PN order by order for $p_\phi(r)$. Once we have computed $p_\phi$, or equivalently $p_t(r)$, we can then compute 
\begin{equation}
\Omega=\left( \frac{\partial H}{\partial p_\phi}\right)_{p_r=0}
\label{27}
\end{equation}
and obtain an expression for the orbital frequency as a function of $r$. For completeness, we can also obtain an expression for the ADM mass defined by
\begin{equation}
M_{ADM}=M+H,
\label{28}
\end{equation}
where $M$ is the total mass and $H$ is the $3.5$PN Hamiltonian in ADMTT gauge. 
 
Taking into account Eqs.  \eqref{eq13.1}, and \eqref{26} - \eqref{28} we obtain explicit expressions for the orbital frequency, tangential momentum and ADM mass as a function of the orbital separation $r$. These expressions can be found in Appendix \ref{sec:AppendixA} and are given by Eqs. \eqref{29},  \eqref{30}, and  \eqref{31}.
The expression for the initial tangential momentum in terms of the orbital separation, Eq. \eqref{30}, is obtained from the conservative part of the dynamics. It remains to specify a value for the radial component of the momentum vector, $p_r$. The inclusion of the radiation reaction through the gravitational wave flux of energy allows us to derive an expression to compute $p_r$, following the procedure described in \cite{Buonanno:2005xu}. First, we consider the definition of the ADM mass given in  \eqref{28} for circular orbits,
\begin{equation}
M_{ADM}=M+H_{\text{circ}},
\label{34}
\end{equation}
where $H_{\text{circ}}$ is the energy corresponding to circular orbits, i.e., the Hamiltonian corresponding to Eq. \eqref{eq13.1} evaluated at the values of $p_r=0$ and $p_t$ derived in Appendix \ref{sec:AppendixA}. Taking a time derivative of \eqref{34} we get 
\begin{equation}
\frac{dM_{\text{ADM}}}{dt}=\frac{dM}{dt}+\frac{dH_{\text{circ}}}{dt}.
\label{35}
\end{equation}
The loss of ADM mass corresponds to a flux of gravitational wave energy leaving the binary, which has to be equal to the energy of the orbital motion plus the change in mass of the black holes. Consequently,
 \begin{equation}
-\frac{dE_{\text{GW}}}{dt}=\frac{dM}{dt}+\frac{dH_{\text{circ}}}{dt},
\label{36}
\end{equation} 
the derivative of the orbital energy can be rewritten as
 \begin{equation}
\frac{dH_{\text{circ}}}{dt}=\left(\frac{dr}{dt}\right)\left(\frac{dH_{\text{circ}}}{dr}\right).
\label{37}
\end{equation}
The expression for $dM/dt$ was derived in \cite{Brown:2007jx} for the spin-aligned or antialigned with respect to the orbital angular momentum. We use that expression taking into account the contribution related to the change in mass of the two black holes because the leading order term of $dM/dt$ is comparable in magnitude to a relative 2.5PN spin effect in the flux.  
The expression for the gravitational wave flux \cite{Blanchet2014, Fluxspinterms} in terms of the basic dynamical variables in ADM coordinates for quasicircular orbits can be found in Appendix \ref{sec:AppendixA}.

Then, we can use Hamilton's equations to compute the time derivative of the orbital separation as
\begin{equation}
\begin{split}
\frac{dr}{dt}=\frac{\partial H}{\partial p_r}.
\end{split}
\label{40}
\end{equation}
If we expand the right-hand side of Eq. (\ref{40}) for $p_r$ around $0$, we can solve it for $p_r$ and obtain a first order approximation to the radial momentum.
\begin{widetext}
\begin{equation}
\begin{split}
p_r &= \left[-\frac{dr}{dt} +\frac{1}{r^{7/2}} \left( -\frac{(6 q+13) q^2 \chi _{1x} \chi _{2 y}}{4 (q+1)^4}-\frac{(6 q+1) q^2 \chi _{2 x} \chi _{2 y}}{4 (q+1)^4}+\chi _{1y} \left(-\frac{q (q+6) \chi _{1x}}{4 (q+1)^4}-\frac{q (13 q+6) \chi _{2 x}}{4 (q+1)^4}\right)\right) + \frac{1}{r^4} \left( \chi _{1z} \left(\frac{3 q (5 q+2) \chi _{1x} \chi _{2 y}}{2 (q+1)^4} \right. \right.\right.
 \\
& \left. \left. \left.  -\frac{3 q^2 (2 q+5) \chi _{2 x} \chi _{2 y}}{2 (q+1)^4}\right)+\chi _{1y} \chi _{2 z} \left(\frac{3 q^2 (2 q+5) \chi _{2 x}}{2 (q+1)^4}-\frac{3 q (5 q+2) \chi _{1x}}{2 (q+1)^4}\right)\right)\right] \times \left[ -\frac{(q+1)^2}{q}-\frac{1 \left(-7 q^2-15 q-7\right)}{2 q r}  \right.
 \\
& \left.  -\frac{47 q^4+229 q^3+363 q^2+229 q+47}{8 q (q+1)^2 r^2} -\frac{1}{r^{5/2}}\left( \frac{\left(4 q^2+11 q+12\right) \chi _{1z}}{4 q (q+1)}+\frac{\left(12 q^2+11 q+4\right) \chi _{2 z}}{4 (q+1)} \right)    \right. \\
& \left. \left.- \frac{1}{r^{7/2}}  \left( \frac{\left(-53 q^5-357 q^4-1097 q^3-1486 q^2-842 q-144\right) \chi _{1z}}{16 q (q+1)^4}+\frac{\left(-144 q^5-842 q^4-1486 q^3-1097 q^2-357 q-53\right) \chi _{2 z}}{16 (q+1)^4} \right)\right. \right. \\
& \left. -\frac{1}{r^3} \left(\frac{\left(q^2+9 q+9\right) \chi _{1x}^2}{2 q (q+1)^2}+\frac{\left(3 q^2+5 q+3\right) \chi _{2 x} \chi _{1x}}{(q+1)^2}+\frac{\left(3 q^2+8 q+3\right) \chi _{1y} \chi _{2 y}}{2 (q+1)^2}-\frac{9 q^2 \chi _{2 y}^2}{4 (q+1)}+\frac{\left(3 q^2+8 q+3\right) \chi _{1z} \chi _{2 z}}{2 (q+1)^2}-\frac{9 q^2 \chi _{2 z}^2}{4 (q+1)} \right. \right.
\\
& \left. \left. +\frac{\left(9 q^3+9 q^2+q\right) \chi _{2 x}^2}{2 (q+1)^2}+\frac{-363 q^6-2608 q^5-7324 q^4-10161 q^3-7324 q^2-2608 q-363}{48 q (q+1)^4}-\frac{9 \chi _{1y}^2}{4 q (q+1)}-\frac{9 \chi _{1z}^2}{4 q (q+1)}-\frac{\pi ^2}{16} \right) \right]^{-1}. 
\end{split}
\label{41}
\end{equation}
\end{widetext}
The expression for $dr/dt$ can be computed combining Eqs. (\ref{36}) and (\ref{37}):
\begin{equation}
\begin{split}
\frac{dr}{dt}= \left[ \frac{dE_{\text{GW}}}{dt}\right] \left[ \frac{dH_{\text{circ}}}{dr}\right]^{-1}.
\end{split}
\label{41.2}
\end{equation}
The procedure to obtain a postcircular expression for the radial momentum can be summarized in the following algorithm:
\begin{enumerate}
\item[1)] Compute the circular expression for $p_t(r)$.
\item[2)] Use the expression for $p_t(r)$ and $p_r=0$ to compute $dH_{\text{circ}}/dr$.
\item[3)] Combine $dH_{\text{circ}}/dr$ with the gravitational wave flux for the quasicircular orbits, $dE_{\text{GW}}/dt$, to obtain $dr/dt$.
\item[4)] Use Hamilton's equations to compute $dr/dt=\partial H/\partial p_r$. Taylor expand at first order in $p_r$ around $p_r=0$ of the right-hand side and isolate $p_r$ as a function of $dr/dt$.
\item[5)] From step 4 compute an expression of $p_r$ using the value of $dr/dt$ calculated in step 3.
\end{enumerate}

\subsection{Post-postcircular approximation} \label{sec:ppc}

The post-postcircular approximation, first presented in \cite{Damour:2007yf}, keeps the value of the tangential momentum $p_t$ from the PC approximation, but applies  a further correction to the radial momentum $p_r$, and has been extensively used to construct initial data for EOB dynamics. 
We start with the postcircular values for $p_t$ and $p_r$ derived in the previous Sec. \ref{sec:pc} and define a bookkeeping parameter $\epsilon$ to arrange the orders of approximation, writing the tangential and radial momenta as
\begin{equation}
p_t =p^0_t+ \epsilon^2 p^2_t+ O \left( \epsilon^4\right),
\label{42}
\end{equation}
\begin{equation}
p_r = \epsilon p^1_r+ O\left(\epsilon^3\right).
\label{43}
\end{equation}
Here $p^0_t$ is the circular approximation, $p^1_r$ is the postcircular approximation, and $p^2_t$ is the post-postcircular value that we want to compute. The parameter $\epsilon$ is also related to the order of the radiation reaction terms of the $\phi$ coordinate in the PN equations of motion. 
 
Hamilton's equation for the radial momentum reads
\begin{equation}
\frac{dp_r}{dt}=-\frac{ \partial H}{\partial r}.
\label{44}
\end{equation} 
The left-hand side of Eq. \eqref{44}  can be approximated using the chain rule and the postcircular solution to 
\begin{equation}
\frac{dp_r}{dt}= \frac{dp_r}{dr}\frac{dr}{dt} \approx \frac{dp^1_r}{dr}\frac{dr}{dt}=\frac{dp^1_r}{dr}\frac{ \partial H}{\partial p_r}.
\label{45}
\end{equation}
Then, combining Eqs. \eqref{44}  and \eqref{45} we obtain
\begin{equation}
-\left[\frac{ \partial H}{\partial r}\right]_{p_r=p^1_r} \approx \left(\frac{dp^1_r}{dr}\right) \left[\frac{ \partial H}{\partial p_r} \right]_{p_r=p^1_r}.
\label{46}
\end{equation}
Given the values of the radial momentum $p_r$, the separation $r$, the masses of the particles $m_1$ and $m_2$ and the dimensionless spin vectors $\vec{\chi}$; one can solve Eq. \eqref{46} for $p_t$ using a numerical root finding method.

\section{Eccentricity reduction iteration} \label{sec:eccRedtheory}

In order to reduce the eccentricity further beyond the postcircular or post-postcircular initial data, we will now develop two methods that  iteratively reduce the eccentricity.
The first method corrects the initial momenta by factors $(\lambda_t$,$\lambda_r)$ such that $(p_t,p_r)\rightarrow (\lambda_t p_t,\lambda_r p_r)$, the 
second method corrects the initial separation by $\delta r$ such that $r \rightarrow  r+ \delta r  $, and the radial momentum $p_r$ as for the first method.
We will provide analytical expressions to compute $\lambda_t$, $\lambda_r$, and $\delta r$ in terms of the measured eccentricity  and an initial phase of the oscillations that characterize eccentricity; thus both methods are very straightforward to apply.

\subsection{Quasi-Keplerian 1PN equations of motion}\label{sec:QK1PN}

At 1PN order, bound orbits in the center of mass frame \cite{1985AIHS...43..107D} are described by
\begin{equation}
\begin{split}
n_t (t-t_0) &=u-e_t \sin u, \\
(\phi-\phi_0) &= (1+k) A_{e_\phi}(u), \\
A_{e_\phi}(u) & =2 \arctan\left[ \left( \frac{1+e_\phi}{1-e_\phi}\right)^{1/2} \tan\left(\frac{u}{2}\right)\right], \\
r &= a_r (1-e_r \cos u ), \\
\end{split}
\label{eq1}
\end{equation}
where $e_t$, $e_r$, and $e_\phi$ are the temporal, radial and angular eccentricities, $n_t$ is called the mean anomaly, $u$ is the true anomaly, and $k$ is the fractional periastron advance per orbit.

The frequency of the radial oscillations is directly related to the mean anomaly by
\begin{equation}
n_t = \Omega_r = 2 \pi / P_r,
\label{eq8.2}
\end{equation}
where $P_r$ is the time between two consecutive periastron passages. The average orbital frequency can be related to the radial oscillations by the expression
\begin{equation}
\Omega_\phi = (1+k) \Omega_r.
\label{eq8.3}
\end{equation}

The orbital quantities can be written in terms of the reduced energy, $E_n=E/\mu$, and angular momentum, $h=J/\mu$, where $E$ and $J=|\textbf{J}|$ are the respective dimensionful quantities and  $\mu=m_1 m_2/M$ is the reduced mass. Moreover,  defining $\gamma = c^{-2}$
at 1PN order the orbital elements can be written as
\begin{equation}
e^2_t=1+2 E_n \left(\gamma  E_n \left(\frac{17}{2}-\frac{7 \eta }{2}\right)+1\right) \left(h^2+\gamma  (2-2 \eta ) \right),
\label{eq122}
\end{equation}
\begin{equation}
e^2_\phi= 1+ 2 E_n \left(\gamma  E_n \left(\frac{\eta }{2}-\frac{15}{2}\right)+1\right) \left(h^2-6 \gamma  \right),
\label{eq222}
\end{equation}
\begin{equation}
e^2_r= 1+ 2 E_n \left(\gamma  E_n \left(\frac{5 \eta }{2}-\frac{15}{2}\right)+1\right) \left(h^2+\gamma  (\eta -6) \right),
\label{eq32}
\end{equation}
\begin{equation}
a_r= -\frac{ \left(1-\frac{1}{2} \gamma  E_n  (\eta -7)\right)}{2 E_n }.
\label{eq4}
\end{equation}
\begin{equation}
n_t= 2 \sqrt{2} (-E_n)^{3/2} \left( 1-\gamma \frac{E_n}{4} (\eta -15) \right).
\label{eq1000}
\end{equation}

The eccentricities $e_t$, $e_r$, and $e_\phi$ can be related to each other in terms of the fractional periastron advance, 
\begin{equation}
e_\phi = e_t \left[ 1-\frac{1}{3} \left(1-e_{t}^2\right) (\eta -4) k \right],
\label{eq5}
\end{equation}
\begin{equation}
e_r = e_t \left[1+ \frac{1}{6} \left(1-e_{t}^2\right) (8-3 \eta ) k \right],
\label{eq6}
\end{equation}
where the fractional periastron advance $k$ is defined as
\begin{equation}
k=\frac{h}{\sqrt{h^2-6 \gamma }}-1.
\label{eq7}
\end{equation}
Combining Eqs. \eqref{eq1000} and \eqref{eq7} we can get a relation between the mean anomaly and the fractional periastron advance,
\begin{equation}
k=3\gamma \frac{n^{2/3}_t}{1-e^2_t}.
\label{eq8}
\end{equation}
Note that this 1PN  parametrization does not take into account the spins of the particles, which only enter at higher PN order.

\subsection{Eccentricity measurement} \label{sec:eccMes}

The eccentricities $e_t, e_r, e_\phi$ introduced in \eqref{eq1} determine the amplitude of oscillations in the orbital quantities relative to the noneccentric values. At Newtonian order the three eccentricities agree, but they differ in general, starting at 1PN.  For general solutions, such as those obtained from numerical relativity, it is useful to define eccentricity estimators as time dependent functions that measure the relative deviation from the noneccentric case, normalized to agree with the eccentricities $e_t, e_r, e_\phi$  at Newtonian order.
For this work, for simplicity, we will only use the eccentricity estimator for the orbital frequency,
\begin{equation}
e_\Omega = \frac{\Omega(t)-\Omega(e=0)}{2 \Omega(e=0)}.
\label{001}
\end{equation}
Here $ \Omega(t)=d\phi/dt$ can be obtained from the coordinate motion of the orbiting objects, and $ \Omega(e=0)$ refers to the orbital frequency setting the eccentricities to zero.
For examples of using eccentricity estimators for other quantities, related to the orbital dynamics or gravitational wave signal, see \cite{Mroue:2010re}, and for a discussion of eccentricity estimators, in particular the differences between using the strain or Newman-Penrose scalar $\Psi_4$, see  \cite{Purrer:2012wy}.

In this work, we choose an orbital quantity as our eccentricity estimator for simplicity and to save computational resources for numerical relativity simulations: Using wave quantities such as the strain or $\psi_4$ requires longer numerical evolutions to allow the waves to travel to the extraction sphere. Also, obtaining a clean wave signal for the first few orbits, where eccentricity reduction is typically applied, may require significant computational effort to carry out the simulations, or effort to postprocess and denoise the signal \cite{Purrer:2012wy}. However, the methods developed in this paper can easily be reused together with other eccentricity estimators. Among  quantities related to the orbital dynamics, the orbital frequency is convenient due to its weak gauge dependence, e.g., compared to the separation.

In the context of numerical data, obtained from a numerical relativity simulation or numerical evolution of the PN EOM,  $\Omega (e=0) $ could be represented by data from a simulation corresponding to negligible eccentricity (which is straightforward to achieve for PN solutions by starting at a very large separation), or be determined by a fit to the numerical data, 
$\Omega^{0}_{\text{fit}}(t)$, which does not contain oscillating terms corresponding to eccentricity (which is common practice in numerical relativity).

A simple way to fit the secular orbital frequency evolution as a function of time, averaging out oscillations due to eccentricity, coordinate gauge, or numerical artifacts, over a small number of cycles is to use a low-order polynomial of coordinate time, however, such fits typically look pathological outside of the fitting interval and are prone to pick up the oscillations due to eccentricity, gauge effects,  or spin evolution,  when using too many terms in the attempt of creating an accurate fit.
A natural  ansatz that avoids these problems uses the orbital frequency evolution of a noneccentric binary  in the form of the TaylorT3 quasicircular PN approximant \cite{Buonanno:2009zt,taylorT3spinning}. For the same reasons a similar fitting strategy has been used in \cite{Buonanno:2010yk}.  There however, only two PN-like terms are used, with all coefficients  
determined by the fit. Here instead we use all known PN terms up to third PN order, and our ansatz $A_0$  for the quasicircular frequency  evolution is
\begin{equation}
A_0= \frac{a \, \theta ^3}{8}  \left(1+ b_1 \theta ^2+ b_2 \theta ^3+ b_3 \theta ^4+ b_4 \theta ^5+ b_5 \theta ^6\right),
\label{eq550}
\end{equation}
where the known coefficients $b_i$ as determined by PN theory are listed in Appendix \ref{sec:AppendixB},  $\theta$ is defined as 
\begin{equation}
\theta =\left[ \frac{\eta}{5}  \left| t_{\text{max}} t_0-t\right| \right]^{-1/8},
\label{eq551}
\end{equation}
and we fit two parameters, $a$ and  $t_0$. To accelerate the convergence of the fit, $t_{\text{max}}$ is chosen of the order of the merger time of the numerical simulation, thus  $t_0$ is of order unity. The parameter $a$ would be unity in PN theory, and fitting it leads to an unphysical low frequency behavior, which would be inappropriate for waveform modeling purposes.
For our application however, we are only interested in the timescale corresponding to a numerical simulation, no inconsistency arises, and we find that our choice of fitting parameters leads to robust and accurate fits.

Once we have obtained a noneccentric fit to our numerical data, we can measure eccentricity by fitting the data using an extended ansatz $A_e$, which adds a sinusoidal function to the noneccentric ansatz $A_0$,
\begin{equation}
A_e= A_0+ e (1+ \left| k_1 \right| t) \cos \left[   (1+ t \left| k_2 \right|) \Omega_1 \Omega_0  t+t_1\right].
\label{eq552}
\end{equation}
Here $\Omega_0$ is the quasicircular value given by Eq. \eqref{29}, and the coefficients to fit  are $a$, $t_0$, $e$, $\Omega_1$, $k_1$, $k_2$ and  $t_1$. The coefficients $k_1$ and $k_2$ have been added to capture the decreasing eccentricity during the inspiral. In this work the fits have been performed using the function \texttt{NonlinearModelFit} from \texttt{Mathematica} with a global minimization method to avoid problems related to fitting the behavior corresponding to local minima of the data. We have found the differential evolution method of the \texttt{NonlinearModelFit} function to result in particularly robust fits. 

Furthermore, we have tested this procedure to measure the eccentricity of genuinely eccentric NR simulations, and we found accurate measurements up to eccentricities $e_t=0.1$. For higher eccentricities the measurements are inaccurate due to the fact that the single harmonic function of the ansatz of Eq. \eqref{eq552} is not able to reproduce the high peak amplitudes in the orbital frequency. As a solution one should replace the single harmonic function in Eq. \eqref{eq552}  by a sum of different harmonics in order to correctly capture the amplitude of those peaks. However, for the purposes of the present paper we found an ansatz with a single harmonic function sufficiently accurate, and we leave extensions of this measurement procedure to the high eccentricity limit for future work.

\subsection{Tangential momentum correction from quasi-Keplerian parametrization} \label{sec:tangentialQK1PN}

In order to reduce the eccentricity resulting from the choice of initial momenta, we need to know how much the momentum changes from its quasicircular value as a function of eccentricity.
We can split the momentum into a tangential and radial part, and will first compute the dependence of the tangential momentum component on the orbital eccentricity  $e_\Omega$ at 1PN order. 

We start by using Eq.~\eqref{eq1} to compute $e_\Omega$ as a function of the eccentricities $e_t$ and $e_\phi$ defined in Eqs.~\eqref{eq122} and ~\eqref{eq222},
\begin{equation}
e_\Omega = f\left(e_t, e_\phi \right).
\label{002}
\end{equation}
From the equations of motion \eqref{eq1} it is straightforward to write at 1PN the radial coordinate, $r$, and the orbital  frequency, $\Omega=\dot{\phi}$ up to linear order in eccentricity as
\begin{equation}
r = a_r (1-e_r \cos\left[ \Omega_r t \right] ),
\label{eq9}
\end{equation}
\begin{equation}
\Omega \equiv \dot{\phi} = \Omega_\phi \left( 1+(e_\phi + e_t) \cos\left[ \Omega_r t \right]  \right).
\label{eq10}
\end{equation}
Combining Eqs. \eqref{eq10} and \eqref{001}, we get the following expression for the orbital frequency estimator
\begin{equation}
e_\Omega  = \frac{e_\phi+e_t}{2}.
\label{eq511}
\end{equation}

We now proceed as follows:
\begin{enumerate}
\item[a)] In Eq. \eqref{eq511} write the eccentricities $e_\phi$, $e_t$ in terms of the energy and the angular momentum using the quasi-Keplerian solution of the compact binaries in eccentric orbits.
\item[b)]  Write the eccentricities, energy, and angular momentum in terms of the $p_r$ and $p_t$ using the Hamiltonian and the angular momentum expressions in ADM coordinates.
\item[c)] Multiply the momenta by the factors  $\lambda_t$ and $\lambda_r$.
\item[d)] Substitute the values of $p_t$ and $p_r$ by the circular ones.
\item[e)] Taylor expand Eq. \eqref{eq511} in powers of $(\lambda_t-1)$ and $(\lambda_r-1)$ up to linear order in $(\lambda_t-1)$ and $(\lambda_r-1)$.
\item[f)] Solve for $\lambda_t$, setting $\lambda_r=1$. 
\end{enumerate}

Using the fact that the energy and the total angular momentum can be written in terms of the momenta $p_t$ and $p_r$, and inserting those expressions into the definitions of Eqs. \eqref{eq222} and \eqref{eq122} we get at 1PN order
\begin{widetext}
\begin{align}
e_t  &=\sqrt{\frac{\eta ^4+r p_t^2 \left[r \left(p_r^2+p_t^2\right)-2 \eta ^2\right]}{\eta ^4}}+\gamma\left[\frac{\eta ^4+r p_t^2 \left(r \left(p_r^2+p_t^2\right)-2 \eta ^2\right)}{\eta ^4}\right]^{-1/2} \times \left[-\frac{(\eta -4) r^2 p_t^6}{2 \eta ^6}\right.\nonumber \\
& \left.-\frac{(\eta -4) r^2 p_r^4 p_t^2}{2 \eta ^6}+p_r^2 \left(\frac{1-\eta }{\eta ^2}-\frac{(\eta -4) r^2 p_t^4}{\eta ^6}+\frac{5 (\eta -4) r p_t^2}{2 \eta ^4}\right)+\frac{(3 \eta -10) r p_t^4}{\eta ^4}+\frac{(20-9 \eta ) p_t^2}{2 \eta ^2} \right. \nonumber\\ 
& \left. +\frac{2 (\eta -1)}{r}\right],
\label{eq512}
\end{align}
\begin{align}
e_\phi  &= \sqrt{\frac{\eta ^4+r p_t^2 \left[r \left(p_r^2+p_t^2\right)-2 \eta ^2\right]}{\eta ^4}} +\gamma\left[ \frac{\eta ^4+r p_t^2 \left[r \left(p_r^2+p_t^2\right)-2 \eta ^2\right]}{\eta ^4} \right]^{-1/2} \times \left[\frac{(\eta -4) r^2 p_t^6}{2 \eta ^6}\right. \nonumber \\
& \left.+\frac{(\eta -4) r^2 p_r^4 p_t^2}{2 \eta ^6}+p_r^2 \left(-\frac{3}{\eta ^2}+\frac{(\eta -4) r^2 p_t^4}{\eta ^6}-\frac{3 (\eta -4) r p_t^2}{2 \eta ^4}\right)-\frac{(\eta -6) r p_t^4}{\eta ^4}+\frac{(\eta -20) p_t^2}{2 \eta ^2}+\frac{6}{r}\right].
\label{eq513}
\end{align}
\end{widetext}
Then, we make the substitutions
\begin{equation}
p_t \rightarrow \lambda_t p_t, \quad \quad p_r \rightarrow \lambda_r p_r.
\label{eq514}
\end{equation}
If we replace Eqs. \eqref{eq512} and \eqref{eq513} into \eqref{eq511}, Taylor expand around $\lambda^0_t=1$, and use the circular value solutions of $p_t$ and $p_r=0$, we obtain at 1PN order
\begin{equation}
e_\Omega= 2 (\lambda_t-1)+\gamma  (\lambda_t-1) \left(\frac{2 \eta }{r}+\frac{4}{r}\right).
\label{eq516}
\end{equation}
We can invert Eq. \eqref{eq516} to obtain an expression for $\lambda_t$ in terms of the eccentricity estimator
\begin{equation}
\lambda_t= 1+\frac{e_\Omega}{2}-\gamma \frac{ e_\Omega }{2 r}(\eta +2).
\label{eq517}
\end{equation}
Equation \eqref{eq517} directly relates the eccentricity of the simulation to the correction factor of $p_t$, at 1PN order, and linear in eccentricity, we can thus read off the momentum correction factor $\lambda_t$ directly from the value of the measured eccentricity.

Although this equation has been derived in the low eccentricity limit, it can be used to generate approximate eccentric initial data for NR simulations. Given a configuration described by the masses of the particles, the spins, the initial linear momenta and the orbital separation, one can choose an initial eccentricity of the simulation and then obtain how much one has to change the tangential momentum to generate that eccentric simulation. 

The computation of $\lambda_t$ in \eqref{eq517} solves the one parameter problem of correcting $p_t$ to reduce the eccentricity. However, the reduction of the eccentricity is a two-dimensional problem in the absence of precession. In the precessing case, eccentricity reduction is in principle a  three-dimensional problem; however, it appears that no correction to the small out-of-the orbital-plane momentum is necessary at the current level of accuracy, so we restrict ourselves to a two-dimensional method. We have previously used
a different two-dimensional method that uses PN information (see \cite{Purrer:2012wy}); however, our new method is significantly simpler to apply. There is a threshold of how much the eccentricity can be reduced correcting only $p_t$, which we find typically around $10^{-3}$ for the cases we consider. Hence, one needs to correct not only $p_t$ but also $p_r$ if one wants to efficiently reduce the eccentricity, and we develop a two-parameter method in the next section.

\subsection{Correcting both tangential and radial momenta from 1PN residuals} \label{sec:momentaRes}

We will describe the relative oscillations in the orbital frequency by the ansatz 
\begin{equation}
\begin{split}
\mathcal{R}_\Omega= A+B \cos \left( \Omega_r t + \Psi \right),
\end{split}
\label{eq534}
\end{equation}
where $\Omega_r$ is the frequency of the radial oscillations, and $A$, $B$ and $\Psi$ are coefficients to be determined. 

We will now derive explicit formulas in terms of the amplitude $B$ and the phase $\Psi$ of the ansatz \eqref{eq534}
to rescale both the tangential momentum by $\lambda_t$, and the radial momentum by a factor $\lambda_r$, in order to reduce the eccentricity
resulting from the choice of initial data. In order to do that we compute the residual of the orbital frequency, i.e., the difference between the configuration perturbing $p_r$ and $p_t$ and the unperturbed configuration with zero eccentricity. To our knowledge, the effects of perturbing such a residual were first studied in \cite{Purrer:2012wy}.

We will assume that the total residual is a linear combination of the residual, $\mathcal{R}^{\lambda_t p_t^0}_{\Omega} $, computed perturbing only $p_t^0$, the residual, $\mathcal{R}^{\lambda_r p_r^0}_{\Omega}$, calculated just perturbing $p_r^0$; and the residual, $ \mathcal{R}^{\lambda_t p_t^0,\lambda_r p_r^0}_{\Omega}$, computed perturbing both momenta. In the rest of the section we are going to compute these three residuals and obtain from them analytical expressions for the correction factors $ \left(\lambda_t,\lambda_r \right)$.

We start writing the residual corresponding to a perturbation $\lambda_t$ of the initial tangential momentum $p^0_t$, 
\begin{equation}
\mathcal{R}^{\lambda_t p^0_t}_\Omega= \Omega^{\lambda_t p^0_t}- \Omega^{p^0_t}.
\label{eq518}
\end{equation}
In Eq. \eqref{eq518}, $\Omega \equiv \Omega (t)$  refers to Eq. \eqref{eq10}, the analytical 1PN solution at linear order in eccentricity. The magnitude of the eccentricities we are working with, usually well below $10^{-2}$, justifies taking just the linear order in eccentricity in the equations of motion.

Note that $\Omega_\phi$ in Eq. \eqref{eq10} also depends on $p_t$. Therefore, we begin computing the effect of perturbing $p_t$ in $\Omega_\phi$. Combining Eqs. \eqref{eq8}, \eqref{eq8.2}, and \eqref{eq8.3} we obtain the following expression
\begin{equation}
\Omega_\phi= \left(1+3\gamma \frac{n^{2/3}_t}{1-e^2_t} \right) n_t.
\label{eq519}
\end{equation}
We can now use Eqs. \eqref{eq122} and \eqref{eq1000} to write $\Omega_\phi$ in terms of the energy and the angular momentum, which at the same time can be written in terms of the radial and tangential momenta. Then, we perturb the tangential momentum a factor $\lambda_t$ and we Taylor expand up to linear order in $\lambda_t$ around $\lambda^0_t=1$. As a result we obtain
\begin{equation}
\Omega^{\lambda_t}_\phi=  \gamma  \left(\frac{(-5 \eta -9) \lambda_t}{2 r_0^{5/2}}+\frac{6 \eta +6}{2 r_0^{5/2}}\right)-\frac{3 \lambda_t}{r_0^{3/2}}+\frac{4}{r_0^{3/2}}.
\label{eq520}
\end{equation}
Defining $\Omega_0 = r^{-3/2}_0$ as the Newtonian-like orbital frequency we can rewrite \eqref{eq520} as
\begin{equation}
\Omega^{\lambda_t}_\phi=  \gamma  \Omega _0 \left(\frac{3 (\eta +1)}{r_0}-\frac{(5 \eta +9) \lambda _t}{2 r_0}\right)+\Omega _0 \left(4-3 \lambda _t\right).
\label{eq521}
\end{equation}
For the expression of the unperturbed $\Omega_\phi$ we will use the analytical circular solution, Eq. \eqref{29}, which coincides with the unperturbed expression of the orbital frequency $\Omega$, assuming $p^0_t$ and $p^0_r$ are given by the circular values,
\begin{equation}
 \Omega^0_\phi=\Omega^{p^0_t}=\Omega_0 \left[1+\frac{\gamma  (\eta -3)}{2 r_0}\right].
\label{eq522}
\end{equation}
The perturbed configuration is calculated replacing \eqref{eq521} in \eqref{eq10} to obtain
\begin{equation}
\begin{split}
\Omega^{\lambda_t p^0_t} &=\Omega _0\left[1+\left(\lambda _t-1\right) \left(4 \cos (\Omega_r t)-3\right)\right]+\gamma \Omega _0 \left[\frac{(\eta -3) }{2 r_0} \right.
\\
& \left. + \left(\lambda _t-1\right) \left(\frac{(6 \eta +2) \cos (\Omega_r t)}{r_0}-\frac{5 \eta +9}{2 r_0}\right)\right] \\
&+ \mathcal{O}\left((\lambda_t-1)^2 \right).
\end{split}
\label{eq524}
\end{equation}
Replacing Eqs. \eqref{eq522} and \eqref{eq524} in Eq. \eqref{eq518}, we finally obtain
\begin{equation}
\begin{split}
\mathcal{R}^{\lambda_t p^0_t}_\Omega &=\Omega _0 \left(\lambda _t-1\right) (4 \cos (\Omega_r t)-3)
\\
 &+\frac{\gamma  \Omega _0 \left(\lambda _t-1\right) (4 (3 \eta +1) \cos (\Omega_r t)-5 \eta -9)}{2 r_0} \\
 &+ \mathcal{O}\left((\lambda_t-1)^2 \right).
\end{split}
\label{eq525}
\end{equation}
We can follow the same procedure to obtain the residual corresponding to just perturbing $p^0_r$. 
We will expand now in powers of $(\lambda_r-1)$ and we will maintain $p^0_r$ in the expressions for a better comparison with the formulas of \cite{Purrer:2012wy}. In practical computations, $p^0_r$ will be replaced by its postcircular value. Note that in the following derivation of the residual, $\Omega_\phi$ does not depend on $p_r$. Another important fact is that Eqs. \eqref{eq1} assume that the motion starts at the periastron, $\phi_0=0$, and this condition combined with the negative value of $p^0_r$ that the postcircular approximation yields causes a shift of the periastron by $\pi/2$. Consequently, the radial perturbations will be dominated by a sine mode \cite{Purrer:2012wy}.

As in Eq. \eqref{eq518} we can write the residual as 
\begin{equation}
\mathcal{R}^{\lambda_r p^0_r}_\Omega=\Omega^{\lambda_r p^0_r}_0 -\Omega^{p^0_t, p^0_r}_0.
\label{eq526}
\end{equation}
In Eq. \eqref{eq526}, $\Omega^{p^0_r}_0$ is given by the unperturbed configuration assuming a nonzero value of $p^0_r$. 

The calculations to obtain $\Omega^{p^0_t, p^0_r}_0$ are the following:
\begin{enumerate}
\item[1)] Write $e_t$ and $e_\phi$ in Eq. \eqref{eq10} in terms of $E_n$ and $h$.
\item[2)] Write $E_n$ and $h$ in terms of $ p^0_t$ and $p^0_r$.
\item[3)] Substitute the value of $ p^0_t$ by Eq. \eqref{30}.
\end{enumerate}
The result of applying steps $(1) - (3)$ is
\begin{equation}
\Omega^{p^0_t, p^0_r}_0=\Omega_0 \left( 1-\frac{ 2 r_0^{1/2}  \left| p^0_r\right|}{\eta  } \text{sin} (\Omega_r t) \right)+\gamma\frac{2   \Omega _0  \left| p^0_r\right| }{\eta  r^{1/2}_0} \text{sin} (\Omega_r t).
\label{eq527}
\end{equation}

The recipe to obtain the perturbed configuration is quite similar with some additional steps:
\begin{enumerate}
\item[a)] Write $e_t$ and $e_\phi$ in Eq. \eqref{eq10} in terms of $E_n$ and $h$.
\item[b)] Write $E_n$ and $h$ in terms of $ p^0_t$ and $\lambda_r p^0_r$.
\item[c)] Substitute the value of $ p^0_t$ by Eq. \eqref{30}.
\item[d)] Taylor expand up to linear order in $(\lambda_r-1)$.
\end{enumerate}
As a result of performing steps $(\text{a}) - (\text{d})$ we obtain 
\begin{equation}
\begin{split}
\Omega^{\lambda_r p^0_r}_0 &=\Omega _0+\gamma\frac{2   \Omega _0 \lambda _r \text{sin}  (\Omega_r t) \left| p^0_r\right| }{\eta  r^{1/2}_0}-\frac{2 r^{1/2}_0 \Omega _0 \lambda _r \text{sin}  (\Omega_r t) \left| p^0_r\right|}{\eta }\\
 &+ \mathcal{O}\left( (\lambda_r-1)^2 \right).
 \end{split}
\label{eq528}
\end{equation}
Combining Eqs. \eqref{eq527} and \eqref{eq528} we now get
\begin{equation}
\begin{split}
\mathcal{R}^{\lambda_r p^0_r}_\Omega &= \frac{2\Omega_0 \left| p^0_r\right|  }{\eta} \left(  r^{1/2}_0  -\gamma r^{-1/2}_0\right)  \left(\lambda _r-1\right)  \text{sin} (\Omega_r t)\\
 &+ \mathcal{O}\left( (\lambda_r-1)^2 \right) .
\end{split}
\label{eq529}
\end{equation}

The next step of the calculation is computing the residual produced by the simultaneous perturbation of $p^0_t$ and $p^0_r$. The procedure to follow is quite similar to the algorithms presented so far. The residual we want to calculate is
\begin{equation}
\mathcal{R}^{\lambda_t p^0_t,\lambda_r p^0_r}_\Omega= \Omega^{\lambda_t p^0_t, \lambda_r p^0_r}_0 -\Omega^{p^0_t, p^0_r}_0,
\label{eq530}
\end{equation}
where $\Omega^{p^0_t, p^0_r}_0$ is given by Eq. \eqref{eq527}. The procedure we follow to compute the residual is summarized as
\begin{itemize}
\item[A)] Write $e_t$ and $e_\phi$ in Eq. \eqref{eq10} in terms of $E_n$ and $h$.
\item[B)] Write $E_n$ and $h$ in terms of $\lambda_t p^0_t$ and $\lambda_r p^0_r$.
\item[C)] Substitute the value of $ p^0_t$ by Eq. \eqref{30} and maintain the value of $p^0_r$.
\item[D)] Taylor expand up to linear order in $(\lambda_t-1)$ and $(\lambda_r-1)$.
\end{itemize} 
After following steps $(\text{A}) - (\text{D})$ we obtain
\begin{equation}
\begin{split}
\Omega^{\lambda_t p^0_t, \lambda_r p^0_r}_0 &=\Omega _0+\frac{2 \sqrt{r_0} \Omega _0 \lambda _r \lambda _t  \left| p^0_r\right| }{\eta }\text{sin} (\Omega_r t)+ \gamma \frac{2 \Omega _0 \lambda _r \left| p^0_r\right|}{ \eta  r^{1/2}_0} \left[ (\eta \right. 
\\
& \left.  +1) \lambda _t  - (\eta +2) \right]\text{sin} (\Omega_r t)+ \mathcal{O}\left( (\lambda_r-1)^2 \right)  \\
& +\mathcal{O}\left( (\lambda_t-1)^2 \right)+ \mathcal{O}\left( (\lambda_t \lambda_r)^2 \right) .
\end{split}
\label{eq531}
\end{equation}

Inserting Eqs. \eqref{eq527} and \eqref{eq531} into \eqref{eq530} gives
\begin{equation}
\begin{split}
\mathcal{R}^{\lambda_t p^0_t,\lambda_r p^0_r}_\Omega & =\frac{2 r^{1/2}_0 \Omega _0}{\eta } \text{sin} (\Omega_r t) \left| p^0_r\right|  \left(\lambda _r \lambda _t-1\right)  \\
& +\gamma \frac{2  \Omega _0 }{\eta  r^{1/2}_0}\text{sin} (\Omega_r t) \left| p^0_r\right| \left[\lambda _r \left(\eta  \left(\lambda _t-1\right)+\lambda _t-2\right)+1\right] .
\end{split}
\label{eq532}
\end{equation}
Finally, the total residual at 1PN can be understood as the sum of \eqref{eq525}, \eqref{eq529}, and \eqref{eq532}, and this is
 
\begin{equation}
\begin{split}
\mathcal{R}^{1PN}_\Omega & =\mathcal{R}^{\lambda_t p^0_t}_\Omega+\mathcal{R}^{\lambda_r p^0_r}_\Omega+\mathcal{R}^{\lambda_t p^0_t,\lambda_r p^0_r}_\Omega  \\
&= -3 \Omega _0 \left(\lambda _t-1\right)-\frac{\gamma  (5 \eta +9) \Omega _0 \left(\lambda _t-1\right)}{2 r_0}+ \text{sin} (\Omega_r t) 2 \Omega _0 \\
&\times \left| p^0_r\right| \left[\frac{\sqrt{r_0}   (\lambda _r-1 )}{\eta}+\gamma  \left(   \frac{ (\lambda _r  (\eta  (\lambda _t-1)+\lambda _t-2)+1}{ \eta  \sqrt{r_0}} \right.\right. \\
& \left. \left. -\frac{ (\lambda _r-1)  }{ \eta \sqrt{r_0}}\right)  +\sqrt{r_0}   \frac{\left(\lambda _r \lambda _t-1\right)}{\eta}\right] +\text{cos} (\Omega_r t) \left[ 4 \Omega _0 \left(\lambda _t-1\right) \right. \\
&  \left. +\gamma  \left(\frac{6 \eta  \Omega _0 \left(\lambda _t-1\right)}{r_0}+\frac{2 \Omega _0 \left(\lambda _t-1\right)}{r_0}\right)\right].
\end{split}
\label{eq533}
\end{equation}

Once we have derived  expression \eqref{eq533} for the residual, we want to compare it to \eqref{eq534} in order to obtain expressions of $\lambda_t$  and $\lambda_r$ in terms of the amplitude and the phase of the residual. We do not take into account the offset terms because the 1PN order is not accurate enough to describe  the full PN dynamics and even less the dynamics of the full Einstein equations dynamics of a NR simulation.

The total residual, Eq. \eqref{eq533}, is a sum of sine and cosine terms that we want to express as a single cosine plus a phase as in Eq. \eqref{eq534}. The result of such a transformation gives two expressions for the amplitude $B$ and the phase $C$ in terms of $\lambda_t$ and $\lambda_r$,
\begin{equation}
B =  \left[ a^2_1+a^2_2\right]^{1/2} ,
\label{eq535}
\end{equation}
\begin{equation}
 \Psi =  \arctan\left(a_1/a_2 \right), 
\label{eq536}
\end{equation}
where $a_1$ and $a_2$ are given by
\begin{equation}
a_1 =4 \Omega _0 \left(\lambda _t-1\right)+ \frac{2 \gamma  (3 \eta +1) \Omega _0 \left(\lambda _t-1\right)}{r_0},
\label{eq537}
\end{equation}
\begin{equation}
\begin{split}
a_2 &=\frac{2 \sqrt{r_0} \Omega _0 \left(\lambda _r-1\right) \left| p^0_r\right| }{\eta }+\gamma \frac{2 \Omega _0 \left| p^0_r\right| }{r^{1/2}_0 \eta } \left[ \lambda _r \left(\eta  \left(\lambda _t-1\right)+\lambda _t \right. \right.
\\
& \left. \left. -2\right)+1 -  \left(\lambda _r-1\right)   \right]+\frac{2 \Omega _0 \left| p^0_r\right|  \left(\lambda _r \lambda _t-1\right)}{\eta  r_0^{-1/2}}.
\end{split}
\label{eq538}
\end{equation}
The solution of the Eqs. \eqref{eq537} and \eqref{eq538} consistently at 1PN order for $\lambda_r$ and $\lambda_t$ provides the  formulas
\begin{equation}
\lambda_t = 1+\left[\frac{B}{4 \Omega _0}-\gamma \frac{B  (3 \eta +1)}{8 r_0 \Omega _0}\right]\cos\Psi,
\label{eq539}
\end{equation}
\begin{equation}
\begin{split}
\lambda_r = 1+  \frac{B \eta}{2 r_0^{1/2} \Omega _0 \left| p^0_r\right|}    \left[  1    +\gamma \frac{  1   }{ r_0 } \right]\sin\Psi .
\end{split}
\label{eq540}
\end{equation}
Equations \eqref{eq539} and \eqref{eq540} can be used to compute the corrections of the momenta from a measured  eccentricity oscillation amplitude $B$ and phase shift $\Psi$.
The accuracy of the procedure is limited by carrying out the computations at 1PN order, but more importantly by the noise in numerical relativity data, due to both numerical and gauge artifacts.

\subsection{Separation correction from 1PN residual}  \label{sec:sepRes}

We will now develop an alternative method of eccentricity reduction, where we replace the correction of the tangential  momentum with a correction of the coordinate separation  where the
NR momentum is identified with the PN momentum. This is motivated  by the fact that the PN and NR coordinates for the initial data only agree to 2PN order 
\cite{Tichy:2002ec,Yunes:2006iw,Yunes:2005nn}, and we will again calculate the required correction  to the initial orbital separation of the binary at 1PN order.

We compute the residual coming from the variation $\delta r$ of the initial separation given by
\begin{equation}
\begin{split}
\mathcal{R}^{\delta r +r_0 }_\Omega= \Omega^{r_0 +\delta r }_0 -\Omega^{r_0}_0.
\end{split}
\label{eq541}
\end{equation}
In Eq. \eqref{eq541},  $\Omega^{r_0}_0$ is the unperturbed configuration, which is computed assuming that $p^0_t$ and $p^0_r$ take the circular values. We obtain
\begin{equation}
\Omega^{r_0}_0= \Omega _0\left[1+\gamma \frac{  (\eta -3)}{2 r_0} \right],
\label{eq542}
\end{equation}
where $\Omega_0=r^{-3/2}_0$ is the Newtonian-like orbital frequency. To compute the perturbed term, we need to calculate first the effect of perturbing the initial separation in $\Omega_\phi$. The calculation is similar to the one performed in Sec. \ref{sec:momentaRes}. We make the replacement
\begin{equation}
r_0 \rightarrow r_0+\delta r,
\label{eq543}
\end{equation}
and expand in Taylor series around $\delta r=0$ up to linear order in $\delta r$. As a result we obtain
\begin{equation}
\begin{split}
\Omega^{r_0+ \delta r }_\phi =\Omega _0  \left[ 1-\frac{3 \delta r}{2 r_0} -\gamma \frac{  (\eta -3) \left(5 \delta r-2 r_0\right)}{4 r_0^2} \right].
\end{split}
\label{eq544}
\end{equation}
Then, for the perturbed configuration we obtain
\begin{equation}
\begin{split}
\Omega^{\delta r +r_0 }_0 & =\Omega _0 \left[1+\frac{\delta r}{r_0} \left(2 \cos \left(\Omega _r t\right)-\frac{3}{2}\right) + \gamma  \left(\frac{\eta -3}{2 r_0} \right. \right.
\\
& \left. \left. +\frac{ \delta r}{4 r_0^2}  \left[12 (\eta +3) \cos \left(\Omega _r t\right)-5 (\eta -3)\right]\right)\right] .
\end{split}
\label{eq545}
\end{equation}
Inserting Eqs. \eqref{eq542} and \eqref{eq545} into \eqref{eq541} we get
\begin{equation}
\begin{split}
\mathcal{R}^{\delta r +r_0 }_\Omega & =  \frac{ \delta r \Omega _0}{r_0} \left[-\frac{3}{2}+2 \cos \left(t \Omega _r\right)+ \frac{ \gamma}{r_0}  \left(-\frac{5 (\eta -3)}{4} \right. \right. 
\\
& \left. \left. +3 (\eta +3) \cos \left(t \Omega _r\right)\right)\right].
\end{split}
\label{eq546}
\end{equation}

As in the previous Sec. \ref{sec:momentaRes}, Eq. \eqref{eq546} can be written as a generic cosine function with an offset, an amplitude and a phase of the form
\begin{equation}
\begin{split}
\mathcal{R}= M + N \cos \left( \Omega_r t + \chi \right).
\end{split}
\label{eq547}
\end{equation}
Again, the amplitude $N$ and the phase $\chi$  can be expressed by the equations
\begin{equation}
N =  \left[ b^2_1+b^2_2\right]^{1/2}, \quad \quad  \chi =  \arctan\left(b_1/b_2 \right), 
\label{eq548}
\end{equation}
where $b_1$ and $b_2$ are given by,
\begin{equation}
b_1 = \frac{N r_0}{2 \Omega_0}-\frac{3 N \gamma  \left(3 q^2+7 q+3\right)}{4 (q+1)^2 \Omega_0}, \quad \quad b_2 = 0.
\label{eq548.1}
\end{equation}
Consistent with the fact that for the separation we have performed a one-parameter analysis toward reducing the eccentricity, we have obtained the result that the phase does not provide information and the whole information is encoded in the amplitude of the residual. Solving Eqs. \eqref{eq548} and \eqref{eq548.1} consistently at 1PN order gives
\begin{equation}
\delta r = \frac{N r_0}{2 \Omega_0}-\gamma\frac{3 N   \left(3 q^2+7 q+3\right)}{4 (q+1)^2  \Omega_0}
\label{eq548.2}
\end{equation}
Equation \eqref{eq548.2} provides an expression to compute a correction to the initial separation of the binary. Note that the applications of the separation correction and the tangential momentum correction are degenerate because both describe the conservative dynamics of the binary.
We could now perform a full two-parameter analysis combining radial separation and radial momentum, in analogy to Sec.~\ref{sec:momentaRes}, but instead we note that we can also extend Eq.~\eqref{eq548.2} to a 2-dimensional iterative scheme by combining the correction for the separation with the correction for the radial momentum derived previously, Eq.~\eqref{eq540}, and we will use this two-dimensional prescription for successful eccentricity reduction in an example case in Sec. \ref{sec:NRresults}.

\section{Eccentricity reduction for numerical data }\label{sec:ecc_red}

In this section we apply the analytical formulas we have previously derived \eqref{eq539}, \eqref{eq540} and \eqref{eq548.2}, relating amplitude and phase of time dependent eccentricity estimators to corrections of the momenta or radial separation, to numerical data obtained from NR simulations, or, as a test case, to numerical post-Newtonian data.
We compute the orbital frequency  $\Omega$ from the position vector $\vec{r}$ in the center of mass frame, with  $r=|\vec{r}|$,  and its time derivative $\vec{v}$ as
\begin{equation}
\Omega=|\vec{\Omega}| =  \frac{|\vec{r} \times \vec{v}|}{r^2}.
\label{eq443.1}
\end{equation}
In the PN simulations $\vec{r}$ and $\vec{v}$ are computed from the motion of the point particles, whereas in the NR simulations they are computed from the coordinate positions of the punctures. Our NR setup is described in Appendix \ref{sec:AppendixC}.
For the NR simulations we use two codes, \texttt{BAM} \cite{Husa:2007hp, Bruegmann:2006at} and the \texttt{EinsteinToolkit} \cite{Loffler:2011ay}, which implement a discretized version of the Baumgarte-Shapiro-Shibata-Nakamura (BSSN) \citep{PhysRevD.52.5428,Baumgarte:1998te} formulation of the Einstein equations. Both codes use the moving puncture approach
\cite{Campanelli:2005dd,Hannam:2006vv,Bruegmann:2006at,Hannam:2008sg} with the ``$1+\log$" slicing and the $\Gamma$-driver shift condition \cite{Alcubierre:2002kk}.
The initial conditions for the evolving coordinate conditions (i.e.~for the lapse and shift), in particular the choice of vanishing shift, lead to gauge transients, which manifest themselves as decaying oscillations in the orbital frequency and separation. As discussed in detail in \cite{Purrer:2012wy} for one binary black hole configuration, these gauge transients complicate reading off the eccentricity, but can be suppressed by choosing a sufficiently small value of the  $\Gamma$-driver ``damping" parameter $\eta$  (not to be confused with the symmetric mass ratio used in Secs.~\ref{sec:PNID}  and \ref{sec:eccRedtheory}), such as $\eta=0.25$. The parameter $\eta$ does in fact have the dimension of inverse mass, and one might expect that for larger mass ratios, a smaller value of $\eta=0.25$ is required to avoid large gauge transients. However, for larger mass ratios gauge transients turn out to be damped out faster in general, possibly related to the faster timescale of the smaller black hole, and in our study we find that the choice $\eta=0.25$ indeed works well for all the simulations we have performed.

We will first apply eccentricity reduction to PN data as a test case, and then apply our methods to different numerical relativity data sets, with and without precession.
As expected, we will find that in PN the PPC prescription for initial data leads to smaller eccentricities than the PC prescription, with the lowest eccentricities obtained with a PN integration starting at a sufficiently large separation \cite{Husa:2007rh}. For NR simulations we will, however, find that PC initial data typically lead to lower eccentricity than the PPC approximation. We also find that for the cases we have studied, a single iteration of our eccentricity reduction procedure is sufficient to obtain an eccentricity below $10^{-3}$.

\subsection{PN example}  \label{sec:PNevolution}

The dynamics of PN particles can be described using Hamilton's equations of motion,
\begin{equation}
\frac{d\textbf{X}}{dt}= \frac{\partial H}{\partial \textbf{P}}, \quad \quad \frac{d\textbf{P}}{dt}= -\frac{\partial H}{\partial \textbf{X}}+ \textbf{F}.
\label{eq12.1}
\end{equation}
with $\textbf{X}$ and $\textbf{P}$ the position and the momentum vectors, respectively, in the CM frame, $H$ the Hamiltonian given by Eq. \eqref{eq13.1} and $\textbf{F}$ the radiation reaction force given by Eq. (3.27) in \cite{Buonanno:2005xu}. The equation of motion for the $i$th spin is
\begin{equation}
\frac{d\textbf{S}_i}{dt}= \frac{\partial H}{\partial \textbf{S}_i}\times \textbf{S}_i.
\label{eq12.2}
\end{equation}
The solution of such a system of equations describes the motion of a binary point-particle system in the inspiral regime. In this section we discuss our method to reduce eccentricity in PN, where the low computational cost of numerical solutions and the avoidance of the initial gauge transients present in NR greatly simplify the analysis.

To illustrate the procedure with an example black hole configuration, we choose mass ratio 4, which is significantly different from unity, and large spins with dimensionless Kerr parameters 
$\vec{\chi}_{1}= (0,0,0.8)$ and $\vec{\chi}_{2}=(0,0,-0.8)$ at an initial separation $D_i=12 M$, where $M$ is the total mass of the binary system. We integrate the PN equations of motion until a minimal separation $D_f=6 M$. We run two PN simulations, with initial momenta computed with the  PC, and alternatively the PPC approximations.

\begin{figure}[ht!]
\centering
\begin{minipage}[b]{\columnwidth}
\captionsetup{justification=centering}
\includegraphics[scale=0.9]{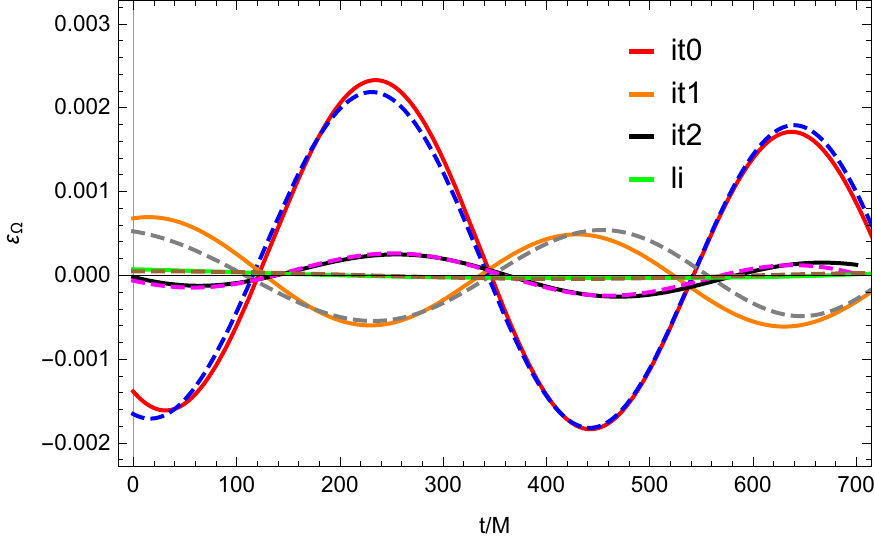}
\end{minipage}
\\
\begin{minipage}[b]{\linewidth}
\captionsetup{justification=centering}
\includegraphics[scale=0.9]{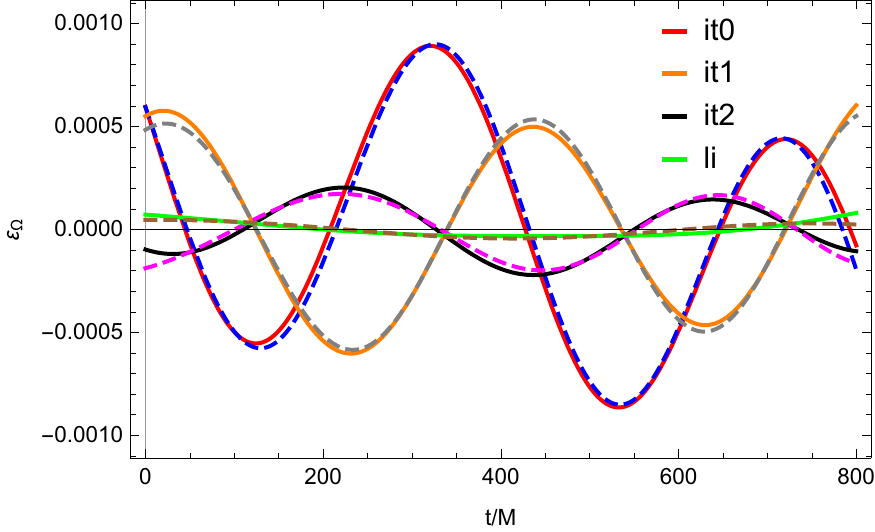}
\end{minipage}
\caption{Eccentricity reduction iterations for the configuration $q=4$, $\chi_{1z}=0.8$, $\chi_{2z}=-0.8$. The upper panel shows the time evolution of $e_\Omega $ specifying PC momenta at iteration 0 (red curve) and the lower panel shows the same quantity specifying PPC momenta at iteration 0. Afterwards, two more iterations are performed (orange and black curves). The continuous curves correspond to the data, and the dashed ones to the fits for each iteration (blue, gray, magenta, and brown curfves). Additionally the result of integrating from a longer separation (li) is shown in each panel.}
\label{fig:PN}
\end{figure}

For both simulations we  measure the eccentricity using a fit to the ansatz \eqref{eq552}  and apply two iterations employing the correction factors for the tangential and radial momenta given by Eqs. \eqref{eq539} and \eqref{eq540}. 
The corresponding eccentricity time evolution of the eccentricity estimators for each iteration are plotted in Fig. \ref{fig:PN}, which shows that the post-postcircular approximation indeed produces a simulation with a smaller eccentricity than the postcircular approximation, as one would expect.
Moreover, in Fig. \ref{fig:PN} we have added the result of initializing the momenta at $D_i=12M$ from another PN evolution starting at a larger initial separation $D_0=30M$ with PC initial momenta, which we have integrated to a separation of $D_i=12M$.  In this case the eccentricity is much smaller,  $e_\Omega= (5 \pm 2)\cdot 10^{-5},$ due to  some initial eccentricity being radiated away during inspiral before reaching $D_i=12M$, and to the high accuracy of PC momenta at $D=30M$.

The eccentricity measurement yields a time dependent result corresponding to the choice of the ansatz \eqref{eq552}. For example, for iteration 0 in the postcircular approximation, one obtains the following expressions for the eccentricity and the amplitude,
\begin{align}
\varepsilon_\Omega &= 0.00197344 - 1.97129\times 10^{-7} t,
\\
A & =0.00008561 - 8.55168 \times 10^{-9} t.
\label{eq12.4}
\end{align}
However, as in this case the time dependent terms are typically very small and can be neglected, and  we simply use the eccentricity values at $t=0$. 

The values of the eccentricity and the different correction factors are shown in Table \ref{tab:tabPN}. In addition, Table \ref{tab:tabPN} contains the values of eccentricity and the corresponding correction factors when one corrects not only the momenta, but also the radial momenta and the distance of the binary. Consistent with Fig.~\ref{fig:PN} we see that
 PPC initial data produce lower eccentricity than PC for the first iteration. The final eccentricities after two iterations are, however, very similar, although the ratio of efficiency gets worse in each iteration due to the fact that a highly accurate measurement of the amplitude and the phase of the residual is required. 
One observes that the method can easily obtain eccentricities of the order $2 \times 10^{-4}$ for a case with a relatively high mass ratio and high spins, and that one can equally well choose to correct the tangential and radial momenta or the orbital separation and the radial momentum.
\\
\begin{table}[ht!]
\begin{center}
\def\arraystretch{1.2 }
\begin{tabular}{|c|c|c|c|c|c|}
\hline  
  \multicolumn{6}{|c|}{  Post-Circular correcting for $(\lambda_t,\lambda_r)$}  \\
 \hline 
Iteration & $( \varepsilon_\Omega \pm \delta \varepsilon_\Omega ) $ $ \times 10^{-3}$  &  $10 \times p_t $  & $p_r \times 10^{3}$ & $\lambda_t $ &  $\lambda_r$   \\
\hline
  0 & $1.973 \pm 0.006  $ &  $0.56477$  & $0.238712$  & $1.00085$  &  $1.19247$\\
\hline
 1 & $0.561 \pm 0.015 $ &  $0.56529$  & $0.284657$  & $0.99974$ &  $0.94794$  \\
\hline
2 & $0.221 \pm 0.007 $ &  $0.56516$  & $0.271206$ &    &  \\
\hline
  \multicolumn{6}{|c|}{  Post-Post-Circular  correcting for   $(\lambda_t,\lambda_r)$}  \\
 \hline 
Iteration & $( \varepsilon_\Omega \pm \delta \varepsilon_\Omega ) $ $ \times 10^{-3}$  &  $10 \times p_t $  & $p_r \times 10^{3}$ & $\lambda_t $ &  $\lambda_r$   \\
\hline
  0 & $0.833 \pm 0.005  $ &  $0.56517$  & $0.238712$  & $1.00013$ &  $1.19737$  \\
\hline
 1 & $0.567 \pm 0.003 $ &  $0.56525$  & $0.285827$  & $0.99974$ &  $0.96201$  \\
\hline
2 & $0.197 \pm 0.005 $ &  $0.56510$  & $0.274971$ &    &  \\
\hline
 \multicolumn{6}{|c|}{ Post-Circular correcting for   $(\delta r ,\lambda_r)$}  \\
 \hline 
Iteration & $( \varepsilon_\Omega \pm \delta \varepsilon_\Omega ) $ $ \times 10^{-3}$  &  $D $  & $p_r \times 10^{3}$ & $\delta r  $&  $\lambda_r$   \\
\hline
  0 & $1.973 \pm 0.006  $ &  $12.0$  & $0.238712$  & $0.01432$  &  $1.19247$\\
\hline
 1 & $0.718 \pm 0.004 $ &  $12.0143$  & $0.284657$ &  $0.00445 $  & $0.999083 $ \\
\hline
2 & $0.230 \pm 0.003 $ &  $12.0099 $  & $0.284396 $ &    &  \\
\hline
 \end{tabular}
\end{center}
\caption{Eccentricity estimator and its corresponding statistical error for the configuration $q=4$, $\chi_{1z}=0.8$, $\chi_{2z}=-0.8$.}
\label{tab:tabPN}
\end{table}

We have also tested our eccentricity reduction method in the PN description of precessing binaries, with similar results: even for high spins we can obtain eccentricities of the order of $10^{-4}$ in one or two iterations. In the precessing case the method of integrating from a longer separation still yields lower eccentricities, but it does not provide control of the initial spin components of the binary at separation $D_i$ due to the fact that the spins also evolve in time during the integration. Controlling the spins at $D_i$  would require one to set up another iteration procedure to define the spins at the larger ``auxiliary separation'' ($D=30M$ in our example) in terms of the desired spins at $D_i$.
Specifying the initial data using the PC or PPC prescription can significantly simplify setting up parameter studies where control of the spin configurations is desired at $D_i$.
As we will see below, this argument is even stronger in NR, where due to the deviations between PN and full GR there is no significant advantage in integrating from a large initial separation as compared with PC or PPC data.


\subsection{Numerical relativity examples} \label{sec:NRresults}

Applying our eccentricity reduction procedure to numerical relativity simulations adds several complications compared with the post-Newtonian example: Apart from the
 increase in computational cost by $6-7$ orders of magnitude, the main technical problems are gauge transients resulting from the procedure of initializing the coordinate conditions of the moving puncture evolutions (in particular the initially vanishing velocity of the punctures). We address this problem by using a small value of the shift parameter $\eta$, of $\eta=0.25$, for the evolutions we report on below, and by cutting away the first $\sim 200M$ of time evolution. 
Black-hole binary puncture initial data also exhibit a burst of junk radiation due to unphysical gravitational wave content in the initial data. Here we do not take into account the resulting small change to initial masses, spins, and momenta, although this may be beneficial  when attempting to construct initial data with even lower eccentricities.
For the cases we have studied so far, our choice of $\eta=0.25$, together with the robust setup of our fitting method to determine eccentricity presented in Sec.~\ref{sec:ecc_red},  provides sufficiently accurate estimates  not only of the eccentricity, but also of the phase shift defined in Eq.~\eqref{eq534}, which is required to determine the change in radial momentum or separation to implement a two-parameter eccentricity reduction algorithm.

We first discuss our procedure for the example of a precessing binary with mass ratio $q=2$ and dimensionless spin vectors $\vec{\chi}_1=(0,0,0)$,  $\vec{\chi}_2=(0.3535,0.3535,0.5)$, 
and initial orbital separation $D=10.8M$. First, we run a simulation with PC initial data with \texttt{BAM} at low resolution with $N=64$ points to measure the eccentricity, fitting the oscillations of
$\Omega$ computed using Eq. {\eqref{eq443.1}. Then, we adjust the values of the tangential and radial momenta according to Eqs. ~\eqref{eq539} and \eqref{eq540} to reduce eccentricity, 
and we run two low resolution simulations with the corrected momenta, one with \texttt{BAM} another with \texttt{ET} with the same numerical resolution and gauge conditions.
The results for the time evolution of the eccentricity estimator for the three simulations are shown in  Fig. \ref{fig:NR1}.
\begin{figure}[ht!]
\centering
\includegraphics[scale=1]{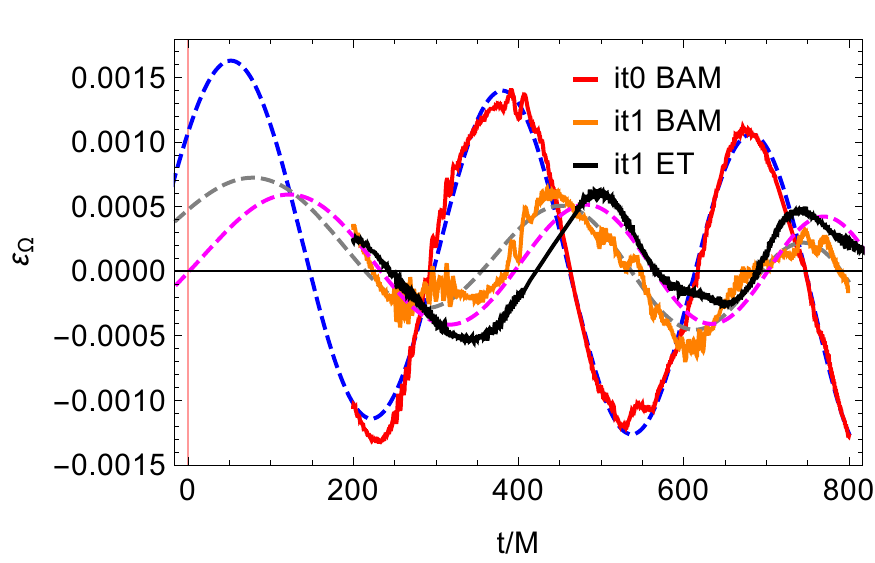}
\caption{Time evolution of the eccentricity estimator for the configuration $q=2$, $\vec{\chi}_1=(0,0,0)$,  $\vec{\chi}_2=(0.3535,0.3535,0.5)$ and $D=10.8M$. The thick curves correspond to the data and the dashed ones to the fits. For the three simulations we have discarded the initial $t=200M$ of evolution time.}
\label{fig:NR1}
\end{figure}

After one iteration the eccentricity has been notably reduced with both codes. The values of eccentricity for iteration 1 in both codes are quite similar. However, the \texttt{ET} residual is cleaner than for the \texttt{BAM} evolution, which contains more high frequency noise that we attribute to different settings for numerical dissipation in this simulation, and which complicates the measurement of the phase and the amplitude of the residual  and leads to different results in iteration 1. The sign of the correction to the tangential momentum is read from the value of the residual at the initial time of the evolution, according to the expression for the residual computed in Sec. \ref{sec:eccRedtheory}: For a positive residual, as is the case in iteration 0, the momentum has to be decreased, while for a negative residual the momenta should be increased. 
The values of the eccentricity as well as the correction factors used are shown in Table \ref{tab:tabNR1}. After a single iteration the eccentricity is well below $10^{-3}$, which we have considered sufficient to neglect eccentricity in our waveform modeling applications, and we have not carried out further iterations. 

For completeness we also show in Fig. \ref{fig:NRorbital1} the time evolution of the orbital separation and the orbital frequency of that configuration. One can observe from the plots that the oscillations remaining after one iteration of the eccentricity reduction procedure cannot be appreciated on that scale of the plot any more.

\begin{table}[ht!]
\begin{center}
\def\arraystretch{1.2 }
\begin{tabular}{|c|c|c|c|c|}
\hline  
Iteration & Code & $( \varepsilon_\Omega \pm \delta \varepsilon_\Omega ) $ $ \times 10^{-3}$ & $\lambda_t $ &  $\lambda_r$   \\
\hline
  0 &\texttt{BAM}  & $1.37 \pm 0.02  $  & $0.9996$  &  $0.8456$\\
\hline
 1 & \texttt{BAM}  & $0.48 \pm 0.02 $ &   &   \\
\hline
1 & \texttt{ET} & $0.51 \pm 0.03 $ &     &  \\
\hline
 \end{tabular}
\end{center}
\caption{Eccentricity estimator and its corresponding statistical error for the configuration $q=2$, $\vec{\chi}_1=(0,0,0)$,  $\vec{\chi}_2=(0.3535,0.3535,0.5)$ and $D=10.8M$.}
\label{tab:tabNR1}
\end{table}

\begin{figure}[ht!]
\centering
\begin{minipage}[b]{\columnwidth}
\captionsetup{justification=centering}
\includegraphics[scale=0.9]{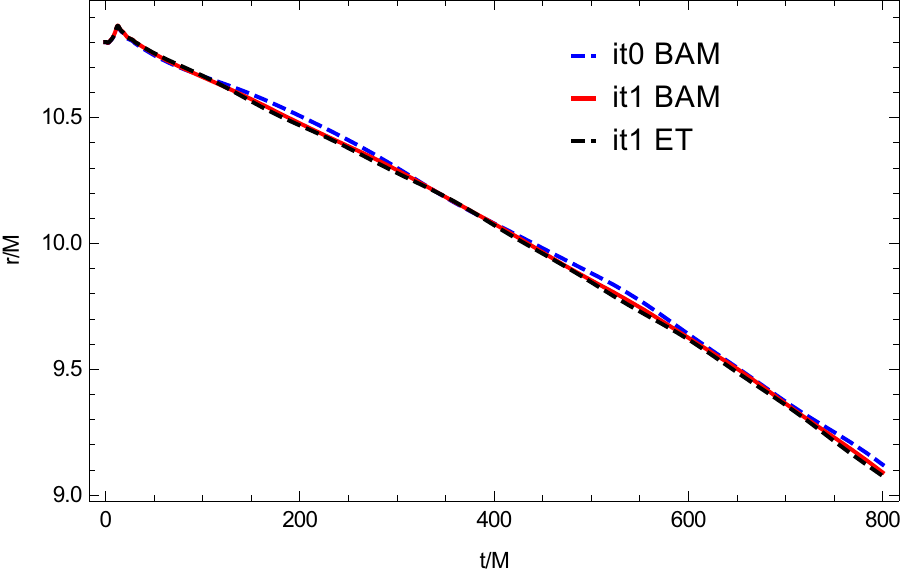}
\end{minipage}
\\
\begin{minipage}[b]{\linewidth}
\captionsetup{justification=centering}
\includegraphics[scale=0.9]{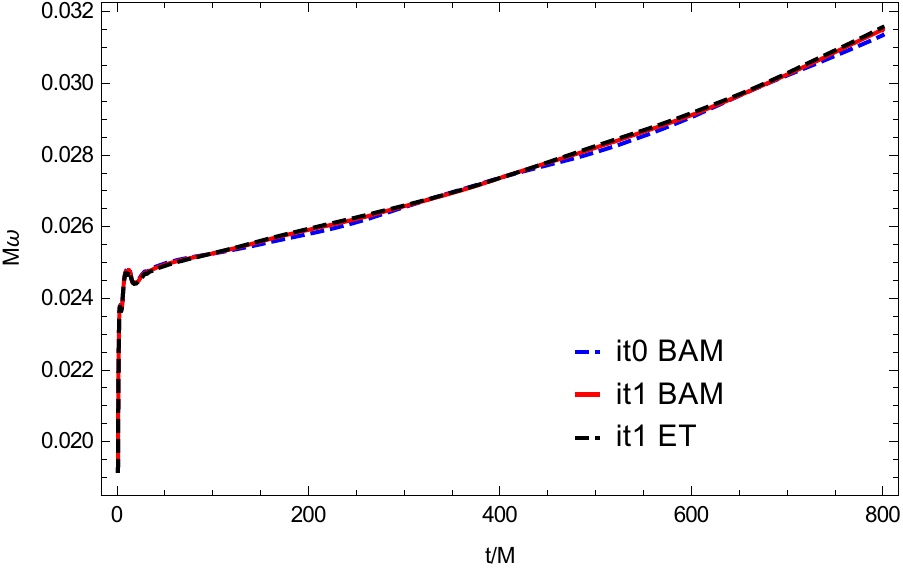}
\end{minipage}
\caption{Time evolution of the orbital quantities for the configuration $q=2$, $\vec{\chi}_1=(0,0,0)$,  $\vec{\chi}_2=(0.3535,0.3535,0.5)$ and initial separation $D=10.8M$. In the upper panel we plot the time evolution of the orbital separation of the binary. In the lower panel the orbital frequency of the binary is plotted for the different iterations. The blue dashed curve corresponds to iteration 0 run with the \texttt{BAM} code and PC initial data. The red curve corresponds to the iteration 1 run the \texttt{BAM} code and the black dashed one to the simulation performed with the \texttt{ET} code.}
\label{fig:NRorbital1}
\end{figure}

In the second example we apply the correction of the separation and radial momentum to a NR simulation, combining the corrections in the radial momentum and the initial orbital separation,
\begin{equation}
p^1_r  = \lambda_r p^0_r, \quad \quad r^1_0 =r^0_0 +\delta r.
\label{eq888}
\end{equation}
We choose the spin-aligned configuration ID13 of Table \ref{tab:tabNR3}, i.e., $q=1$, $\chi_{1z}=-0.5$, $\chi_{2z}=0.5$ with $D=11M$. The results of applying the eccentricity reduction procedure are shown in Table \ref{tab:tabsep}. The eccentricity residual is plotted in Fig. \ref{fig:NRq1sep}. 

Looking at Fig. \ref{fig:NRq1sep} one checks that the eccentricity estimator is dominated by high frequency noise. That is the reason why the quality of the fit is so bad and its statistical error so large. One can also check comparing the value of the eccentricity after one iteration for ID13 from Table \ref{tab:tabNR3} where one corrects the momenta and the value from Table \ref{tab:tabsep} that both results are consistent and similar providing eccentricity of the same magnitude.

\begin{table}[ht!]
\begin{center}
\def\arraystretch{1.2 }
\begin{tabular}{|c|c|c|c|c|c|}
\hline  
Iteration & Code  & $N$  & $ \delta r $ &  $\lambda_r$  & $( \varepsilon_\Omega \pm \delta \varepsilon_\Omega ) $ $ \cdot 10^{-3}$ \\
\hline
  0 & \texttt{BAM}  & 64 &   &  & $ 1.24 \pm 0.03 $\\
\hline
  1 & \texttt{BAM}  & 64 & $-0.0023$  &  $0.8581 $& $ 0.2 \pm 0.2 $\\
\hline
 \end{tabular}
\end{center}
\caption{Eccentricity estimator and its corresponding statistical error for the configuration ID2 of Table \ref{tab:tabNR3}.}
\label{tab:tabsep}
\end{table}

\begin{figure}[ht!]
\centering
\includegraphics[scale=0.7]{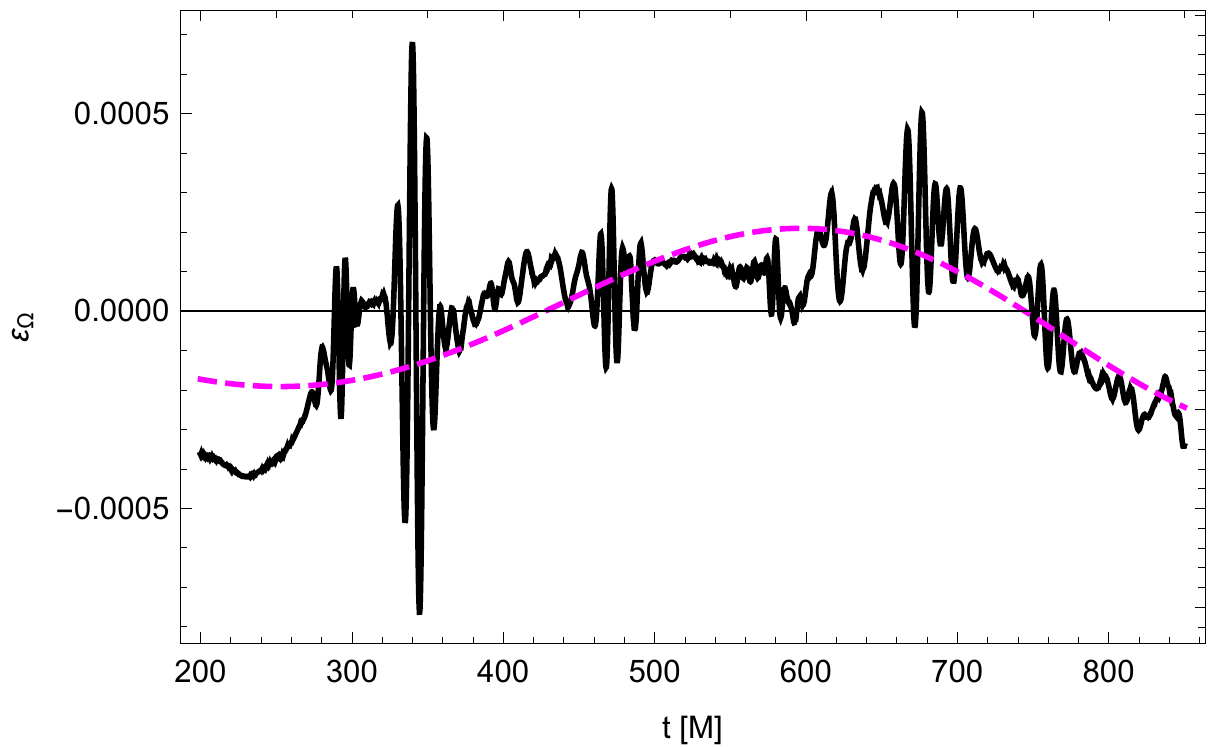}
\caption{Time evolution of the eccentricity estimator for the configuration $q=1$, $\chi_{1z}=-0.5$, $\chi_{2z}=0.5$. The black curve corresponds to the data and the dashed pink line to the fit to the data.}
\label{fig:NRq1sep}
\end{figure}

\subsection{Postcircular and Post-postcircular in NR} \label{sec:postNR}
 
In order to compare PC and PPC initial data, we have run 12 pairs of simulations, ranging from equal mass nonspinning to mass ratio $q=8$ and precessing simulations, using both PC and PPC initial data for each case. The results are shown in Table \ref{tab:tabNR2}.  All the simulations in Table \ref{tab:tabNR2} have been computed using the  \texttt{BAM} code, at low resolution with $N=64$ points in the innermost box, and setting the gauge parameter $\eta=0.25$ as before.

\begin{table}[ht!]
\begin{center}
\resizebox{9cm}{!}{
\def\arraystretch{1.2 }
\begin{tabular}{|c|c|c|c|c|c|c|}
\hline  
ID  & Approx. & $q$  &$\vec{\chi}_1$ & $\vec{\chi}_2$ & $D/M$& $( \varepsilon_\Omega \pm \delta \varepsilon_\Omega )\times 10^{3} $ \\
\hline
1 & PC & 1 & $(0,0,0)$ &$(0,0,0)$ & $11$ & $1.42 \pm 0.02$ \\
\hline
1 & PPC & 1 & $(0,0,0)$ &$(0,0,0)$ & $11$ &$ 1.43 \pm 0.04$ \\
\hline
2  &  PC & 1  & $( 0,0,-0.5)$ &$(0,0,-0.5)$ & $11$  & $ 5.3 \pm 0.4$ \\
\hline
2   & PPC  & 1  & $( 0,0,-0.5)$ &$(0,0,-0.5)$  & $11$  &$9.8 \pm 0.5$ \\
\hline
3   & PC  & 1  & $( 0,0,0.5)$ &$(0,0,-0.5)$ & $11$  &$1.5 \pm 0.05$ \\
\hline
3   & PPC  & 1  & $( 0,0,0.5)$ &$(0,0,-0.5)$ & $11$  &$2.27 \pm 0.04$ \\
\hline
4   & PC  & 2 & $( 0,0,-0.75)$ &$(0,0,-0.75)$ & $12.6$  &$4.22 \pm 0.07$ \\
\hline
4   &  PPC &  2 & $( 0,0,-0.75)$ &$(0,0,-0.75)$ & $12.6$  &$4.61 \pm 0.16$ \\
\hline
5   & PC  &  2 & $( 0,0,0)$ &$\vec{\alpha}$ & $10.8$  &$2.68 \pm 0.17 $\\
\hline
5   & PPC  &  2 & $( 0,0,0)$ &$\vec{\alpha}$ & $10.8$   &$5.43 \pm 0.13$ \\
\hline
6   &  PC &  2 & $( 0,0,0)$ &$\vec{\beta}$ & $10.8$ & $3.61 \pm 0.017$ \\
\hline
6   &  PPC & 2  & $( 0,0,0)$ &$\vec{\beta}$ & $10.8$  &$4.003 \pm 0.018$ \\
\hline
7   & PC  & 4 & $( 0,0,-0.8)$ &$(0,0,0.8)$ & $11$ &$4.05 \pm 0.07$ \\ 
\hline
7   & PPC  & 4 & $( 0,0,-0.8)$ &$(0,0,0.8)$ & $11$  &$7.25 \pm 0.06$ \\
\hline
8   &  PC &  4 & $( 0,0,-0.8)$ &$(0,0,-0.8)$ & $11$  &$17.9 \pm 1.5$ \\
\hline
8   &  PPC & 4 & $( 0,0,-0.8)$ &$(0,0,-0.8)$ & $11$  &$17.5 \pm 1.5$ \\
\hline
9   & PC  &  4 & $( 0,0,0.8)$ &$(0,0,-0.8)$ & $11$  &$17.4 \pm 0.6$ \\
\hline
9   & PPC  & 4  & $( 0,0,0.8)$ &$(0,0,-0.8)$ & $11$  &$15.3 \pm 0.5$ \\
\hline
10   & PC  &  4 & $( 0,0,0.8)$ &$(0,0,0.8)$ & $11$  &$5.5 \pm 0.5 $\\
\hline
 10  & PPC  & 4  & $( 0,0,0.8)$ &$(0,0,0.8)$ & $11$  &$9.9 \pm 0.6$ \\
\hline
 11  & PC  & 8  & $( 0,0,0.5)$ &$(0,0,-0.5)$ & $11$  &$4.64\pm 0.14$ \\
\hline
 11 &  PPC & 8  & $( 0,0,0.5)$ &$(0,0,-0.5)$ & $11$  &$8.0 \pm 0.2$ \\
\hline
 12  & PC  &  8 & $( 0,0,-0.5)$ &$(0,0,-0.5)$ & $11$  &$12.49 \pm 0.18$ \\
\hline
 12  & PPC  & 8  & $( 0,0,-0.5)$ &$(0,0,-0.5)$ & $11$  &$22.9 \pm 0.4$ \\
\hline
 \end{tabular}
 }
\end{center}
\caption{Simulations performed to compare PC and PPC initial data. In the first column an identifier is assigned to each configuration which is run with the PC and PPC approximations. In the following columns the mass ratio, the dimensionless spin vectors of each black hole are specified, with the vectors $\vec{\alpha}=(0.3535,-0.3535,-0.5)$ and $\vec{\beta}=(0.3535,-0.3535,0.5)$. Also shown are the initial orbital separation and the value of the eccentricity estimator and its corresponding statistical error.}
\label{tab:tabNR2}
\end{table}

\begin{figure}[ht!]
\centering
\includegraphics[scale=0.9]{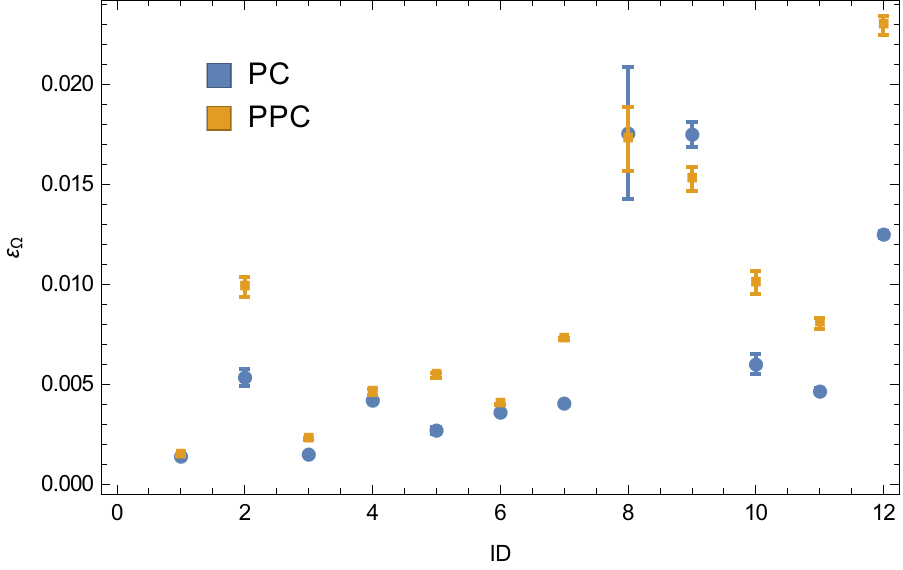}
\caption{Measured eccentricity, with statistical error bars computed from the nonlinear fits, for the 12 configurations reported in Table \ref{tab:tabNR2}, comparing PC initial data
 (rounded blue points)  and PPC  data (yellow squares).}
\label{fig:NRpc}
\end{figure}

Figure \ref{fig:NRpc} shows a graphical representation of Table \ref{tab:tabNR2}. Overall, the PC initial data seem to work better in NR than PPC, except for   configuration 9, where PPC initial data lead to a lower eccentricity than PC data. This apparently counterintuitive result is not particularly surprising: the numerical relativity evolutions differ from PN not only because of missing higher order PN terms  but also because the ADMTT \cite{SchaeferADMTT} gauge underlying our post-Newtionan results differs from the gauge used in  our numerical relativity code beyond 2PN order \cite{Tichy:2002ec}. While in post-Newtonian theory the PPC approximation is indeed superior, the deviation of PC data could either lead to momenta that are closer to NR or indeed show larger eccentricities than PPC.

\subsection{Eccentricity reduction for postcircular initial data for a range of numerical relativity simulations} \label{sec:resultsNR}

In Table \ref{tab:tabNR3} we present results from single step eccentricity reduction for a variety of configurations, using both the \texttt{BAM} and \texttt{ET} codes, and starting with PC initial momenta, which as we have seen in Sec. \ref{sec:postNR} typically yield smaller eccentricities than PPC momenta for numerical relativity gauge and initial separations we use.
All the simulations using PC initial data, labeled as iteration 0 of the eccentricity reduction procedure, are carried out with gauge parameter $\eta=0.25$ and low numerical resolution of $64^3$ grid points for the innermost grid (containing the black holes). While we have used the same setup for some of the iteration 1 simulations, for others we use our typical setup for productions runs: a higher resolution of $80^3$  or $96^3$ points, and a gauge parameter of $\eta=1$, which increases initial gauge transients, but tends to reduce high frequency noise.
For all the cases shown, a single eccentricity reduction step reduces the eccentricity to below $10^{-3}$.

\begin{figure}[ht!]
\centering
\includegraphics[scale=0.9]{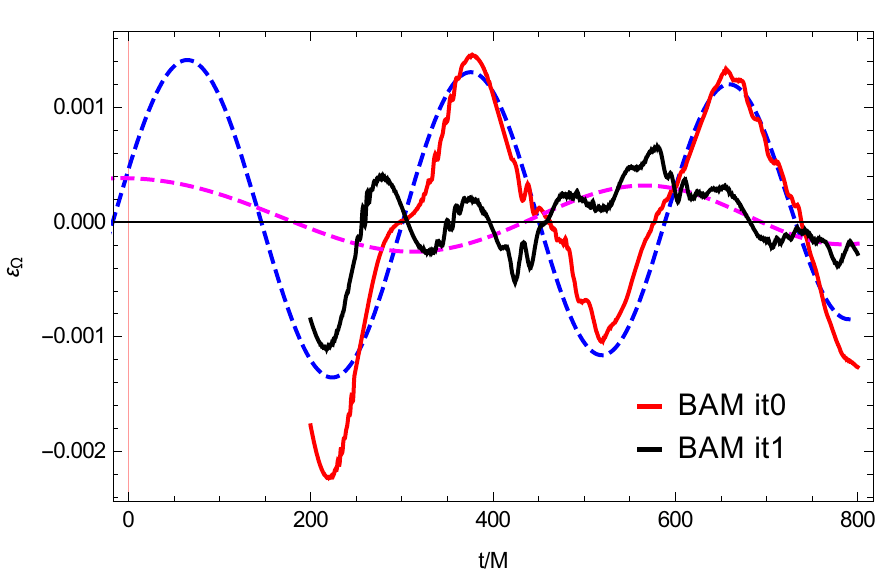}
\caption{Time evolution of the eccentricity estimator for the configuration $q=3$ nonspinning. The red curve corresponds to iteration 0 and the black one to iteration 1. The dashed lines are fits to the data. Both simulations were run with $\eta=1$.}
\label{fig:NRq3}
\end{figure}

However, we show that the $\eta$  parameter can also be set to $1$ in the first iteration, and one can also get an important reduction of the eccentricity, as happens with the case ID19. The residuals of such a configuration are shown in Fig. \ref{fig:NRq3}. For that configuration one can also observe the poor quality of the fit in iteration 1, which is consistent with the high value of the error of the eccentricity in Table \ref{tab:tabNR3}. 

\begin{table*}[ht!]
\begin{center}
\def\arraystretch{1.2 }
\begin{tabular}{|c|c|c|c|c|c|c|c|c|c|c|c|c|c|}
\hline  
ID & Iteration  & Code & N & $\eta$ & q  &$\vec{\chi}_1$ & $\vec{\chi}_2$ & $D/M$&$10 \cdot p_t $& $10^3 \cdot p_r$ &  $\lambda_t $& $\lambda_r$ & $( \varepsilon_\Omega \pm \delta \varepsilon_\Omega )\cdot 10^{3} $ \\
\hline
 \multirow{2}{*}{1}& 0  &  \texttt{BAM} & 64 & 0.25  & 1 & $(0,0,0)$ &$(0,0,0)$ & $12$ & $0.850941$ & $0.53833$ & 0.9997 & 0.8695 & $1.42 \pm 0.02$ \\
\cline{2-14}
 & 1  & \texttt{BAM}   & 96 & 0.25 & 1  & $(0,0,0)$ &$(0,0,0)$ & $12$ & $0.850686$ & $0.468113$ & &  & $0.22 \pm 0.02$ \\
\hline
 \multirow{2}{*}{13} &0 & \texttt{BAM}& 64  & 0.25 &   1 & $(0,0,-0.5)$ &$(0,0,0.5)$ & $11$ & 0.901836 & 0.722706 & 0.9998 & 0.8581 & $1.24 \pm 0.03$ \\
\cline{2-14}
&1 &  \texttt{BAM} & 64  & 0.25 &  1 & $(0,0,-0.5)$ &$(0,0,0.5)$ & $11$ & 0.901688 & 0.620187 & &  & $0.27 \pm 0.02$ \\
\hline
 \multirow{2}{*}{14} &0 & \texttt{BAM} & 64  & 0.25   & 1 & $(0,0,0.5)$ &$(0,0,0.5)$ & $11$ & 0.874251 & 0.601797 & 0.999237 & 0.9346 & $1.64 \pm 0.03$ \\
\cline{2-14}
&1 &    \texttt{BAM} & 64  & 0.25  & 1 & $(0,0,0.5)$ &$(0,0,0.5)$ & $11$ &  0.873583 & 0.562465 & &  & $0.39 \pm 0.03$ \\
\hline
 \multirow{2}{*}{15} &0   &  \texttt{ET} & 64  & 0.24 & 1.5 & $(0,0,-0.6)$ &$(0,0,0.6)$ & $10.8$ & 0.868557 & 0.699185 & 0.999737 & 0.9168 & $1.12 \pm 0.05$ \\
\cline{2-14}
&1 &    \texttt{ET} & 80  & 0.24  &   1.5 & $(0,0,-0.6)$ &$(0,0,0.6)$ & $10.8$ &  0.856941 & 0.641051 & &  & $0.84 \pm 0.165$ \\
\hline
 \multirow{2}{*}{16} &0   &   \texttt{ET} & 64  & 0.2314  & 1.75 & $(0,0,0.6)$ &$(0,0,-0.6)$ & $10.8$ & 0.856941 & 0.685199 & 0.999643 & 0.8525 & $1.52 \pm 0.08$ \\
\cline{2-14}
&1 &  \texttt{ET} & 80  & 0.2314   &    1.75 & $(0,0,0.6)$ &$(0,0,-0.6)$ & $10.8$ & 0.856636 & 0.584173 & &  & $0.43 \pm 0.07$ \\
\hline
\multirow{2}{*}{17} &0 &   \texttt{ET} & 64  & 0.2314    &  1.75 & $(0,0,-0.6)$ &$(0,0,0.6)$ & $10.8$ & 0.834827 & 0.649957 & 0.999903 & 0.8941 & $1.12 \pm 0.14$ \\
\cline{2-14}
&1 &   \texttt{ET} & 80  & 0.2314  &   1.75 & $(0,0,-0.6)$ &$(0,0,0.6)$ & $10.8$ &  0.834746 & 0.581178 & &  & $0.66 \pm 0.13$ \\
\hline
\multirow{2}{*}{18} &0   &  \texttt{BAM} & 64  & 0.2222  & 2 & $(0,0,0.75)$ &$(0,0,0.75)$ & $11.1117$ & 0.760924 & 0.450647 & 0.999937 & 0.6566 & $2.38 \pm 0.07$ \\
\cline{2-14}
&1 &  \texttt{BAM} & 96  & 0.2222  &   2 & $(0,0,0.75)$ &$(0,0,0.75)$ & $11.1117$ &  0.760876 & 0.295898 & &  & $0.47 \pm 0.05$ \\
\hline
\multirow{2}{*}{19} &0  &   \texttt{BAM} & 80  & 0.1875  & 3 & $(0,0,0)$ &$(0,0,0)$ & $10$ & 0.72377 & 0.575703 & 0.999914 & 0.8629 & $1.41 \pm 0.07$ \\
\cline{2-14}
&1 &   \texttt{BAM} & 64  & 0.1875  &  3 & $(0,0,0)$ &$(0,0,0)$ & $10$ &  0.723708 & 0.496774 & &  & $0.29 \pm 0.24$ \\
\hline
\multirow{2}{*}{7} &0 &\texttt{BAM} & 64  & 0.16 &  4 & $(0,0,-0.8)$ &$(0,0,0.8)$ & $11$ & 0.559207 & 0.336564 & 0.998501 & 0.7341 & $4.05 \pm 0.07$ \\
\cline{2-14}
&1 &  \texttt{BAM} & 64  & 0.16  &  4 & $(0,0,-0.8)$ &$(0,0,0.8)$ & $11$ &  0.558369 & 0.24708 & &  & $0.45 \pm 0.4$ \\
\hline
\multirow{2}{*}{20} &0  &  \texttt{BAM} & 64  & 0.0987  & 8 & $(0,0,0.5)$ &$(0,0,0.5)$ & $11$ & 0.102969 & 0.345755 & 1.00066 & 1.3512 & $2.2 \pm 0.4$ \\
\cline{2-14}
&1 &   \texttt{BAM} & 64  & 0.0987  &  8 & $(0,0,0.5)$ &$(0,0,0.5)$ & $11$  &  0.139132 & 0.345985 & &  & $0.45 \pm 0.4$ \\
\hline
\multirow{2}{*}{21} &0   &  \texttt{BAM} & 64  & 0.2222  & 2 & $(0,0,0)$ &$(0.4949,0.4949, 0)$ & $10.8$ & 0.811783  & 0.649957  & 0.999788  &  0.9802 & $ 6.4 \pm 1.7$ \\
\cline{2-14}
&1  &    \texttt{ET} & 80  & 0.2222  &  2  & $(0,0,0)$ & $(0.4949,0.4949, 0)$ & $10.8$  &  0.811611  & 0.581178 &   &  & $ 0.40 \pm 0.05 $ \\
\hline
\multirow{2}{*}{22} &0   &   \texttt{BAM} & 64  & 0.2222 & 2 & $(0,0,0)$ &$(0.1767,0.1767,0)$ & $10.8$ &  0.812379 &  0.610965 & 0.999534  & 0.9009  & $ 1.46 \pm 0.02 $ \\
\cline{2-14}
&1 &   \texttt{ET} & 80  & 0.2222  & 2  & $(0,0,0)$ & $(0.1767,0.1767,0)$ & $10.8$  & 0.812001   & 0.550427 & &  & $ 0.54 \pm 0.05 $ \\
\hline
\multirow{2}{*}{23} &0   &   \texttt{ET} & 64  & 0.2222 & 2 & $(0,0,0)$ &$(-0.1767,0.1767,0.5)$ & $10.8$ &  0.793749 & 0.53149 &0.99994  & 0.881549  & $ 1.88 \pm 0.01 $ \\
\cline{2-14}
&1 &   \texttt{ET} & 80  & 0.2222  & 2  & $(0,0,0)$ &$(-0.1767,0.1767,0.5)$ & $10.8$  & 0.793701  & 0.468535  & &  & $ 0.28  \pm 0.05  $ \\
\hline
\multirow{2}{*}{24} &0   &   \texttt{ET} & 64  & 0.2222 & 2 & $(0,0,0)$ &$(-0.3535, 0.3535, 0.5)$ & $10.8$ &  0.7935 & 0.531374  &  0.999772   &  0.843376  & $ 2.13  \pm  0.03 $ \\
\cline{2-14}
&1 &    \texttt{ET} & 80  & 0.2222  & 2  & $(0,0,0)$ &$(-0.3535, 0.3535, 0.5)$ & $10.8$  & 0.79332  &  0.448148 & &  & $ 0.48  \pm 0.05  $ \\
\hline
\multirow{2}{*}{25} &0   &   \texttt{ET}  & 64  & 0.2222 & 2 & $(0,0,0)$ &$(-0.3535, 0.3535, 0.)$ & $10.8$ &  0.812118 & 0.611108   & 0.999848   &  0.895657  & $ 1.78  \pm  0.07 $ \\
\cline{2-14}
&1 &   \texttt{ET} & 80  & 0.2222  & 2  & $(0,0,0)$ &$(-0.3535, 0.3535, 0.)$ & $10.8$  & 0.811994  & 0.547343  & &  & $  0.69 \pm 0.07  $ \\
\hline
\end{tabular}
\end{center}
\caption{ Summary of the eccentricity reduced simulations. In the first column we label each configuration, and the second one specifies the iteration. The code used and the number of points $N$ used in the innermost level of the codes are displayed, as well as the value of the parameter $\eta$ appearing in the $\Gamma$-driver shift condition. Then, the mass ratio $q=m_2/m_1$, and the dimensionless spin vectors, $\vec{\chi}_1$, $\vec{\chi}_2$, the orbital separation $D/M$, the tangential momenta $p_t$ multiplied by 10 and the radial momentum $p_r$ multiplied by a factor of $10^3$ are shown. The correction factors $\lambda_t$ and $\lambda_r$ computed from iteration zero are described. The values of the eccentricity estimators $\varepsilon_\Omega$ and their corresponding statistical error $\delta \varepsilon_\Omega$ from the fit are also given.}
\label{tab:tabNR3}
\end{table*}

\begin{figure*}[ht!]
\centering
\begin{minipage}[b]{\columnwidth}
\captionsetup{justification=centering}
\includegraphics[scale=0.8]{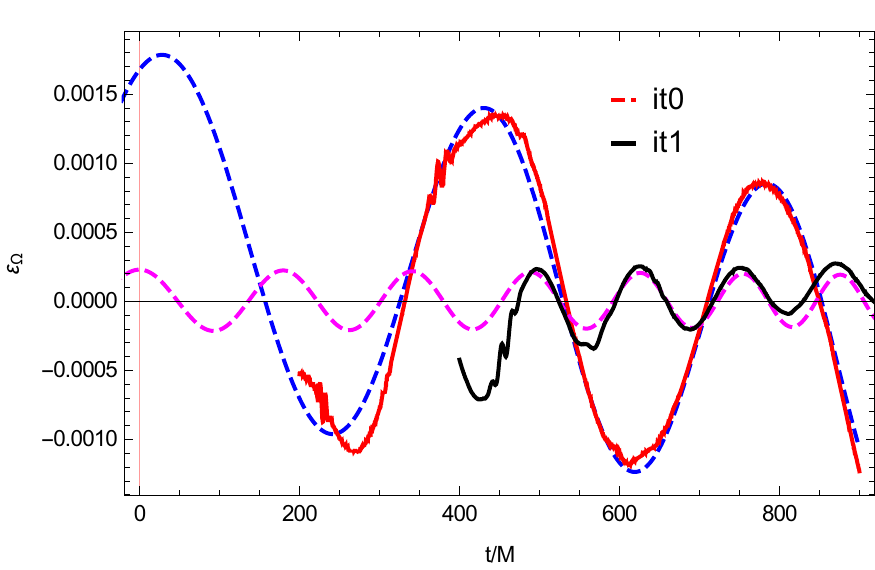}
\end{minipage}
\begin{minipage}[b]{\columnwidth}
\captionsetup{justification=centering}
\includegraphics[scale=0.8]{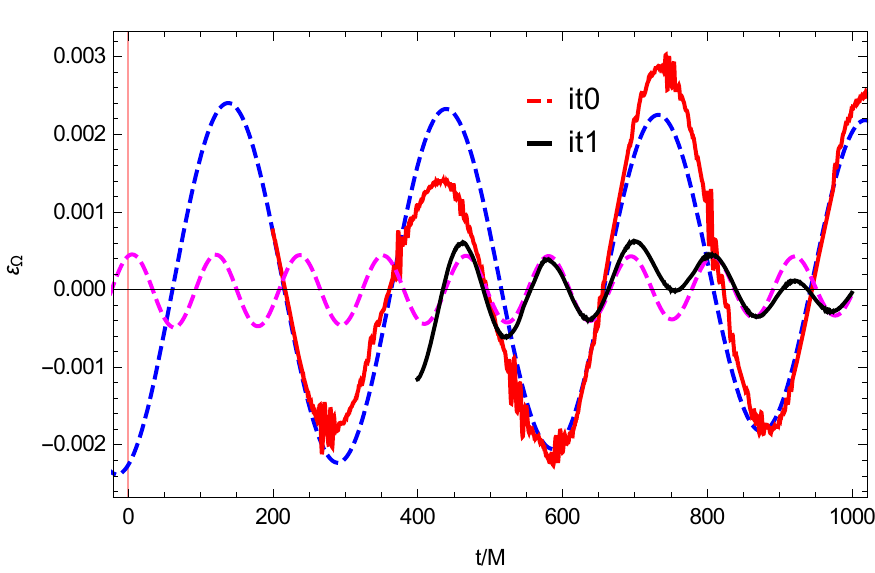}
\end{minipage}
\begin{minipage}[b]{\columnwidth}
\captionsetup{justification=centering}
\includegraphics[scale=0.8]{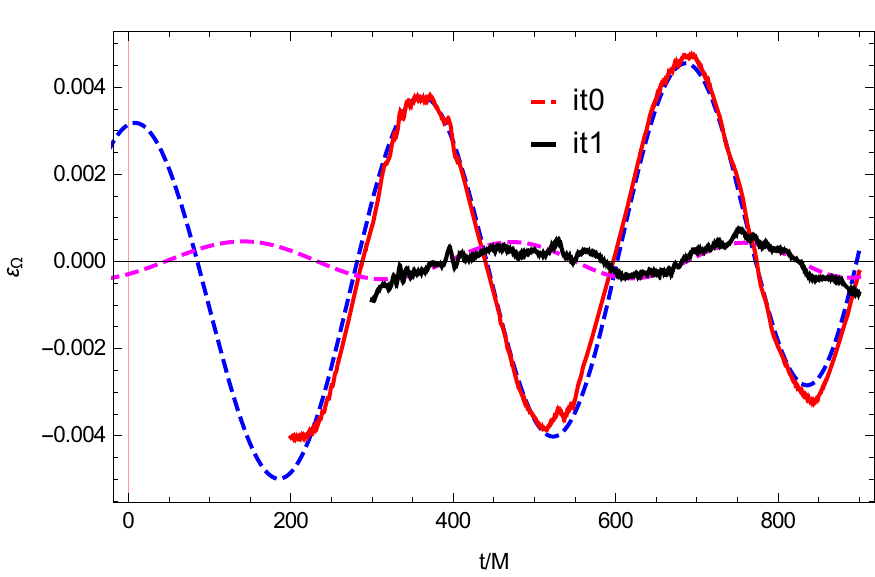}
\end{minipage}
\begin{minipage}[b]{\columnwidth}
\captionsetup{justification=centering}
\includegraphics[scale=0.8]{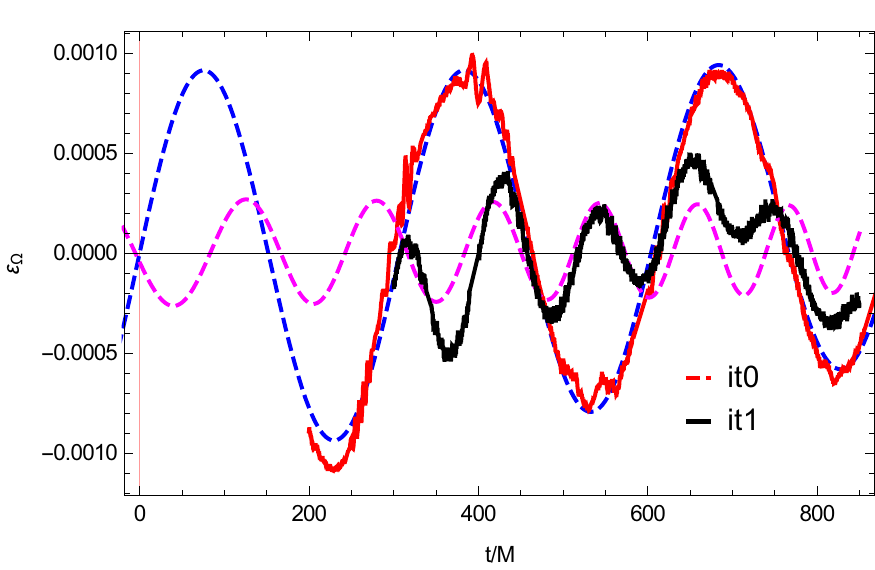}
\end{minipage}
\caption{Time evolution of the eccentricity estimators. In the top left panel one has the configuration ID1, in the top right panel ID18, in the bottom left plot ID7, and in the bottom right picture ID23 from Table \ref{tab:tabNR3}.  The red curves correspond to the data of iteration 0 and the black ones to the data of iteration 1. The dashed lines correspond to fits to the eccentricity estimators.}
\label{fig:NR2}
\end{figure*}

The lower the value of the eccentricity, the more difficult becomes the eccentricity measurement because some features due to the lack of resolution of the code can appear, such as high frequency noise coming from the finite difference scheme. Furthermore, it becomes difficult to disentangle gauge oscillations from eccentricity oscillations, as one can observe in Fig. \ref{fig:NR2}  where the eccentricity estimators of the configurations ID1, ID7, ID18, and ID23 from Table \ref{tab:tabNR3} are plotted. 

Finally, the results of Table \ref{tab:tabNR3} allow one to discuss which PN order in the PN expressions for the initial momenta $(p_t,p_r)$ is closer to the corrected momenta that provide low eccentric initial data. The results are displayed in Fig. \ref{fig:ptPNorder}. We have computed the difference in absolute values between the corrected tangential or radial momentum ($p_t^{\text{ref}}$, $p_r^{\text{ref}}$) and the PC and PPC values at a given PN order ($p_t^i$, $p_r^i$), with $i=0,1,1.5,2,2.5,3,3.5$. 

On the one hand, the upper and intermediate plots of Fig.  \ref{fig:ptPNorder} show that in order to have low-eccentricity initial data one requires the knowledge of high PN orders for the tangential momentum. In addition, when comparing the top and intermediate panels of Fig. \ref{fig:ptPNorder} one can check that the PPC approximation has larger values than the PC, and also one observes that for the PC the difference between 3PN and $3.5$PN is very small.

On the other hand, the lower panel of Fig. \ref{fig:ptPNorder} reveals that the use of higher PN orders for the radial momentum does not help significantly to reduce the eccentricity. In fact, the lower PN orders seem to provide lower differences. This is in agreement with some of the statements of \cite{Tichy:2010qa} with respect to the use of low PN order expressions in eccentricity reduction procedures and explains the success of their method.  However, note that small changes in the tangential momentum translate into large changes in the eccentricity, while the eccentricity is less sensitive to changes in the radial momentum \cite{Purrer:2012wy}; this is due to the fact that $\partial e_t / \partial \lambda_t \gg \partial e_r / \partial \lambda_r $. In addition, the small difference between the different PN orders implies that the use of different PN orders for the radial momentum provides very similar results. Therefore, while the differences between the values of $p_r$ at different PN do not have a large effect on the eccentricity, the smaller differences for $p_t$ between the PN orders are large enough to directly affect the eccentricity.

\begin{figure}[ht!]
\centering
\begin{minipage}[b]{\columnwidth}
\captionsetup{justification=centering}
\includegraphics[scale=0.65]{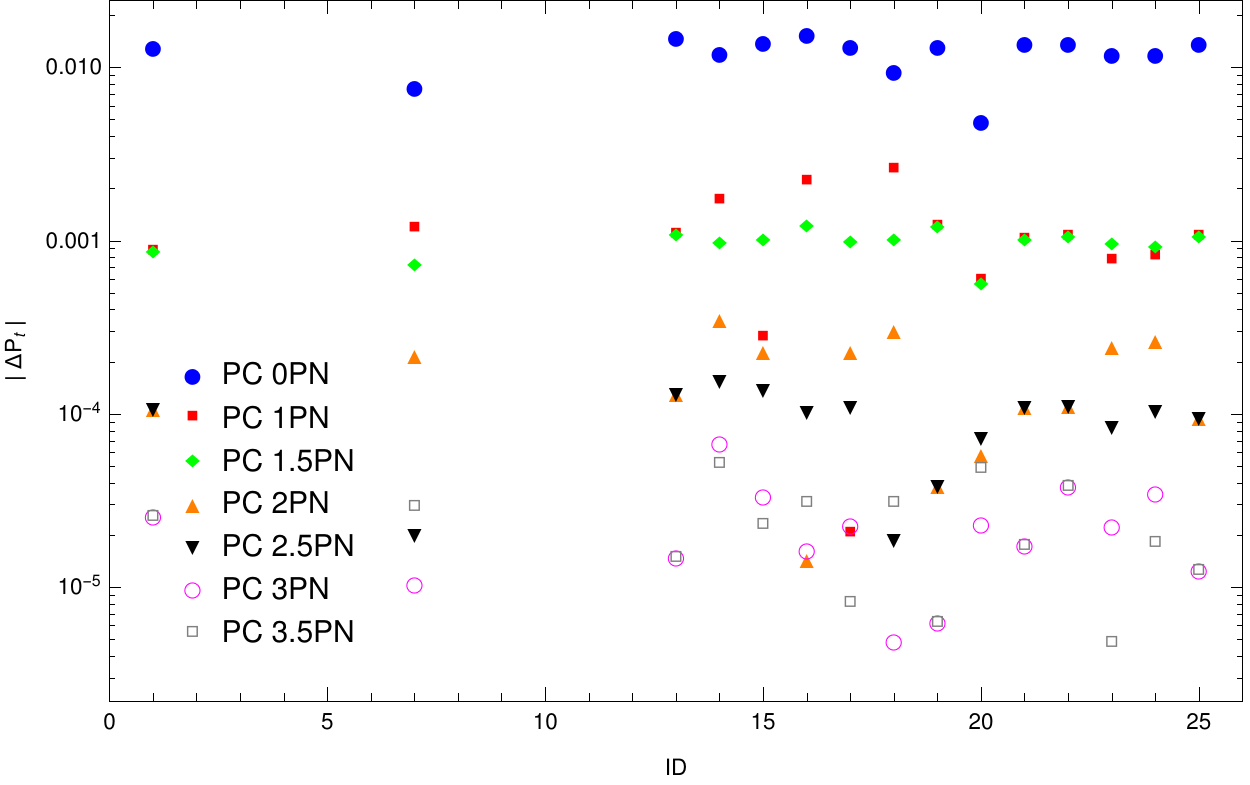}
\end{minipage}
\begin{minipage}[b]{\columnwidth}
\captionsetup{justification=centering}
\includegraphics[scale=0.65]{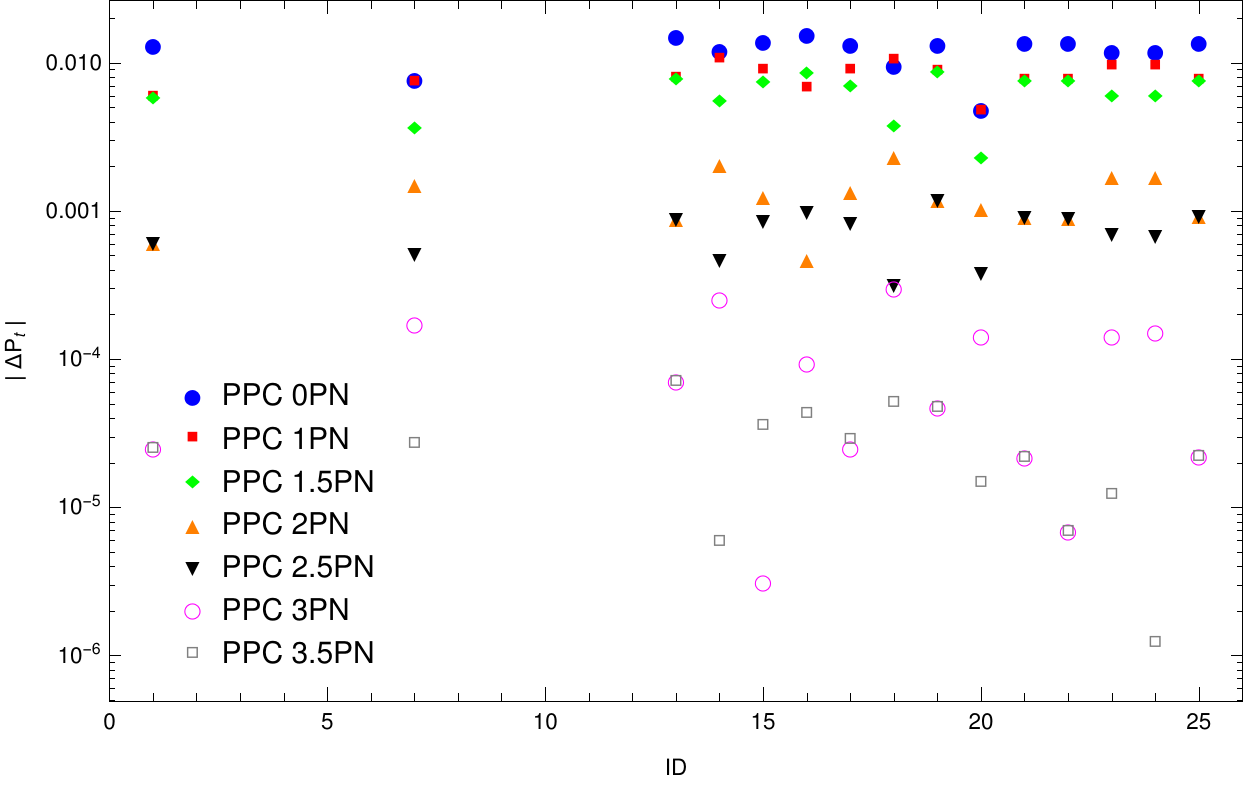}
\end{minipage}
\begin{minipage}[b]{\columnwidth}
\captionsetup{justification=centering}
\includegraphics[scale=0.65]{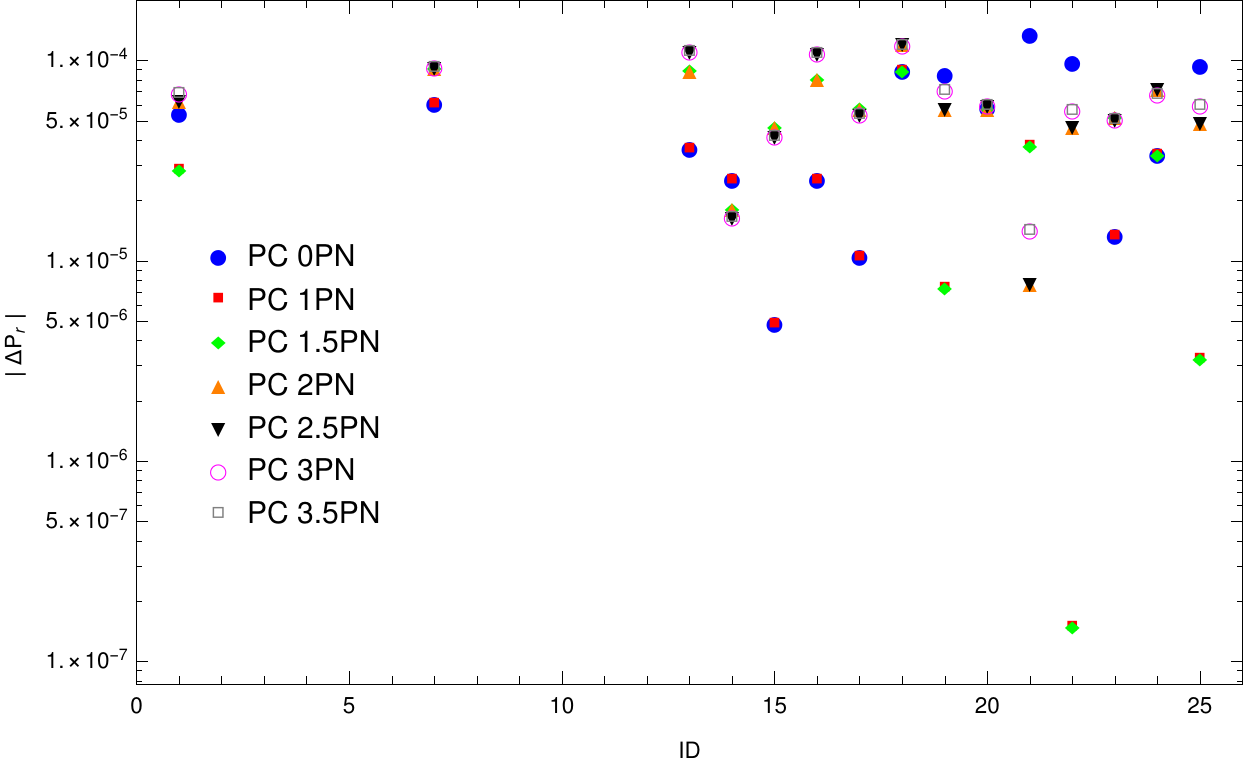}
\end{minipage}
\caption{Absolute difference between the low eccentric tangential or radial momentum value, ($p_t^{\text{ref}}$, $p_r^{\text{ref}}$), from Table \ref{tab:tabNR3} and the momentum at a given PN order, ($p_t^i$, $p_r^i$), with $i=0,1,1.5,2,2.5,3,3.5$ for the configurations of Table \ref{tab:tabNR3}. In the upper panel the absolute difference for the values of the PC tangential momentum at different PN orders are shown, in the middle one the absolute differences for the PPC tangential momentum, and in the lower panel the absolute differences for the radial momentum.  The IDs in the three plots correspond to those of Table \ref{tab:tabNR3}.}
\label{fig:ptPNorder}
\end{figure}

\section{Summary and conclusions} \label{sec:summary}

In this paper we have developed a suite of methods that use post-Newtonian approximations to produce low eccentricity initial data for binary black hole evolutions in numerical relativity.
The methods rely on working with sufficiently large numerical separations to allow for several orbits before merger, so that an accurate fit can be performed to determine the eccentricity of the numerical data, and to avoid a breakdown of the post-Newtonian approximations that we use. These requirements are consistent with the usual requirements for waveform modeling, where e.g., waveforms need to be long enough to be able to glue NR data to PN data and construct a PN-NR hybrid waveform. Length requirements for numerical relativity waveforms have been discussed, e.g., in \cite{PhysRevD.82.064016,Boyle:2011dy,Ohme:2011zm}.

We have first compared three alternatives to set initial momenta from PN calculations: numerical integration from a large distance, and the PPC and PC approximations. We have found that, as expected, integration from a large distance indeed leads to PN evolutions with negligible eccentricity, and that PPC initial data yield smaller eccentricity than PC initial data for PN evolutions. When using the same prescriptions for the initial momenta in NR evolutions however, PC initial data typically lead to smaller eccentricities. The fact that PC initial data result in particularly low eccentricities of puncture initial data for NR simulations has previously been noted in \cite{Healy:2017zqj}, and we extend their explicit formulas for the momenta in the postcircular approximation to $3.5$PN order.

We have also discussed the post-postcircular approximation, which provides an analytical correction to the tangential momentum, maintaining the radial momentum from the PC approximation. We have explicitly shown the success of the PPC approximation in PN, and the ability to generate low eccentric PN initial data without any further iteration. However, we have also checked performing 24 simulations corresponding to 12 configurations that PPC momenta do not provide lower eccentric initial data than PC in NR. This is mainly due to the fact that PPC corrections do not provide the appropriate correction in NR, because the difference due to the fact that PN and NR have different coordinate systems up to $2.5$PN overshoots the correction.

The key idea of our eccentricity reduction procedure is to derive explicit formulas to the correction of either the tangential and radial momentum, or alternatively the separation and radial momentum, in terms of the measured eccentricity and the initial phase of the oscillations related to eccentricity. We have found that fitting the orbital frequency evolution to the TaylorT3 approximant provides a robust method to determine the eccentricity and initial phase with sufficient accuracy to be able to reduce the eccentricity below $10^{-3}$ in a single iteration. 
Reducing the eccentricity below $10^{-4}$ for our moving puncture evolutions will require one to reduce the numerical noise with improved choices for numerical dissipation, on which we will report elsewhere, and will also require a discussion of spin oscillations in the context of spin precession. Such a study has been performed in  \cite{Buonanno:2010yk}, where eccentricities below $10^{-4}$ have been achieved for precessing simulations in four iterations, while we can reach eccentricities of the order $\mathcal{O}(10^{-4})$ in one iteration. 
We also note that in  \cite{Buonanno:2010yk} the test cases start at separation $d=16M$, which  would improve the performance of the PC approximation and of the PN expressions on which we base our eccentricity reduction method; however, here we want to show that the method works well for simulations of intermediate length, of typically between 5 and 10 orbits, which can be performed with moderate computational cost and are still very beneficial for waveform modeling purposes.

When only moderately low eccentricities are desired, or as the first step in an iterative procedure, it is possible to only correct the tangential momentum, using Eq.~\eqref{eq517}. In this case it is important to accurately determine the eccentricity, but not the phase $\Psi$ in Eq.~\eqref{eq539}. The two-dimensional schemes, where the radial momentum is also changed, rely on an accurate extrapolation of the residual \eqref{eq534} to the initial time $t=0$ of the simulation. This is made possible by fitting the frequency evolution to the TaylorT3 approximant. This ansatz avoids artifacts outside of the numerical fitting region, which are characteristic for polynomial fits.

In this paper we use the orbital frequency, which is coordinate dependent, to measure eccentricity. In order to suppress initial gauge transients we use a small value of  the $\eta$ parameter appearing in the  $\Gamma$-driver shift condition, $\eta=0.25$, as has been studied in some detail in \cite{Purrer:2012wy}. Here we show that this method works well for a variety of cases, including precessing ones. As an alternative to measuring the eccentricity from the orbital frequency one could use the wave frequency \cite{Purrer:2012wy}, employing methods to denoise the wave frequency such as those employed in \cite{Purrer:2012wy}. For our setup of numerical relativity simulations, abstaining from an accurate determination of the gravitational wave signal, however, saves computational cost for the low resolution simulations used to compute the corrected momenta or separation.
The method should also apply to numerical relativity codes based on different methods and in particular coordinate gauges, e.g. the \texttt{SpEC} code \cite{Buchman:2012dw}. 
We also hope that the simplicity of the procedure benefits extension to binary systems containing matter, in particular neutron stars or boson stars.

A coordinate dependence that is more problematic than the one for the orbital frequency arises from mapping PN momenta at some coordinate separation in the PN ADMTT gauge to the same value of the coordinate separation of the punctures in the coordinates corresponding to Bowen-York initial data, which only agree with ADMTT up to second PN order \cite{Tichy:2002ec}. We have addressed this problem by developing two versions of our iterative scheme to correct the initial parameters of the simulation to reduce the inherent eccentricity: In the ``traditional'' version we correct our initial guesses for the tangential and radial momenta $(p_t,p_r)$. In the alternative version we correct for the initial separation and $p_r$. The second version, which appears logically more consistent, is hoped to provide advantages when constructing hybrid PN-NR dynamics and waveforms, e.g. for precessing configurations, where not only the waveforms but also the spin evolutions should be glued together. This will be explored in future work.

The corrections $p_t \rightarrow \lambda_t p_t$ \eqref{eq539}, $p_r \rightarrow \lambda_r p_r$  \eqref{eq540}, and $r_0 \rightarrow r_0+  \delta r$  \eqref{eq548.2} can be applied iteratively;
we find, however, that when combining the procedure with PC initial momenta for iteration 0, for the cases we have studied, which include  mass ratios as high as $8$ and also some precessing simulations, a single iteration was sufficient to obtain eccentricities below $10^{-4}$. For those cases where we applied a second iteration, eccentricities dropped at least by an additional factor of 2.
However, there may be parts of the parameter space, especially high mass ratios and high spins, where the initial PN formulas will produce significantly larger eccentricities of the order 
$\mathcal{O}(10^{-2})$ requiring in those cases more than iteration to reach a value of the eccentricity of the order $\mathcal{O}(10^{-4})$.

Our implementation of the eccentricity reduction procedure with analytical formulas relating the eccentricity and the correction of the momenta needed to eliminate it provides real control in the eccentricity of a PN or NR simulation. As shown in this article this can be used to reduce the amount of eccentricity in the simulation, but it can also be used to perform eccentric simulations. This can be used to generate eccentric NR and PN simulations, which can be glued into hybrid waveforms that are the fundamental inputs for waveform modeling.
\\\\ 
\section{Acknowledgements}  \label{sec:Acknowledgements}

We   acknowledge Francisco Jiménez Forteza for initially using Eq. (\ref{eq516}) at Newtonian order for eccentricity reduction, which has been the starting point of this project, and we thank Maria Haney for valuable discussions. This work was supported by the Spanish Ministry of Education, Culture and Sport Grant No. FPU15/03344, the Spanish Ministry of Economy and Competitiveness Grant No. FPA2016-76821-P, the Agencia estatal de Investigación, the RED CONSOLIDER CPAN  FPA2017-90687-REDC, RED CONSOLIDER MULTIDARK: Multimessenger Approach for Dark Matter Detection, No. FPA2017-90566-REDC, Red nacional de astropartículas (RENATA), No. FPA2015-68783-REDT, European Union FEDER funds, Vicepresidència i Conselleria d'Innovació, Recerca i Turisme, Conselleria d'Educació, i Universitats del Govern de les Illes Balears i Fons Social Europeu, Gravitational waves, black holes and fundamental physics, CA COST Action CA16104 (EU Framework Programme Horizon 2020, H2020-MSCA-IF-2016).

\texttt{BAM} and \texttt{ET} simulations were carried out on the BSC MareNostrum computer under PRACE and RES (Red Española de Supercomputación) allocations and on the FONER computer at the University of the Balearic Islands. 

\appendix

\section{PN initial data formulas} \label{sec:AppendixA}
We present the formulas for the orbital frequency, the tangential momentum and the ADM mass as a function of the separation at $3.5$PN order, 
\begin{widetext}
\begin{align}
M\Omega &=\frac{1}{r^{3/2}}\left(1-\frac{1}{r}\left[ \frac{  \left(3 q^2+5 q+3\right)}{2 (q+1)^2}\right]+ \frac{1}{r^{3/2}} \left[-\frac{(3 q+4) \chi _{1z}}{4 (q+1)^2 }-  \frac{q (4 q+3) \chi _{2z}}{4 (q+1)^2 }\right] +\frac{1}{r^2} \left[-\frac{3 q^2 \chi _{2x}^2}{2 (q+1)^2}+ \frac{3 q^2 \chi _{2y}^2}{4 (q+1)^2 } \right.\right. \nonumber 
\\
& \left.\left.+\frac{3 q^2 \chi _{2z}^2}{4 (q+1)^2 }+\frac{24 q^4+103 q^3+164 q^2+103 q+24}{16 (q+1)^4 }-\frac{3 \chi _{1x}^2}{2 (q+1)^2 }-\frac{3 q \chi _{1x}\chi _{2x}}{(q+1)^2 }+\frac{3 \chi _{1y}^2}{4 (q+1)^2 }+\frac{3 q \chi _{1y} \chi _{2y}}{2 (q+1)^2 }\right. \right. \nonumber 
\\
&\left. \left. +\frac{3 \chi _{1z}^2}{4 (q+1)^2 }+\frac{3 q \chi _{1z} \chi _{2z}}{2 (q+1)^2}\right]+\frac{1}{r^{5/2}} \left[\frac{3 \left(13 q^3+34 q^2+30 q+16\right) \chi _{1z}}{16 (q+1)^4}+  \frac{3 q \left(16 q^3+30 q^2+34 q+13\right) \chi _{2z}}{16 (q+1)^4 }\right]  \right. \nonumber
\\
&  \left.+\frac{1}{r^3} \left[\frac{\left(155 q^2+180 q+76\right) \chi _{1x}^2}{16 (q+1)^4 r^3}+\frac{q \left(120 q^2+187 q+120\right) \chi _{1x} \chi _{2x}}{8 (q+1)^4 r^3}-\frac{\left(55 q^2+85 q+43\right) \chi _{1y}^2}{8 (q+1)^4 r^3}\right.\right. \nonumber
\\
&  \left. \left.-\frac{q \left(54 q^2+95 q+54\right) \chi _{1y} \chi _{2y}}{4 (q+1)^4 r^3}-\frac{q \left(96 q^2+127 q+96\right) \chi _{1z} \chi _{2z}}{16 (q+1)^4 r^3}+\frac{q^2 \left(76 q^2+180 q+155\right) \chi _{2x}^2}{16 (q+1)^4 r^3}\right.\right. \nonumber
\\
&  \left. \left. -\frac{q^2 \left(43 q^2+85 q+55\right) \chi _{2y}^2}{8 (q+1)^4 r^3}-\frac{q^2 (2 q+5) (14 q+27) \chi _{2z}^2}{32 (q+1)^4 r^3} -\frac{(5 q+2) (27 q+14) \chi _{1z}^2}{32 (q+1)^4 r^3}\right.\right. \nonumber
\\
&\left. \left. +\frac{501 \pi ^2 q (q+1)^4-4 \left(120 q^6+2744 q^5+10049 q^4+14820 q^3+10049 q^2+2744 q+120\right)}{384 (q+1)^6 r^3}\right]+\frac{1}{r^{7/2}} \left[ \frac{3 (4 q+1) q^3 \chi _{2 x}^2 \chi _{2 z}}{2 (q+1)^4}\right.\right. \nonumber
\\
&\left. \left.-\frac{3 (4 q+1) q^3 \chi _{2 y}^2 \chi _{2 z}}{8 (q+1)^4}-\frac{3 (4 q+1) q^3 \chi _{2 z}^3}{8 (q+1)^4}+\chi _{1x} \left(\frac{9 (2 q+1) q^2 \chi _{2 x} \chi _{2 z}}{4 (q+1)^4}+\frac{9 (q+2) q \chi _{2 x} \chi _z}{4 (q+1)^4}\right)+\chi _{1y} \left(\frac{9 q^2 \chi _{2 y} \chi _{1z}}{4 (q+1)^4}+\frac{9 q^2 \chi _{2 y} \chi _{2 z}}{4 (q+1)^4}\right)\right.\right. \nonumber
\\
&\left. \left.+\chi _{1z} \left(\frac{9 q^2 (2 q+3) \chi _{2 x}^2}{4 (q+1)^4}-\frac{9 q^2 (q+2) \chi _{2 y}^2}{4 (q+1)^4}-\frac{9 q^2 \chi _{2 z}^2}{4 (q+1)^3}-\frac{135 q^5+385 q^4+363 q^3+377 q^2+387 q+168}{32 (q+1)^6}\right)\right.\right. \nonumber
\\
&\left. \left.-\frac{\left(168 q^5+387 q^4+377 q^3+363 q^2+385 q+135\right) q \chi _{2 z}}{32 (q+1)^6}+\chi _{1x}^2 \left(\frac{3 (q+4) \chi _{1z}}{2 (q+1)^4}+\frac{9 q (3 q+2) \chi _{2 z}}{4 (q+1)^4}\right)\right.\right. \nonumber
\\
&\left. \left.+\chi _{1y}^2 \left(-\frac{3 (q+4) \chi _{1z}}{8 (q+1)^4}-\frac{9 q (2 q+1) \chi _{2 z}}{4 (q+1)^4}\right)-\frac{9 q \chi _{1z}^2 \chi _{2 z}}{4 (q+1)^3}-\frac{3 (q+4) \chi _{1z}^3}{8 (q+1)^4}\right]\right),
\label{29}
\end{align}

\begin{align}
p_t &= \frac{q}{(1+q)^2}\frac{1}{r^{1/2}}\left(1+ \frac{2 }{r} +
\frac{1}{r^{3/2}} \left[-\frac{3 \left(4 q^2+3 q\right) \chi _{2z}}{4 (q+1)^2}-\frac{3 (3 q+4) \chi _{1z}}{4 (q+1)^2}\right]+ \frac{1 }{r^{2}}\left[ -\frac{3 q^2 \chi _{2x}^2}{2 (q+1)^2} +\frac{3 q^2 \chi _{2y}^2}{4 (q+1)^2}+\frac{3 q^2 \chi _{2z}^2}{4 (q+1)^2} \right.\right.\nonumber
\\
& \left. \left. +\frac{42 q^2+41 q+42}{16 (q+1)^2}-\frac{3 \chi _{1x}^2}{2 (q+1)^2}-\frac{3 q  \chi _{1x} \chi _{2x}}{(q+1)^2}+\frac{3 \chi _{1y}^2}{4 (q+1)^2}+\frac{3 q \chi _{1y}\chi _{2y}}{2 (q+1)^2}+\frac{3 \chi _{1z}^2}{4 (q+1)^2}+\frac{3 q \chi _{1z} \chi _{2z}}{2 (q+1)^2}\right]  \right.\nonumber
\\
& \left. +\frac{1 }{r^{5/2}} \left[ \frac{\left(13 q^3+60 q^2+116 q+72\right) \chi _{1z}}{16 (q+1)^4}+\frac{\left(-72 q^4-116 q^3-60 q^2-13 q\right) \chi _{2z}}{16 (q+1)^4}  \right]+\frac{1 }{r^{3}} \left[\frac{\left(472 q^2-640\right) \chi _{1x}^2}{128 (q+1)^4}\right.\right.\nonumber
\\
&\left.\left. + \frac{\left(-512 q^2-640 q-64\right) \chi _{1y}^2}{128 (q+1)^4}+\frac{\left(-108 q^2+224 q+512\right) \chi _{1z}^2}{128 (q+1)^4}+\frac{\left(472 q^2-640 q^4\right) \chi _{2x}^2}{128 (q+1)^4}+\frac{\left(192 q^3+560 q^2+192 q\right) \chi _{1x} \chi _{2x}}{128 (q+1)^4} \right. \right. \nonumber
\\
&\left.\left.+\frac{\left(-864 q^3-1856 q^2-864 q\right) \chi _{1y} \chi _{2y}}{128 (q+1)^4}+\frac{\left(480 q^3+1064 q^2+480 q\right) \chi _{1z} \chi _{2z}}{128 (q+1)^4}+\frac{\left(-64 q^4-640 q^3-512 q^2\right) \chi _{2y}^2}{128 (q+1)^4}\right. \right. \nonumber
\\
&\left.\left.+\frac{\left(512 q^4+224 q^3-108 q^2\right) \chi _{2z}^2}{128 (q+1)^4} +\frac{480 q^4+163 \pi ^2 q^3-2636 q^3+326 \pi ^2 q^2-6128 q^2+163 \pi ^2 q-2636 q+480}{128 (q+1)^4} \right] \right. \nonumber
\\
&\left. + \frac{1 }{r^{7/2}} \left[ \frac{5 (4 q+1) q^3 \chi _{2 x}^2 \chi _{2 z}}{2 (q+1)^4}-\frac{5 (4 q+1) q^3 \chi _{2 y}^2 \chi _{2 z}}{8 (q+1)^4}-\frac{5 (4 q+1) q^3 \chi _{2 z}^3}{8 (q+1)^4}+\chi _{1x} \left(\frac{15 (2 q+1) q^2 \chi _{2 x} \chi _{2 z}}{4 (q+1)^4}+\frac{15 (q+2) q \chi _{2 x} \chi _{1z}}{4 (q+1)^4}\right)\right.\right. \nonumber
\\
&\left.\left.+\chi _{1y} \left(\frac{15 q^2 \chi _{2 y} \chi _{1z}}{4 (q+1)^4}+\frac{15 q^2 \chi _{2 y} \chi _{2 z}}{4 (q+1)^4}\right)+\chi _{1z} \left(\frac{15 q^2 (2 q+3) \chi _{2 x}^2}{4 (q+1)^4}-\frac{15 q^2 (q+2) \chi _{2 y}^2}{4 (q+1)^4}-\frac{15 q^2 \chi _{2 z}^2}{4 (q+1)^3} \right.\right.\right. \nonumber
\\
& \left.\left.\left. -\frac{103 q^5+145 q^4-27 q^3+252 q^2+670 q+348}{32 (q+1)^6}\right)-\frac{\left(348 q^5+670 q^4+252 q^3-27 q^2+145 q+103\right) q \chi _{2 z}}{32 (q+1)^6}+\chi _{1x}^2 \left(\frac{5 (q+4) \chi _{1z}}{2 (q+1)^4}\right.\right.\right.\nonumber
\\
& \left.\left.\left.+\frac{15 q (3 q+2) \chi _{2 z}}{4 (q+1)^4}\right)+\chi _{1y}^2 \left(-\frac{5 (q+4) \chi _{1z}}{8 (q+1)^4}-\frac{15 q (2 q+1) \chi _{2 z}}{4 (q+1)^4}\right)-\frac{15 q \chi _{1z}^2 \chi _{2 z}}{4 (q+1)^3}-\frac{5 (q+4) \chi _{1z}^3}{8 (q+1)^4} \right] \right),
\label{30}
\end{align}

\begin{align} 
\frac{M_{\text{ADM}}}{M} &= 1- \frac{  q}{2 r (q+1)^2}+ \frac{1}{r^2}  \frac{q \left(7 q^2+13 q+7\right)}{8 (q+1)^4 }+\frac{1}{r^{5/2}} \left[-\frac{(4 q+3) q^2  \chi _{2z}}{4 (q+1)^4}-\frac{(3 q+4) q  \chi _{1z}}{4 (q+1)^4} \right]+\frac{1}{r^{3}}   \left[ -\frac{q^3 \chi _{2x}^2}{2 (q+1)^4} \right. \nonumber
\\
& \left. +\frac{q^3 \chi _{2z}^2}{4 (q+1)^4}-\frac{q^2 \chi _{1x}\chi _{2x}}{(q+1)^4}+\frac{q^2 \chi _{1y} \chi _{2y}}{2 (q+1)^4}+\frac{q^2 \chi _{1z} \chi _{2z}}{2 (q+1)^4}+\frac{\left(9 q^4+16 q^3+13 q^2+16 q+9\right) q}{16 (q+1)^6}-\frac{q \chi _{1x}^2}{2 (q+1)^4}  +\frac{q \chi _{1y}^2}{4 (q+1)^4}\right.  \nonumber
\\
&\left.+\frac{q \chi _{1z}^2}{4 (q+1)^4}\right]+\frac{1}{r^{7/2}}  \left[-\frac{\left(q^3+14 q^2+42 q+32\right) q  \chi _{1z}}{16 (q+1)^6}-\frac{\left(32 q^3+42 q^2+14 q+1\right) q^2  \chi _{2z}}{16 (q+1)^6}\right]+\frac{1}{r^{4}}  \left[\frac{179 q^7}{128 (q+1)^8}-\frac{3497 q^6}{384 (q+1)^8} \right.  \nonumber
\\
& \left. -\frac{18707 q^5}{384 (q+1)^8}-\frac{9787 q^4}{128 (q+1)^8}+\frac{9 q^3 \chi _{1x} \chi _{2 x}}{8 (q+1)^6}-\frac{18707 q^3}{384 (q+1)^8}+\frac{\left(25 q^2-12 q-52\right) q \chi _{1x}^2}{16 (q+1)^6}-\frac{3 \left(4 q^2+9 q+4\right) q^2 \chi _{1y} \chi _{2 y}}{4 (q+1)^6}\right. \nonumber
\\
& \left.+\frac{3 \left(10 q^2+21 q+10\right) q^2 \chi _z \chi _{2 z}}{8 (q+1)^6}+\frac{\left(3 q^2+38 q+50\right) q \chi _z^2}{16 (q+1)^6}-\frac{3497 q^2}{384 (q+1)^8}-\frac{\left(52 q^2+12 q-25\right) q^3 \chi _{2 x}^2}{16 (q+1)^6}\right. \nonumber
\\
& \left.+\frac{\left(q^2-17 q-15\right) q^3 \chi _{2 y}^2}{8 (q+1)^6}+\frac{\left(-15 q^3-17 q^2+q\right) \chi _y^2}{8 (q+1)^6}+\frac{\left(50 q^2+38 q+3\right) q^3 \chi _{2 z}^2}{16 (q+1)^6}+\pi ^2 \left(\frac{81 q^6}{128 (q+1)^8}+\frac{81 q^5}{32 (q+1)^8}\right. \right. \nonumber
\\
& \left.\left.+\frac{243 q^4}{64 (q+1)^8}+\frac{81 q^3}{32 (q+1)^8}+\frac{81 q^2}{128 (q+1)^8}\right)+\frac{179 q}{128 (q+1)^8}\right]+\frac{1}{r^{9/2}} \left[ \frac{3 (20 q+7) q^4 \chi _{2 x}^2 \chi _{2 z}}{8 (q+1)^6}-\frac{3 (12 q+5) q^4 \chi _{2 y}^2 \chi _{2 z}}{16 (q+1)^6}\right. \nonumber
\\
& \left.-\frac{3 (12 q+5) q^4 \chi _{2 z}^3}{16 (q+1)^6}+\chi _{1x}^2 \left(\frac{3 (22 q+15) q^2 \chi _{2 z}}{8 (q+1)^6}+\frac{3 (7 q+20) q \chi _{1z}}{8 (q+1)^6}\right)+\chi _{1y}^2 \left(-\frac{3 (28 q+15) q^2 \chi _{2 z}}{16 (q+1)^6}-\frac{3 (5 q+12) q \chi _z}{16 (q+1)^6}\right)\right. \nonumber
\\
& \left.-\frac{3 (22 q+23) q^2 \chi _{1z}^2 \chi _{2 z}}{16 (q+1)^6}+\chi _{1x} \left(\frac{3 (5 q+3) q^3 \chi _{2 x} \chi _{2 z}}{2 (q+1)^6}+\frac{3 (3 q+5) q^2 \chi _{2 x} \chi _{1z}}{2 (q+1)^6}\right)+\chi _{1y} \left(\frac{3 (3-4 q) q^3 \chi _{2 y} \chi _{2 z}}{8 (q+1)^6}+\frac{3 (3 q-4) q^2 \chi _{2 y} \chi _z}{8 (q+1)^6}\right)\right. \nonumber
\\
& \left.+\chi _{1z} \left(\frac{3 (15 q+22) q^3 \chi _{2 x}^2}{8 (q+1)^6}-\frac{3 (15 q+28) q^3 \chi _{2 y}^2}{16 (q+1)^6}-\frac{3 (23 q+22) q^3 \chi _{2 z}^2}{16 (q+1)^6}-\frac{\left(128 q^5+181 q^4-88 q^3+81 q^2+544 q+312\right) q}{64 (q+1)^8}\right)\right. \nonumber
\\
& \left.-\frac{\left(312 q^5+544 q^4+81 q^3-88 q^2+181 q+128\right) q^2 \chi _{2 z}}{64 (q+1)^8}-\frac{3 (5 q+12) q \chi _z^3}{16 (q+1)^6} \right].
\label{31}
\end{align}
\end{widetext}

In this article we have chosen to specify as initial condition the orbital separation $r$. Another possible choice is to specify the initial orbital frequency where we want to start our simulation. Then, Eq. \eqref{29} can be inverted to obtain the relation $r(\Omega)$, and then write the separation, the tangential momentum and the ADM mass in terms of the orbital frequency. The resulting equations are,

\begin{widetext}
\begin{align}
\frac{r}{M} & = \Omega ^{-2/3}-\frac{3 q^2+5 q+3}{3 (q+1)^2}+\left[-\frac{(3 q+4) \chi _{1z}}{6 (q+1)^2}-\frac{q (4 q+3) \chi _{2 z}}{6 (q+1)^2}\right] \Omega^{1/3}+\left[-\frac{\chi _{1x}^2}{(q+1)^2}-\frac{2 q \chi _{2 x} \chi _{1x}}{(q+1)^2}+\frac{\chi _{1y}^2}{2 (q+1)^2}+\frac{q^2 \chi _{2 y}^2}{2 (q+1)^2}\right. \nonumber
 \\
& \left.+\frac{\chi _{1z}^2}{2 (q+1)^2}+\frac{q^2 \chi _{2 z}^2}{2 (q+1)^2}+\frac{-18 q^4+9 q^3+62 q^2+9 q-18}{72 (q+1)^4}+\frac{q \chi _{1y} \chi _{2 y}}{(q+1)^2}+\frac{q \chi_{1z} \chi _{2 z}}{(q+1)^2}-\frac{q^2 \chi _{2 x}^2}{(q+1)^2}\right] \Omega ^{2/3}  \nonumber
 \\
&+\left[\frac{q \left(3 q^2-6 q-26\right) \chi _{1z}}{24 (q+1)^4}-\frac{q \left(26 q^2+6 q-3\right) \chi _{2 z}}{24 (q+1)^4}\right] \Omega +\left[\frac{\left(71 q^2+40 q-8\right) \chi _{1x}^2}{24 (q+1)^4}+\frac{q \left(36 q^2+47 q+36\right) \chi _{2 x} \chi _{1x}}{12 (q+1)^4}\right. \nonumber
  \\
& \left.+\frac{q^2 \left(-8 q^2+40 q+71\right) \chi _{2 x}^2}{24 (q+1)^4}+\frac{\left(-27 q^2-15 q+7\right) \chi _{1z}^2}{18 (q+1)^4}+\frac{q^2 \left(7 q^2-15 q-27\right) \chi _{2 z}^2}{18 (q+1)^4}-\frac{q \left(11 q^2+20 q+11\right) \chi _{1y} \chi _{2 y}}{2 (q+1)^4}\right. \nonumber
 \\
& \left.-\frac{\left(17 q^2+25 q+11\right) \chi _{1y}^2}{6 (q+1)^4}-\frac{q^2 \left(11 q^2+25 q+17\right) \chi _{2 y}^2}{6 (q+1)^4}-\frac{q \left(15 q^2+17 q+15\right) \chi _{1z} \chi _{2 z}}{9 (q+1)^4}  +\frac{167 \pi ^2 q}{192 (q+1)^2} \right. \nonumber
 \\
& \left.-\frac{324 q^6+16569 q^5+65304 q^4+98086 q^3+65304 q^2+16569 q+324}{1296 (q+1)^6}\right] \Omega ^{4/3} +\left[\frac{(4 q+9) \chi _{2 z}^3 q^3}{12 (q+1)^4}+\frac{(4 q-3) \chi _{2 x}^2 \chi _{2 z} q^3}{3 (q+1)^4}\right. \nonumber
 \\
& \left.+\frac{(4 q+9) \chi _{2 y}^2 \chi _{2 z} q^3}{12 (q+1)^4}+\frac{(11 q+13) \chi _{1z}^2 \chi _{2 z} q}{6 (q+1)^4}+\frac{\left(-72 q^5+1629 q^4+6731 q^3+7197 q^2+2331 q+81\right) \chi _{2 z} q}{432 (q+1)^6}+\frac{(9 q+4) \chi _{1z}^3}{12 (q+1)^4}\right. \nonumber
 \\
& \left.+\left(\frac{(4-3 q) \chi _{1z}}{3 (q+1)^4}+\frac{q (11 q+6) \chi _{2 z}}{6 (q+1)^4}\right) \chi _{1x}^2+\left(\frac{(9 q+4) \chi _{1z}}{12 (q+1)^4}-\frac{q (10 q+3) \chi _{2 z}}{6 (q+1)^4}\right) \chi _{1y}^2+\left(-\frac{(14 q+15) \chi _{2 x} \chi _{2 z} q^2}{6 (q+1)^4}\right. \right.\nonumber
 \\
& \left.\left.-\frac{(15 q+14) \chi _{2 x} \chi _{1z} q}{6 (q+1)^4}\right) \chi _{1x}+\left(\frac{(16 q+21) \chi _{2 y} \chi _{2 z} q^2}{6 (q+1)^4}+\frac{(21 q+16) \chi _{2 y} \chi _{1z} q}{6 (q+1)^4}\right) \chi _{1y}+\left(\frac{q^2 (6 q+11) \chi _{2 x}^2}{6 (q+1)^4}+\frac{q^2 (13 q+11) \chi _{2 z}^2}{6 (q+1)^4}\right. \right.\nonumber
 \\*
& \left.\left.+\frac{81 q^5+2331 q^4+7197 q^3+6731 q^2+1629 q-72}{432 (q+1)^6}\right) \chi _{1z}\right] \Omega ^{5/3}.
\label{a1}
\end{align}

\begin{align}
\frac{p_t}{M} & = \frac{q \Omega^{1/3}}{(q+1)^2}+\frac{q \left(15 q^2+29 q+15\right) \Omega }{6 (q+1)^4}+\left[-\frac{2 (4 q+3) \chi _{2 z} q^2}{3 (q+1)^4}-\frac{2 (3 q+4) \chi _z q}{3 (q+1)^4}\right] \Omega ^{4/3}+\left[\frac{\chi _{2 y}^2 q^3}{2 (q+1)^4}+\frac{\chi _{2 z}^2 q^3}{2 (q+1)^4}-\frac{\chi _{2 x}^2 q^3}{(q+1)^4}\right.\nonumber
 \\*
& \left.+\frac{\chi _{1y} \chi _{2 y} q^2}{(q+1)^4}+\frac{\chi _{1z} \chi _{2 z} q^2}{(q+1)^4}-\frac{2 \chi _{1x} \chi _{2 x} q^2}{(q+1)^4}+\frac{\chi _{1y}^2 q}{2 (q+1)^4}+\frac{\chi _{1z}^2 q}{2 (q+1)^4}+\frac{\left(441 q^4+1440 q^3+1997 q^2+1440 q+441\right) q}{72 (q+1)^6}\right.\nonumber
 \\*
& \left.-\frac{\chi _{1x}^2 q}{(q+1)^4}\right] \Omega ^{5/3}+\left[-\frac{\left(16 q^3+29 q^2+22 q+7\right) \chi _{2 z} q^2}{2 (q+1)^6}-\frac{\left(7 q^3+22 q^2+29 q+16\right) \chi _{1z} q}{2 (q+1)^6}\right] \Omega ^2+\left(\frac{\left(-116 q^2+4 q+53\right) \chi _{2 x}^2 q^3}{24 (q+1)^6}\right.\nonumber
 \\*
& \left.+\frac{\left(5 q^2-41 q-31\right) \chi _{2 y}^2 q^3}{12 (q+1)^6}+\frac{53 \chi _{1x} \chi _{2 x} q^3}{12 (q+1)^6}-\frac{\left(q^2+147 q+81\right) \chi _{2 z}^2 q^3}{36 (q+1)^6}+\frac{161 \pi ^2 q^2}{192 (q+1)^4}-\frac{\left(8 q^2+21 q+8\right) \chi _{1y} \chi _{2 y} q^2}{2 (q+1)^6}\right.\nonumber
 \\*
& \left.-\frac{\left(21 q^2+67 q+21\right) \chi _{1z} \chi _{2 z} q^2}{18 (q+1)^6}+\frac{\left(53 q^2+4 q-116\right) \chi _{1x}^2 q}{24 (q+1)^6}-\frac{\left(31 q^2+41 q-5\right) \chi _{1y}^2 q}{12 (q+1)^6}-\frac{\left(81 q^2+147 q+1\right) \chi _{1z}^2 q}{36 (q+1)^6}\right.\nonumber
 \\*
& \left.+\frac{\left(20007 q^6+60489 q^5+67320 q^4+53681 q^3+67320 q^2+60489 q+20007\right) q}{1296 (q+1)^8}\right] \Omega ^{7/3}+\left[\frac{2 (2 q+3) \chi _{2 z}^3 q^4}{3 (q+1)^6} \right.\nonumber
 \\*
& \left. +\frac{(4 q-9) \chi _{2 x}^2 \chi _{2 z} q^4}{3 (q+1)^6}+\frac{2 (2 q+3) \chi _{2 y}^2 \chi _{2 z} q^4}{3 (q+1)^6}+\frac{(32 q+37) \chi _{1z}^2 \chi _{2 z} q^2}{6 (q+1)^6}+\frac{2 (3 q+2) \chi _{1z}^3 q}{3 (q+1)^6}+\left(\frac{(7 q+3) \chi _{2 z} q^2}{3 (q+1)^6}+\frac{(4-9 q) \chi _{1z} q}{3 (q+1)^6}\right) \chi _{1x}^2\right.\nonumber
 \\*
& \left.-\frac{\left(10656 q^5+25560 q^4+24235 q^3+14853 q^2+8550 q+2808\right) \chi _{2 z} q^2}{432 (q+1)^8}+\left(\frac{2 q (3 q+2) \chi _{1z}}{3 (q+1)^6}-\frac{q^2 (16 q+3) \chi _{2 z}}{6 (q+1)^6}\right) \chi _{1y}^2\right.\nonumber
 \\*
& \left.+\left(-\frac{(22 q+21) \chi _{2 x} \chi _{2 z} q^3}{3 (q+1)^6}-\frac{(21 q+22) \chi _{2 x} \chi _{1z} q^2}{3 (q+1)^6}\right) \chi _{1x}+\left(\frac{4 (5 q+6) \chi _{2 y} \chi _{2 z} q^3}{3 (q+1)^6}+\frac{4 (6 q+5) \chi _{2 y} \chi _{1z} q^2}{3 (q+1)^6}\right) \chi _{1y}\right. \nonumber
 \\*
& \left. +\left(\frac{(3 q+7) \chi _{2 x}^2 q^3}{3 (q+1)^6}+\frac{(37 q+32) \chi _{2 z}^2 q^3}{6 (q+1)^6}-\frac{\left(2808 q^5+8550 q^4+14853 q^3+24235 q^2+25560 q+10656\right) q}{432 (q+1)^8}\right.\right. \nonumber
 \\*
& \left. \left.
-\frac{(3 q+16) \chi _{2 y}^2 q^3}{6 (q+1)^6}\right) \chi _{1z}\right] \Omega ^{8/3}
\label{a2}
\end{align}
\begin{align}
\frac{M_{\text{ADM}}}{M} & =1-\frac{q \Omega^{-1}}{2 (q+1)^2 }+\left[-\frac{(4 q+3) q^2 \chi _{2 z}}{4 (q+1)^4}-\frac{(3 q+4) q \chi _{1z}}{4 (q+1)^4}\right]\Omega ^{-5/2} +\frac{\left(7 q^2+13 q+7\right) q \Omega ^{-2}}{8 (q+1)^4 }  \nonumber
 \\
&  +\left[-\frac{\left(32 q^3+42 q^2+14 q+1\right) q^2 \chi _{2 z}}{16 (q+1)^6}-\frac{\left(q^3+14 q^2+42 q+32\right) q \chi _{1z}}{16 (q+1)^6}\right]\Omega ^{-7/2}+\left[-\frac{q^3 \chi _{2 x}^2}{2 (q+1)^4}+\frac{q^3 \chi _{2 y}^2}{4 (q+1)^4}+\frac{q^3 \chi _{2 z}^2}{4 (q+1)^4}\right. \nonumber
 \\
& \left.-\frac{q^2 \chi _{1x} \chi _{2 x}}{(q+1)^4}+\frac{q^2 \chi _y \chi _{2 y}}{2 (q+1)^4}+\frac{q^2 \chi _z \chi _{2 z}}{2 (q+1)^4}+\frac{\left(9 q^4+16 q^3+13 q^2+16 q+9\right) q}{16 (q+1)^6}-\frac{q \chi _{1x}^2}{2 (q+1)^4}+\frac{q \chi _{1y}^2}{4 (q+1)^4}+\frac{q \chi _{1z}^2}{4 (q+1)^4} \right]\Omega ^{-3} \nonumber
 \\
& +\left[  \frac{9 q^3 \chi _{1x} \chi _{2 x}}{8 (q+1)^6}+\frac{\left(25 q^2-12 q-52\right) q \chi _{1x}^2}{16 (q+1)^6}-\frac{3 \left(4 q^2+9 q+4\right) q^2 \chi _{1y} \chi _{2 y}}{4 (q+1)^6}+\frac{3 \left(10 q^2+21 q+10\right) q^2 \chi _{1z} \chi _{2 z}}{8 (q+1)^6}\right. \nonumber
 \\
& \left.+\frac{\left(3 q^2+38 q+50\right) q \chi _{1z}^2}{16 (q+1)^6}+\frac{81 \pi ^2 q^2}{128 (q+1)^4}-\frac{\left(52 q^2+12 q-25\right) q^3 \chi _{2 x}^2}{16 (q+1)^6}+\frac{\left(q^2-17 q-15\right) q^3 \chi _{2 y}^2}{8 (q+1)^6}+\frac{\left(-15 q^3-17 q^2+q\right) \chi _{1y}^2}{8 (q+1)^6}\right. \nonumber
 \\
& \left.+\frac{\left(50 q^2+38 q+3\right) q^3 \chi _{2 z}^2}{16 (q+1)^6}+\frac{\left(537 q^6-3497 q^5-18707 q^4-29361 q^3-18707 q^2-3497 q+537\right) q}{384 (q+1)^8}\right]\Omega ^{-4}  \nonumber
 \\
&  + \left[ \frac{3 (20 q+7) q^4 \chi _{2 x}^2 \chi _{2 z}}{8 (q+1)^6}-\frac{3 (12 q+5) q^4 \chi _{2 y}^2 \chi _{2 z}}{16 (q+1)^6}-\frac{3 (12 q+5) q^4 \chi _{2 z}^3}{16 (q+1)^6}+\chi _{1x}^2 \left(\frac{3 (22 q+15) q^2 \chi _{2 z}}{8 (q+1)^6}+\frac{3 (7 q+20) q \chi _{1z}}{8 (q+1)^6}\right) \right. \nonumber
 \\
&  \left. +\chi _{1y}^2 \left(-\frac{3 (28 q+15) q^2 \chi _{2 z}}{16 (q+1)^6}-\frac{3 (5 q+12) q \chi _{1z}}{16 (q+1)^6}\right)-\frac{3 (22 q+23) q^2 \chi _{1z}^2 \chi _{2 z}}{16 (q+1)^6}+\chi _{1x} \left(\frac{3 (5 q+3) q^3 \chi _{2 x} \chi _{2 z}}{2 (q+1)^6}+\frac{3 (3 q+5) q^2 \chi _{2 x} \chi _{1z}}{2 (q+1)^6}\right)\right.  \nonumber
 \\
& \left. +\chi _{1y} \left(\frac{3 (3-4 q) q^3 \chi _{2 y} \chi _{2 z}}{8 (q+1)^6}+\frac{3 (3 q-4) q^2 \chi _{2 y} \chi _{1z}}{8 (q+1)^6}\right)+\chi _{1z} \left(\frac{3 (15 q+22) q^3 \chi _{2 x}^2}{8 (q+1)^6}-\frac{3 (15 q+28) q^3 \chi _{2 y}^2}{16 (q+1)^6}-\frac{3 (23 q+22) q^3 \chi _{2 z}^2}{16 (q+1)^6}\right.\right. \nonumber
 \\
& \left.\left.-\frac{\left(128 q^5+181 q^4-88 q^3+81 q^2+544 q+312\right) q}{64 (q+1)^8}\right)-\frac{\left(312 q^5+544 q^4+81 q^3-88 q^2+181 q+128\right) q^2 \chi _{2 z}}{64 (q+1)^8}\right. \nonumber
 \\
& \left.-\frac{3 (5 q+12) q \chi _{1z}^3}{16 (q+1)^6} \right]\Omega ^{-9/2}
\label{a3}
\end{align}

The expression used in this paper for the gravitational wave energy flux \cite{Blanchet2014,Fluxspinterms} is 
\begin{align}
\frac{dE_{\text{GW}}}{dt}& =\frac{32}{5} \eta^2  \Omega^{10/3} 	\left( 1+\left[-\frac{35 \eta }{12}-\frac{1247}{336}\right] \Omega ^{2/3}+\left[4 \pi -\frac{5 \delta  }{4}\Sigma_l-4 S_l \right]  \Omega+\left[\frac{65 \eta ^2}{18}+\frac{9271 \eta }{504}-\frac{44711}{9072}-\frac{89 \delta  \chi_a \chi_s}{48}+\frac{287 \delta  \chi_a \chi_s}{48} \right.\right.  \nonumber
\\
& \left. \left. +\left(\frac{33}{16}-8 \eta \right) \chi_a^2 - \left( \frac{33}{16}-\frac{\eta }{4} \right) \chi_s^2 \right] \Omega ^{4/3}+\left[ \pi  \left(-\frac{583 \eta }{24}-\frac{8191}{672} \right) +\frac{43 \delta  \eta   \Sigma_l}{4}-\frac{13 \delta   \Sigma_l}{16}+\frac{272 \eta  S_l}{9}-\frac{9 S_l}{2} \right] \Omega ^{5/3} \right. \nonumber
\\
& \left.\left. + \Omega ^2 \left[ -\frac{4843497781}{69854400}-\frac{775 \eta ^3}{324}-\frac{94403 \eta ^2}{3024}+\left(\frac{8009293}{54432}-\frac{41 \pi ^2}{64}\right) \eta  +\frac{287 \pi ^2}{192}+\frac{1712}{105} \left(-\gamma +\frac{35 \pi ^2}{107}  -\frac{1}{2} \log \left(16 \Omega ^{2/3}\right) \right) \right.\right. \right. \nonumber
\\
& \left. \left.\left.-\frac{31 \pi  \delta  \Sigma_l}{6}-16 \pi  S_l+\delta  \left(\frac{611}{252}-\frac{809 \eta }{18}\right) \chi_a \chi_s+\left(43 \eta ^2-\frac{8345 \eta }{504}+\frac{611}{504}\right) \chi_a^2+\left(\frac{173 \eta ^2}{18}-\frac{2393 \eta }{72}+\frac{611}{504}\right) \chi_s^2  \right]  + \left[ -\frac{31 \pi  \delta  \Sigma_l}{6} \right. \right. \right. \nonumber
\\
& \left. \left. \left. -16 \pi  S_l\right] \Omega
^{3}+\left[\pi  \left(\frac{193385 \eta ^2}{3024}+\frac{214745 \eta }{1728}-\frac{16285}{504}\right)
\right.-\frac{1501}{36} \delta  \eta ^2 \Sigma_l +\frac{1849 \delta  \eta  \Sigma_l}{126}+\frac{9535 \delta  \Sigma_l}{336}-\frac{2810 \eta ^2 S_l}{27} +\frac{6172 \eta  S_l}{189} \right.\right. \nonumber
\\
& \left.\left.+\frac{476645 S_l}{6804}+ \delta  \left(\frac{34 \eta }{3}-\frac{71}{24}\right) \chi_a^3+\delta  \left(\frac{109 \eta }{6}-\frac{71}{8}\right) \chi_a \chi_s^2+\left(-\frac{104 \eta ^2}{3}+\frac{263 \eta }{6}-\frac{71}{8}\right) \chi_a^2 \chi_s+\left(-\frac{2 \eta ^2}{3}+\frac{28 \nu }{3} \right. \right.\right. \nonumber
\\
& \left. \left.\left.  -\frac{71}{24}\right) \chi_s^3 \right]\Omega ^{7/3} + \left[ \frac{130583 \pi  \delta  \eta  \Sigma_l}{2016}-\frac{7163 \pi  \delta  \Sigma_l}{672}  +\frac{13879 \pi  \eta  S_l}{72}-\frac{3485 \pi  S_l}{96} \right]\Omega ^{8/3} \right)  
\label{a4}
\end{align}
\end{widetext}
where we have the following definitions:
\begin{align}
\vec{\lambda}  & =\frac{\vec{L}}{|\vec{L}|} ,
\\ 
\delta &=\frac{m_2-m_1}{m_1+m_2},
 \\
 \chi_1 &=S_{1z}/m_1^2,
 \\
 \chi_2 &= S_{2z}/m_2^2,
\\
 \chi_a &=\frac{ \chi_1-\chi_2}{2},
\\ 
 \chi_s &=\frac{\chi_1+ \chi_2}{2},
 \\
 S_l &= m_1^2  \chi_1+m_2^2 \chi_2,
 \\
 \Sigma_l &= m_2 \chi_2- m_1 \chi_1,
 \\
S_l & =(\vec{S}_1+\vec{S}_2 ) \cdot \vec{\lambda} ,
 \\
\Sigma_l & =(m_1+m_2) \left(\frac{\vec{S}_2}{m_2}-\frac{\vec{S}_1}{m_1}\right) \cdot \vec{\lambda} .
 \label{a5}
\end{align}

\section{Ansatz coefficients} \label{sec:AppendixB}
The coefficients of the ansatz of the nonspinning fit described in Sec. \ref{sec:eccMes} are derived using the energy given by Eq. \eqref{eq13.1} and the gravitational wave energy flux given by Eq. \eqref{a4}. The coefficients are 
\begin{widetext}
\begin{align}
b_1 & = \frac{11 \eta }{32}+\frac{743}{2688} ,
\\ 
b_2 &=\frac{1}{320} \left(-113 \left[\left(\sqrt{1-4 \eta }-1\right) \chi_{1z}-\left(\sqrt{1-4 \eta }+1\right) \chi_{2z}\right]-96 \pi \right)-\frac{19}{80} \eta  (\chi_{1z}+\chi_{2z}),
 \\
 b_3 &= \frac{371 \eta ^2}{2048}+\frac{\eta  \left(61236 \text{s1z}^2-119448 \chi_{1z} \chi_{2z}+61236 \chi_{2z}^2+56975\right)}{258048}+\frac{1}{14450688} \left[1714608 \left(\sqrt{1-4 \eta }-1\right) \chi_{1z}^2 \right. \nonumber
  \\
& \left. -1714608 \left(\sqrt{1-4 \eta }+1\right) \chi_{2z}^2+1855099\right],
 \\
 b_4 &= -\frac{1}{128} 17 \eta ^2 (\chi_{1z}+\chi_{2z})+\frac{\eta  \left(117 \pi -2 \left(\left(63 \sqrt{1-4 \eta }+1213\right) \chi_{1z}+\left(1213-63 \sqrt{1-4 \eta }\right) \chi_{2z}\right)\right)}{2304}  \nonumber
  \\
& +\frac{-146597 \left(\left(\sqrt{1-4 \eta }-1\right)\chi_{1z}-\left(\sqrt{1-4 \eta }+1\right) \chi_{2z}\right)-46374 \pi }{129024},
\\
b_5 &= \frac{235925 \eta ^3}{1769472}+\eta ^2 \left[\frac{335129 \chi_{1z}^2}{2457600}-\frac{488071 \text{s1z} \chi_{2z}}{1228800}+\frac{335129 \chi_{2z}^2}{2457600}-\frac{30913}{1835008}\right] \nonumber
\\
& +\eta  \left[\frac{\left(23281001-6352738 \sqrt{1-4 \eta }\right) \chi_{1z}^2}{68812800}+\chi_{1z} \left(\frac{1051 \pi }{3200}-\frac{377345 \chi_{2z}}{1376256}\right)+\frac{\left(6352738 \sqrt{1-4 \eta }+23281001\right) \chi_{2z}^2}{68812800} \right. \nonumber
\\
& \left. +\frac{1051 \pi \chi_{2z}}{3200}-\frac{451 \pi ^2}{2048}+\frac{25302017977}{4161798144}\right]+\frac{6127 \pi  \left(\sqrt{1-4 \eta }-1\right) \chi_{1z}}{12800}-\frac{16928263 \left(\sqrt{1-4 \eta }+1\right) \chi_{2z}^2}{137625600} \nonumber
\\
& +\frac{16928263 \left(\sqrt{1-4 \eta }-1\right) \chi_{1z}^2}{137625600}-\frac{6127 \pi  \left(\sqrt{1-4 \eta }+1\right) \chi_{2z}}{12800}+\frac{53 \pi ^2}{200}-\frac{720817631400877}{288412611379200} +\frac{107}{280} \left[ \gamma +  \log (2 \theta ) \right] 
 \label{a6}
\end{align}
\end{widetext}

\section{Numerical Relativity setup} \label{sec:AppendixC}
\subsection{BAM}
Here we describe the numerical setup for NR simulations produced using the \texttt{BAM} code. The numerical setup is similar to that in \cite{Purrer:2012wy} but we present the details here for completeness. The \texttt{BAM} code starts with black-hole binary puncture initial data \cite{Bowen1980,Brandt1997} and evolves them using the $\chi$ variant of the moving puncture \cite{Baker:2005vv,Campanelli:2005dd} version of the BSSN \citep{PhysRevD.52.5428,Baumgarte:1998te} formulation of the Einstein equations. The black-hole punctures are initially placed on the $y$ axis at positions $y_1=-qD/(1+q)$ and $y_2=D/(1+q)$, where $D$ is the coordinate distance between the two punctures and the mass ratio is $q=m_2/m_1 > 1$. The punctures are provided initial momenta $\textbf{p} = \left( \mp p_t, \pm p_r,0 \right) $. The spin parameter of a  BH is defined as $\chi_i=S_i/m_i^2$. 

The code uses sixth-order spatial finite-difference derivatives, a fourth-order Runge-Kutta algorithm, and Kreiss-Oliger (KO) dissipation terms that converge at fifth order. Moreover, the code utilizes 16 mesh-refinement buffer points and the base configuration consists of $n_1$ nested mesh-refinement boxes with $N^3$ points surrounding each black hole, and $n_2$ nested boxes with $(2N)^3$ points surrounding the entire system. On the levels where the extraction of gravitational radiation is performed $(4N)^3$ points are used in order to extract more accurately the gravitational waves emitted by the binary. These waves are computed from the Newman-Penrose scalar $\Psi_4$ \cite{Bruegmann:2006at}. In addition, in order to reduce gauge oscillations in the orbital quantities we set the value of the parameter $\eta$ appearing in the $\Gamma$-driver shift condition to $0.25$ for simulations used to reduce the eccentricity and we use $\eta=1$ for higher resolution production simulations, that will be used in future waveform modeling and LIGO data analysis.

\subsection{Einstein Toolkit}
The \texttt{Einstein Toolkit} (\texttt{ET}) is an open source code suite for relativistic astrophysics simulations built around the Cactus framework, where individual modules are denoted \textit{thorns}. The numerical setup of the simulations is similar to that used in \cite{Pollney:2009yz} but we present the details here for completeness. 

The simulations use standard Bowen-York initial data \cite{Bowen1980,Brandt1997} computed using the \texttt{TwoPunctures} thorn \cite{Ansorg2004}. Time evolution is performed using the $W$ variant \cite{Marronetti:2007wz} of the BSSN formulation \citep{PhysRevD.52.5428,Baumgarte:1998te} of the Einstein equations by \texttt{McLachlan} \cite{Brown:2008sb}, in which the BHs are evolved using the standard moving puncture gauge conditions \cite{Baker:2005vv,Campanelli:2005dd}. The lapse is evolved according to the `$1+\log$' condition \cite{Bona:1994dr} and the shift evolved using the hyperbolic $\tilde{\Gamma}$-driver equation \cite{Alcubierre:2002kk}.

The simulations were performed using eighth order accurate finite differencing along with the appropriate KO dissipation terms. Adaptive mesh refinement is provided by \texttt{Carpet}, with the near zone being computed with high resolution Cartesian grids that track the motion of the BHs and the wave extraction zone being computed on spherical grids using the \texttt{LLAMA} multipatch infrastructure \cite{Pollney:2009yz}. By using grids adapted to the spherical topology of the wave extraction zone, we are able to efficiently compute high-accuracy waveforms at large extraction radii relative to standard Cartesian grids. The apparent horizons are computed using \texttt{AHFinderDirect} \cite{Thornburg:2003sf} and a calculation of the spins is performed in the dynamical horizon formalism using the \texttt{QuasiLocalMeasures} thorn \cite{Dreyer:2002mx}. In contrast to BAM, the two punctures are initially placed on the $x$ axis at positions $x_1=D/(1+q)$ and $x_2=-qD/(1+q)$, in which $D$ is the coordinate distance separation and we assume $m_1 > m_2$. Initial momenta are chosen such that $\textbf{p}=\left( \mp p_r, \pm p_t,0 \right)$. As with BAM, the parameter $\eta$ that appears in the $\Gamma$-driver shift condition, which is denoted $\textit{BetaDriver}$ in the \texttt{McLachlan} code, is set to $0.25$ for low-resolution simulations and set to $1$ for the higher resolution production runs. 

The gravitational waves are computed using \texttt{WeylScal4} and the GW strain $h$ calculated from $\Psi_4$ using fixed-frequency integration \cite{Reisswig:2010di}. The thorns \texttt{McLachlan} and \texttt{WeylScal4} are generated using the \texttt{Kranc} \cite{Husa:2004ip} automated-code-generation package. The \texttt{ET} simulations are managed using \texttt{Simulation Factory} \cite{SimulationFactory} and the analysis and postprocessing of \texttt{ET} waveforms were performed using the open source Mathematica package \texttt{Simulation Tools} \cite{SimulationTools}.

\renewcommand{\refname}{}

\bibliography{uib}

\begin{thebibliography}{74}
\expandafter\ifx\csname natexlab\endcsname\relax\def\natexlab#1{#1}\fi
\expandafter\ifx\csname bibnamefont\endcsname\relax
  \def\bibnamefont#1{#1}\fi
\expandafter\ifx\csname bibfnamefont\endcsname\relax
  \def\bibfnamefont#1{#1}\fi
\expandafter\ifx\csname citenamefont\endcsname\relax
  \def\citenamefont#1{#1}\fi
\expandafter\ifx\csname url\endcsname\relax
  \def\url#1{\texttt{#1}}\fi
\expandafter\ifx\csname urlprefix\endcsname\relax\def\urlprefix{URL }\fi
\providecommand{\bibinfo}[2]{#2}
\providecommand{\eprint}[2][]{\url{#2}}

\bibitem[{\citenamefont{Abbott}(2016{\natexlab{a}})}]{PhysRevLett.116.061102}
\bibinfo{author}{\bibfnamefont{B.~P. e.~a.} \bibnamefont{Abbott}}
  (\bibinfo{collaboration}{LIGO Scientific Collaboration and Virgo
  Collaboration}), \bibinfo{journal}{Phys. Rev. Lett.}
  \textbf{\bibinfo{volume}{116}}, \bibinfo{pages}{061102}
  (\bibinfo{year}{2016}{\natexlab{a}}).

\bibitem[{\citenamefont{Aasi et~al.}(2015)}]{TheLIGOScientific:2014jea}
\bibinfo{author}{\bibfnamefont{J.}~\bibnamefont{Aasi}} \bibnamefont{et~al.}
  (\bibinfo{collaboration}{LIGO Scientific}), \bibinfo{journal}{Class. Quant.
  Grav.} \textbf{\bibinfo{volume}{32}}, \bibinfo{pages}{074001}
  (\bibinfo{year}{2015}).

\bibitem[{\citenamefont{Abbott}(2016{\natexlab{b}})}]{PhysRevLett.116.241103}
\bibinfo{author}{\bibfnamefont{B.~P. e.~a.} \bibnamefont{Abbott}}
  (\bibinfo{collaboration}{LIGO Scientific Collaboration and Virgo
  Collaboration}), \bibinfo{journal}{Phys. Rev. Lett.}
  \textbf{\bibinfo{volume}{116}}, \bibinfo{pages}{241103}
  (\bibinfo{year}{2016}{\natexlab{b}}).

\bibitem[{\citenamefont{Abbott et~al.}(2017{\natexlab{a}})}]{Abbott:2017vtc}
\bibinfo{author}{\bibfnamefont{B.~P.} \bibnamefont{Abbott}}
  \bibnamefont{et~al.} (\bibinfo{collaboration}{VIRGO, LIGO Scientific}),
  \bibinfo{journal}{Phys. Rev. Lett.} \textbf{\bibinfo{volume}{118}},
  \bibinfo{pages}{221101} (\bibinfo{year}{2017}{\natexlab{a}}),
  \eprint{1706.01812}.

\bibitem[{\citenamefont{Abbott et~al.}(2017{\natexlab{b}})}]{Abbott:2017gyy}
\bibinfo{author}{\bibfnamefont{B.~P.} \bibnamefont{Abbott}}
  \bibnamefont{et~al.} (\bibinfo{collaboration}{Virgo, LIGO Scientific}),
  \bibinfo{journal}{Astrophys. J.} \textbf{\bibinfo{volume}{851}},
  \bibinfo{pages}{L35} (\bibinfo{year}{2017}{\natexlab{b}}),
  \eprint{1711.05578}.

\bibitem[{\citenamefont{Abbott et~al.}(2017{\natexlab{c}})}]{Abbott:2017oio}
\bibinfo{author}{\bibfnamefont{B.~P.} \bibnamefont{Abbott}}
  \bibnamefont{et~al.} (\bibinfo{collaboration}{Virgo, LIGO Scientific}),
  \bibinfo{journal}{Phys. Rev. Lett.} \textbf{\bibinfo{volume}{119}},
  \bibinfo{pages}{141101} (\bibinfo{year}{2017}{\natexlab{c}}).

\bibitem[{\citenamefont{Abbott
  et~al.}(2017{\natexlab{d}})}]{TheLIGOScientific:2017qsa}
\bibinfo{author}{\bibfnamefont{B.~P.} \bibnamefont{Abbott}}
  \bibnamefont{et~al.} (\bibinfo{collaboration}{Virgo, LIGO Scientific}),
  \bibinfo{journal}{Phys. Rev. Lett.} \textbf{\bibinfo{volume}{119}},
  \bibinfo{pages}{161101} (\bibinfo{year}{2017}{\natexlab{d}}).

\bibitem[{\citenamefont{Abbott et~al.}(2016)}]{TheLIGOScientific:2016wfe}
\bibinfo{author}{\bibfnamefont{B.~P.} \bibnamefont{Abbott}}
  \bibnamefont{et~al.} (\bibinfo{collaboration}{Virgo, LIGO Scientific}),
  \bibinfo{journal}{Phys. Rev. Lett.} \textbf{\bibinfo{volume}{116}},
  \bibinfo{pages}{241102} (\bibinfo{year}{2016}), \eprint{1602.03840}.

\bibitem[{\citenamefont{Peters and Mathews}(1963)}]{PhysRev.131.435}
\bibinfo{author}{\bibfnamefont{P.~C.} \bibnamefont{Peters}} \bibnamefont{and}
  \bibinfo{author}{\bibfnamefont{J.}~\bibnamefont{Mathews}},
  \bibinfo{journal}{Phys. Rev.} \textbf{\bibinfo{volume}{131}},
  \bibinfo{pages}{435} (\bibinfo{year}{1963}).

\bibitem[{\citenamefont{Peters}(1964)}]{PhysRev.136.B1224}
\bibinfo{author}{\bibfnamefont{P.~C.} \bibnamefont{Peters}},
  \bibinfo{journal}{Phys. Rev.} \textbf{\bibinfo{volume}{136}},
  \bibinfo{pages}{B1224} (\bibinfo{year}{1964}).

\bibitem[{\citenamefont{Pretorius}(2005)}]{Pretorius:2005gq}
\bibinfo{author}{\bibfnamefont{F.}~\bibnamefont{Pretorius}},
  \bibinfo{journal}{Phys. Rev. Lett.} \textbf{\bibinfo{volume}{95}},
  \bibinfo{pages}{121101} (\bibinfo{year}{2005}), \eprint{gr-qc/0507014}.

\bibitem[{\citenamefont{Campanelli et~al.}(2006)\citenamefont{Campanelli,
  Lousto, Marronetti, and Zlochower}}]{Campanelli:2005dd}
\bibinfo{author}{\bibfnamefont{M.}~\bibnamefont{Campanelli}},
  \bibinfo{author}{\bibfnamefont{C.~O.} \bibnamefont{Lousto}},
  \bibinfo{author}{\bibfnamefont{P.}~\bibnamefont{Marronetti}},
  \bibnamefont{and}
  \bibinfo{author}{\bibfnamefont{Y.}~\bibnamefont{Zlochower}},
  \bibinfo{journal}{Phys. Rev. Lett.} \textbf{\bibinfo{volume}{96}},
  \bibinfo{pages}{111101} (\bibinfo{year}{2006}).

\bibitem[{\citenamefont{Baker et~al.}(2006)\citenamefont{Baker, Centrella,
  Choi, Koppitz, and van Meter}}]{Baker:2005vv}
\bibinfo{author}{\bibfnamefont{J.~G.} \bibnamefont{Baker}},
  \bibinfo{author}{\bibfnamefont{J.}~\bibnamefont{Centrella}},
  \bibinfo{author}{\bibfnamefont{D.-I.} \bibnamefont{Choi}},
  \bibinfo{author}{\bibfnamefont{M.}~\bibnamefont{Koppitz}}, \bibnamefont{and}
  \bibinfo{author}{\bibfnamefont{J.}~\bibnamefont{van Meter}},
  \bibinfo{journal}{Phys. Rev. Lett.} \textbf{\bibinfo{volume}{96}},
  \bibinfo{pages}{111102} (\bibinfo{year}{2006}), \eprint{gr-qc/0511103}.

\bibitem[{\citenamefont{Mrou\'e et~al.}(2013)\citenamefont{Mrou\'e, Scheel,
  Szil\'agyi, Pfeiffer, Boyle, Hemberger, Kidder, Lovelace, Ossokine, Taylor
  et~al.}}]{PhysRevLett.111.241104}
\bibinfo{author}{\bibfnamefont{A.~H.} \bibnamefont{Mrou\'e}},
  \bibinfo{author}{\bibfnamefont{M.~A.} \bibnamefont{Scheel}},
  \bibinfo{author}{\bibfnamefont{B.}~\bibnamefont{Szil\'agyi}},
  \bibinfo{author}{\bibfnamefont{H.~P.} \bibnamefont{Pfeiffer}},
  \bibinfo{author}{\bibfnamefont{M.}~\bibnamefont{Boyle}},
  \bibinfo{author}{\bibfnamefont{D.~A.} \bibnamefont{Hemberger}},
  \bibinfo{author}{\bibfnamefont{L.~E.} \bibnamefont{Kidder}},
  \bibinfo{author}{\bibfnamefont{G.}~\bibnamefont{Lovelace}},
  \bibinfo{author}{\bibfnamefont{S.}~\bibnamefont{Ossokine}},
  \bibinfo{author}{\bibfnamefont{N.~W.} \bibnamefont{Taylor}},
  \bibnamefont{et~al.}, \bibinfo{journal}{Phys. Rev. Lett.}
  \textbf{\bibinfo{volume}{111}}, \bibinfo{pages}{241104}
  (\bibinfo{year}{2013}),
  \urlprefix\url{https://link.aps.org/doi/10.1103/PhysRevLett.111.241104}.

\bibitem[{\citenamefont{Healy et~al.}(2017{\natexlab{a}})\citenamefont{Healy,
  Lousto, Zlochower, and Campanelli}}]{Healy:2017psd}
\bibinfo{author}{\bibfnamefont{J.}~\bibnamefont{Healy}},
  \bibinfo{author}{\bibfnamefont{C.~O.} \bibnamefont{Lousto}},
  \bibinfo{author}{\bibfnamefont{Y.}~\bibnamefont{Zlochower}},
  \bibnamefont{and}
  \bibinfo{author}{\bibfnamefont{M.}~\bibnamefont{Campanelli}},
  \bibinfo{journal}{Class. Quant. Grav.} \textbf{\bibinfo{volume}{34}},
  \bibinfo{pages}{224001} (\bibinfo{year}{2017}{\natexlab{a}}),
  \eprint{1703.03423}.

\bibitem[{\citenamefont{Jani et~al.}(2016)\citenamefont{Jani, Healy, Clark,
  London, Laguna, and Shoemaker}}]{Jani:2016wkt}
\bibinfo{author}{\bibfnamefont{K.}~\bibnamefont{Jani}},
  \bibinfo{author}{\bibfnamefont{J.}~\bibnamefont{Healy}},
  \bibinfo{author}{\bibfnamefont{J.~A.} \bibnamefont{Clark}},
  \bibinfo{author}{\bibfnamefont{L.}~\bibnamefont{London}},
  \bibinfo{author}{\bibfnamefont{P.}~\bibnamefont{Laguna}}, \bibnamefont{and}
  \bibinfo{author}{\bibfnamefont{D.}~\bibnamefont{Shoemaker}},
  \bibinfo{journal}{Class. Quant. Grav.} \textbf{\bibinfo{volume}{33}},
  \bibinfo{pages}{204001} (\bibinfo{year}{2016}), \eprint{1605.03204}.

\bibitem[{\citenamefont{Br\"ugmann et~al.}(2008)\citenamefont{Br\"ugmann,
  Gonz\'alez, Hannam, Husa, Sperhake, and Tichy}}]{Bruegmann:2006at}
\bibinfo{author}{\bibfnamefont{B.}~\bibnamefont{Br\"ugmann}},
  \bibinfo{author}{\bibfnamefont{J.~A.} \bibnamefont{Gonz\'alez}},
  \bibinfo{author}{\bibfnamefont{M.}~\bibnamefont{Hannam}},
  \bibinfo{author}{\bibfnamefont{S.}~\bibnamefont{Husa}},
  \bibinfo{author}{\bibfnamefont{U.}~\bibnamefont{Sperhake}}, \bibnamefont{and}
  \bibinfo{author}{\bibfnamefont{W.}~\bibnamefont{Tichy}},
  \bibinfo{journal}{Phys. Rev. D} \textbf{\bibinfo{volume}{77}},
  \bibinfo{pages}{024027} (\bibinfo{year}{2008}).

\bibitem[{\citenamefont{Husa et~al.}(2016)\citenamefont{Husa, Khan, Hannam,
  P\"urrer, Ohme, Forteza, and Boh\'e}}]{Husa:2015iqa}
\bibinfo{author}{\bibfnamefont{S.}~\bibnamefont{Husa}},
  \bibinfo{author}{\bibfnamefont{S.}~\bibnamefont{Khan}},
  \bibinfo{author}{\bibfnamefont{M.}~\bibnamefont{Hannam}},
  \bibinfo{author}{\bibfnamefont{M.}~\bibnamefont{P\"urrer}},
  \bibinfo{author}{\bibfnamefont{F.}~\bibnamefont{Ohme}},
  \bibinfo{author}{\bibfnamefont{X.~J.} \bibnamefont{Forteza}},
  \bibnamefont{and} \bibinfo{author}{\bibfnamefont{A.}~\bibnamefont{Boh\'e}},
  \bibinfo{journal}{Phys. Rev. D} \textbf{\bibinfo{volume}{93}},
  \bibinfo{pages}{044006} (\bibinfo{year}{2016}).

\bibitem[{\citenamefont{Khan et~al.}(2016)\citenamefont{Khan, Husa, Hannam,
  Ohme, P\"urrer, Forteza, and Boh\'e}}]{Khan:2015jqa}
\bibinfo{author}{\bibfnamefont{S.}~\bibnamefont{Khan}},
  \bibinfo{author}{\bibfnamefont{S.}~\bibnamefont{Husa}},
  \bibinfo{author}{\bibfnamefont{M.}~\bibnamefont{Hannam}},
  \bibinfo{author}{\bibfnamefont{F.}~\bibnamefont{Ohme}},
  \bibinfo{author}{\bibfnamefont{M.}~\bibnamefont{P\"urrer}},
  \bibinfo{author}{\bibfnamefont{X.~J.} \bibnamefont{Forteza}},
  \bibnamefont{and} \bibinfo{author}{\bibfnamefont{A.}~\bibnamefont{Boh\'e}},
  \bibinfo{journal}{Phys. Rev. D} \textbf{\bibinfo{volume}{93}},
  \bibinfo{pages}{044007} (\bibinfo{year}{2016}).

\bibitem[{\citenamefont{Damour et~al.}(2008{\natexlab{a}})\citenamefont{Damour,
  Nagar, Hannam, Husa, and Br\"ugmann}}]{Damour:2008te}
\bibinfo{author}{\bibfnamefont{T.}~\bibnamefont{Damour}},
  \bibinfo{author}{\bibfnamefont{A.}~\bibnamefont{Nagar}},
  \bibinfo{author}{\bibfnamefont{M.}~\bibnamefont{Hannam}},
  \bibinfo{author}{\bibfnamefont{S.}~\bibnamefont{Husa}}, \bibnamefont{and}
  \bibinfo{author}{\bibfnamefont{B.}~\bibnamefont{Br\"ugmann}},
  \bibinfo{journal}{Phys. Rev. D} \textbf{\bibinfo{volume}{78}},
  \bibinfo{pages}{044039} (\bibinfo{year}{2008}{\natexlab{a}}).

\bibitem[{\citenamefont{Boh\'e et~al.}(2017)\citenamefont{Boh\'e, Shao,
  Taracchini, Buonanno, Babak, Harry, Hinder, Ossokine, P\"urrer, Raymond
  et~al.}}]{Bohe:2016gbl}
\bibinfo{author}{\bibfnamefont{A.}~\bibnamefont{Boh\'e}},
  \bibinfo{author}{\bibfnamefont{L.}~\bibnamefont{Shao}},
  \bibinfo{author}{\bibfnamefont{A.}~\bibnamefont{Taracchini}},
  \bibinfo{author}{\bibfnamefont{A.}~\bibnamefont{Buonanno}},
  \bibinfo{author}{\bibfnamefont{S.}~\bibnamefont{Babak}},
  \bibinfo{author}{\bibfnamefont{I.~W.} \bibnamefont{Harry}},
  \bibinfo{author}{\bibfnamefont{I.}~\bibnamefont{Hinder}},
  \bibinfo{author}{\bibfnamefont{S.}~\bibnamefont{Ossokine}},
  \bibinfo{author}{\bibfnamefont{M.}~\bibnamefont{P\"urrer}},
  \bibinfo{author}{\bibfnamefont{V.}~\bibnamefont{Raymond}},
  \bibnamefont{et~al.}, \bibinfo{journal}{Phys. Rev. D}
  \textbf{\bibinfo{volume}{95}}, \bibinfo{pages}{044028}
  (\bibinfo{year}{2017}).

\bibitem[{\citenamefont{Blackman et~al.}(2017)\citenamefont{Blackman, Field,
  Scheel, Galley, Ott, Boyle, Kidder, Pfeiffer, and
  Szil\'agyi}}]{Blackman:2017pcm}
\bibinfo{author}{\bibfnamefont{J.}~\bibnamefont{Blackman}},
  \bibinfo{author}{\bibfnamefont{S.~E.} \bibnamefont{Field}},
  \bibinfo{author}{\bibfnamefont{M.~A.} \bibnamefont{Scheel}},
  \bibinfo{author}{\bibfnamefont{C.~R.} \bibnamefont{Galley}},
  \bibinfo{author}{\bibfnamefont{C.~D.} \bibnamefont{Ott}},
  \bibinfo{author}{\bibfnamefont{M.}~\bibnamefont{Boyle}},
  \bibinfo{author}{\bibfnamefont{L.~E.} \bibnamefont{Kidder}},
  \bibinfo{author}{\bibfnamefont{H.~P.} \bibnamefont{Pfeiffer}},
  \bibnamefont{and}
  \bibinfo{author}{\bibfnamefont{B.}~\bibnamefont{Szil\'agyi}},
  \bibinfo{journal}{Phys. Rev. D} \textbf{\bibinfo{volume}{96}},
  \bibinfo{pages}{024058} (\bibinfo{year}{2017}).

\bibitem[{\citenamefont{Blanchet}(2014)}]{Blanchet2014}
\bibinfo{author}{\bibfnamefont{L.}~\bibnamefont{Blanchet}},
  \bibinfo{journal}{Living Reviews in Relativity} \textbf{\bibinfo{volume}{17}}
  (\bibinfo{year}{2014}), ISSN \bibinfo{issn}{1433-8351}.

\bibitem[{\citenamefont{Hannam et~al.}(2007)\citenamefont{Hannam, Husa,
  Pollney, Br\"ugmann, and OMurchadha}}]{Hannam:2006vv}
\bibinfo{author}{\bibfnamefont{M.}~\bibnamefont{Hannam}},
  \bibinfo{author}{\bibfnamefont{S.}~\bibnamefont{Husa}},
  \bibinfo{author}{\bibfnamefont{D.}~\bibnamefont{Pollney}},
  \bibinfo{author}{\bibfnamefont{B.}~\bibnamefont{Br\"ugmann}},
  \bibnamefont{and}
  \bibinfo{author}{\bibfnamefont{N.}~\bibnamefont{OMurchadha}},
  \bibinfo{journal}{Phys. Rev. Lett.} \textbf{\bibinfo{volume}{99}},
  \bibinfo{pages}{241102} (\bibinfo{year}{2007}).

\bibitem[{\citenamefont{Hannam et~al.}(2008)\citenamefont{Hannam, Husa, Ohme,
  Br\"ugmann, and OMurchadha}}]{Hannam:2008sg}
\bibinfo{author}{\bibfnamefont{M.}~\bibnamefont{Hannam}},
  \bibinfo{author}{\bibfnamefont{S.}~\bibnamefont{Husa}},
  \bibinfo{author}{\bibfnamefont{F.}~\bibnamefont{Ohme}},
  \bibinfo{author}{\bibfnamefont{B.}~\bibnamefont{Br\"ugmann}},
  \bibnamefont{and}
  \bibinfo{author}{\bibfnamefont{N.}~\bibnamefont{OMurchadha}},
  \bibinfo{journal}{Phys. Rev. D} \textbf{\bibinfo{volume}{78}},
  \bibinfo{pages}{064020} (\bibinfo{year}{2008}).

\bibitem[{\citenamefont{Bowen and York}(1980{\natexlab{a}})}]{PhysRevD.21.2047}
\bibinfo{author}{\bibfnamefont{J.~M.} \bibnamefont{Bowen}} \bibnamefont{and}
  \bibinfo{author}{\bibfnamefont{J.~W.} \bibnamefont{York}},
  \bibinfo{journal}{Phys. Rev. D} \textbf{\bibinfo{volume}{21}},
  \bibinfo{pages}{2047} (\bibinfo{year}{1980}{\natexlab{a}}).

\bibitem[{\citenamefont{Schaefer}(1985)}]{SchaeferADMTT}
\bibinfo{author}{\bibfnamefont{G.}~\bibnamefont{Schaefer}},
  \bibinfo{journal}{Annals of Physics} \textbf{\bibinfo{volume}{161}},
  \bibinfo{pages}{81} (\bibinfo{year}{1985}).

\bibitem[{\citenamefont{Tichy et~al.}(2003)\citenamefont{Tichy, Br\"ugmann,
  Campanelli, and Diener}}]{Tichy:2002ec}
\bibinfo{author}{\bibfnamefont{W.}~\bibnamefont{Tichy}},
  \bibinfo{author}{\bibfnamefont{B.}~\bibnamefont{Br\"ugmann}},
  \bibinfo{author}{\bibfnamefont{M.}~\bibnamefont{Campanelli}},
  \bibnamefont{and} \bibinfo{author}{\bibfnamefont{P.}~\bibnamefont{Diener}},
  \bibinfo{journal}{Phys. Rev. D} \textbf{\bibinfo{volume}{67}},
  \bibinfo{pages}{064008} (\bibinfo{year}{2003}).

\bibitem[{\citenamefont{Yunes and Tichy}(2006)}]{Yunes:2006iw}
\bibinfo{author}{\bibfnamefont{N.}~\bibnamefont{Yunes}} \bibnamefont{and}
  \bibinfo{author}{\bibfnamefont{W.}~\bibnamefont{Tichy}},
  \bibinfo{journal}{Phys. Rev.} \textbf{\bibinfo{volume}{D74}},
  \bibinfo{pages}{064013} (\bibinfo{year}{2006}).

\bibitem[{\citenamefont{Yunes et~al.}(2006)\citenamefont{Yunes, Tichy, Owen,
  and Br\"ugmann}}]{Yunes:2005nn}
\bibinfo{author}{\bibfnamefont{N.}~\bibnamefont{Yunes}},
  \bibinfo{author}{\bibfnamefont{W.}~\bibnamefont{Tichy}},
  \bibinfo{author}{\bibfnamefont{B.~J.} \bibnamefont{Owen}}, \bibnamefont{and}
  \bibinfo{author}{\bibfnamefont{B.}~\bibnamefont{Br\"ugmann}},
  \bibinfo{journal}{Phys. Rev. D} \textbf{\bibinfo{volume}{74}},
  \bibinfo{pages}{104011} (\bibinfo{year}{2006}).

\bibitem[{\citenamefont{Tichy and Marronetti}(2011)}]{Tichy:2010qa}
\bibinfo{author}{\bibfnamefont{W.}~\bibnamefont{Tichy}} \bibnamefont{and}
  \bibinfo{author}{\bibfnamefont{P.}~\bibnamefont{Marronetti}},
  \bibinfo{journal}{Phys. Rev.} \textbf{\bibinfo{volume}{D83}},
  \bibinfo{pages}{024012} (\bibinfo{year}{2011}).

\bibitem[{\citenamefont{Buonanno and Damour}(1999)}]{Buonanno:1998gg}
\bibinfo{author}{\bibfnamefont{A.}~\bibnamefont{Buonanno}} \bibnamefont{and}
  \bibinfo{author}{\bibfnamefont{T.}~\bibnamefont{Damour}},
  \bibinfo{journal}{Phys. Rev.} \textbf{\bibinfo{volume}{D59}},
  \bibinfo{pages}{084006} (\bibinfo{year}{1999}), \eprint{gr-qc/9811091}.

\bibitem[{\citenamefont{Husa et~al.}(2008{\natexlab{a}})\citenamefont{Husa,
  Hannam, Gonz\'alez, Sperhake, and Br\"ugmann}}]{Husa:2007rh}
\bibinfo{author}{\bibfnamefont{S.}~\bibnamefont{Husa}},
  \bibinfo{author}{\bibfnamefont{M.}~\bibnamefont{Hannam}},
  \bibinfo{author}{\bibfnamefont{J.~A.} \bibnamefont{Gonz\'alez}},
  \bibinfo{author}{\bibfnamefont{U.}~\bibnamefont{Sperhake}}, \bibnamefont{and}
  \bibinfo{author}{\bibfnamefont{B.}~\bibnamefont{Br\"ugmann}},
  \bibinfo{journal}{Phys. Rev. D} \textbf{\bibinfo{volume}{77}},
  \bibinfo{pages}{044037} (\bibinfo{year}{2008}{\natexlab{a}}).

\bibitem[{\citenamefont{Healy et~al.}(2017{\natexlab{b}})\citenamefont{Healy,
  Lousto, Nakano, and Zlochower}}]{Healy:2017zqj}
\bibinfo{author}{\bibfnamefont{J.}~\bibnamefont{Healy}},
  \bibinfo{author}{\bibfnamefont{C.~O.} \bibnamefont{Lousto}},
  \bibinfo{author}{\bibfnamefont{H.}~\bibnamefont{Nakano}}, \bibnamefont{and}
  \bibinfo{author}{\bibfnamefont{Y.}~\bibnamefont{Zlochower}},
  \bibinfo{journal}{Class. Quant. Grav.} \textbf{\bibinfo{volume}{34}},
  \bibinfo{pages}{145011} (\bibinfo{year}{2017}{\natexlab{b}}),
  \eprint{1702.00872}.

\bibitem[{\citenamefont{Damour and Nagar}(2007)}]{Damour:2007xr}
\bibinfo{author}{\bibfnamefont{T.}~\bibnamefont{Damour}} \bibnamefont{and}
  \bibinfo{author}{\bibfnamefont{A.}~\bibnamefont{Nagar}},
  \bibinfo{journal}{Phys. Rev.} \textbf{\bibinfo{volume}{D76}},
  \bibinfo{pages}{064028} (\bibinfo{year}{2007}), \eprint{0705.2519}.

\bibitem[{\citenamefont{Damour and Nagar}(2008)}]{Damour:2007yf}
\bibinfo{author}{\bibfnamefont{T.}~\bibnamefont{Damour}} \bibnamefont{and}
  \bibinfo{author}{\bibfnamefont{A.}~\bibnamefont{Nagar}},
  \bibinfo{journal}{Phys. Rev.} \textbf{\bibinfo{volume}{D77}},
  \bibinfo{pages}{024043} (\bibinfo{year}{2008}), \eprint{0711.2628}.

\bibitem[{\citenamefont{Damour and Deruelle}(1985)}]{1985AIHS...43..107D}
\bibinfo{author}{\bibfnamefont{T.}~\bibnamefont{Damour}} \bibnamefont{and}
  \bibinfo{author}{\bibfnamefont{N.}~\bibnamefont{Deruelle}},
  \bibinfo{journal}{Ann.~Inst.~Henri Poincar{\'e} Phys.~Th{\'e}or., Vol.~43,
  No.~1, p.~107 - 132}  (\bibinfo{year}{1985}).

\bibitem[{\citenamefont{Pratten~et al}(2020)}]{GeraintPhenomX}
\bibinfo{author}{\bibfnamefont{G.}~\bibnamefont{Pratten~et al}}
  (\bibinfo{year}{2020}).

\bibitem[{\citenamefont{Pfeiffer et~al.}(2007)\citenamefont{Pfeiffer, Brown,
  Kidder, Lindblom, Lovelace, and Scheel}}]{Pfeiffer:2007yz}
\bibinfo{author}{\bibfnamefont{H.~P.} \bibnamefont{Pfeiffer}},
  \bibinfo{author}{\bibfnamefont{D.~A.} \bibnamefont{Brown}},
  \bibinfo{author}{\bibfnamefont{L.~E.} \bibnamefont{Kidder}},
  \bibinfo{author}{\bibfnamefont{L.}~\bibnamefont{Lindblom}},
  \bibinfo{author}{\bibfnamefont{G.}~\bibnamefont{Lovelace}}, \bibnamefont{and}
  \bibinfo{author}{\bibfnamefont{M.~A.} \bibnamefont{Scheel}},
  \bibinfo{journal}{Class. Quant. Grav.} \textbf{\bibinfo{volume}{24}},
  \bibinfo{pages}{S59} (\bibinfo{year}{2007}), \eprint{gr-qc/0702106}.

\bibitem[{\citenamefont{Purrer et~al.}(2012)\citenamefont{Purrer, Husa, and
  Hannam}}]{Purrer:2012wy}
\bibinfo{author}{\bibfnamefont{M.}~\bibnamefont{Purrer}},
  \bibinfo{author}{\bibfnamefont{S.}~\bibnamefont{Husa}}, \bibnamefont{and}
  \bibinfo{author}{\bibfnamefont{M.}~\bibnamefont{Hannam}},
  \bibinfo{journal}{Phys. Rev.} \textbf{\bibinfo{volume}{D85}},
  \bibinfo{pages}{124051} (\bibinfo{year}{2012}), \eprint{1203.4258}.

\bibitem[{\citenamefont{Buonanno et~al.}(2006)\citenamefont{Buonanno, Chen, and
  Damour}}]{Buonanno:2005xu}
\bibinfo{author}{\bibfnamefont{A.}~\bibnamefont{Buonanno}},
  \bibinfo{author}{\bibfnamefont{Y.}~\bibnamefont{Chen}}, \bibnamefont{and}
  \bibinfo{author}{\bibfnamefont{T.}~\bibnamefont{Damour}},
  \bibinfo{journal}{Phys. Rev.} \textbf{\bibinfo{volume}{D74}},
  \bibinfo{pages}{104005} (\bibinfo{year}{2006}), \eprint{gr-qc/0508067}.

\bibitem[{\citenamefont{Damour et~al.}(2008{\natexlab{b}})\citenamefont{Damour,
  Jaranowski, and Sch\"afer}}]{Damour:2007nc}
\bibinfo{author}{\bibfnamefont{T.}~\bibnamefont{Damour}},
  \bibinfo{author}{\bibfnamefont{P.}~\bibnamefont{Jaranowski}},
  \bibnamefont{and}
  \bibinfo{author}{\bibfnamefont{G.}~\bibnamefont{Sch\"afer}},
  \bibinfo{journal}{Phys. Rev. D} \textbf{\bibinfo{volume}{77}},
  \bibinfo{pages}{064032} (\bibinfo{year}{2008}{\natexlab{b}}).

\bibitem[{\citenamefont{Hartung and Steinhoff}(2011)}]{Hartung:2011te}
\bibinfo{author}{\bibfnamefont{J.}~\bibnamefont{Hartung}} \bibnamefont{and}
  \bibinfo{author}{\bibfnamefont{J.}~\bibnamefont{Steinhoff}},
  \bibinfo{journal}{Annalen Phys.} \textbf{\bibinfo{volume}{523}},
  \bibinfo{pages}{783} (\bibinfo{year}{2011}), \eprint{1104.3079}.

\bibitem[{\citenamefont{Steinhoff
  et~al.}(2008{\natexlab{a}})\citenamefont{Steinhoff, Hergt, and
  Sch\"afer}}]{Steinhoff:2008ji}
\bibinfo{author}{\bibfnamefont{J.}~\bibnamefont{Steinhoff}},
  \bibinfo{author}{\bibfnamefont{S.}~\bibnamefont{Hergt}}, \bibnamefont{and}
  \bibinfo{author}{\bibfnamefont{G.}~\bibnamefont{Sch\"afer}},
  \bibinfo{journal}{Phys. Rev. D} \textbf{\bibinfo{volume}{78}},
  \bibinfo{pages}{101503} (\bibinfo{year}{2008}{\natexlab{a}}).

\bibitem[{\citenamefont{Steinhoff
  et~al.}(2008{\natexlab{b}})\citenamefont{Steinhoff, Hergt, and
  Sch\"afer}}]{Steinhoff:2007mb}
\bibinfo{author}{\bibfnamefont{J.}~\bibnamefont{Steinhoff}},
  \bibinfo{author}{\bibfnamefont{S.}~\bibnamefont{Hergt}}, \bibnamefont{and}
  \bibinfo{author}{\bibfnamefont{G.}~\bibnamefont{Sch\"afer}},
  \bibinfo{journal}{Phys. Rev. D} \textbf{\bibinfo{volume}{77}},
  \bibinfo{pages}{081501} (\bibinfo{year}{2008}{\natexlab{b}}).

\bibitem[{\citenamefont{Levi and Steinhoff}(2015)}]{Levi:2014gsa}
\bibinfo{author}{\bibfnamefont{M.}~\bibnamefont{Levi}} \bibnamefont{and}
  \bibinfo{author}{\bibfnamefont{J.}~\bibnamefont{Steinhoff}},
  \bibinfo{journal}{JHEP} \textbf{\bibinfo{volume}{06}}, \bibinfo{pages}{059}
  (\bibinfo{year}{2015}), \eprint{1410.2601}.

\bibitem[{\citenamefont{Brown et~al.}(2007)\citenamefont{Brown, Fairhurst,
  Krishnan, Mercer, Kopparapu, Santamaria, and Whelan}}]{Brown:2007jx}
\bibinfo{author}{\bibfnamefont{D.}~\bibnamefont{Brown}},
  \bibinfo{author}{\bibfnamefont{S.}~\bibnamefont{Fairhurst}},
  \bibinfo{author}{\bibfnamefont{B.}~\bibnamefont{Krishnan}},
  \bibinfo{author}{\bibfnamefont{R.~A.} \bibnamefont{Mercer}},
  \bibinfo{author}{\bibfnamefont{R.~K.} \bibnamefont{Kopparapu}},
  \bibinfo{author}{\bibfnamefont{L.}~\bibnamefont{Santamaria}},
  \bibnamefont{and} \bibinfo{author}{\bibfnamefont{J.~T.} \bibnamefont{Whelan}}
  (\bibinfo{year}{2007}), \eprint{0709.0093}.

\bibitem[{\citenamefont{Arun et~al.}(2009)\citenamefont{Arun, Buonanno, Faye,
  and Ochsner}}]{Fluxspinterms}
\bibinfo{author}{\bibfnamefont{K.~G.} \bibnamefont{Arun}},
  \bibinfo{author}{\bibfnamefont{A.}~\bibnamefont{Buonanno}},
  \bibinfo{author}{\bibfnamefont{G.}~\bibnamefont{Faye}}, \bibnamefont{and}
  \bibinfo{author}{\bibfnamefont{E.}~\bibnamefont{Ochsner}},
  \bibinfo{journal}{Phys. Rev. D} \textbf{\bibinfo{volume}{79}},
  \bibinfo{pages}{104023} (\bibinfo{year}{2009}).

\bibitem[{\citenamefont{Mroue et~al.}(2010)\citenamefont{Mroue, Pfeiffer,
  Kidder, and Teukolsky}}]{Mroue:2010re}
\bibinfo{author}{\bibfnamefont{A.~H.} \bibnamefont{Mroue}},
  \bibinfo{author}{\bibfnamefont{H.~P.} \bibnamefont{Pfeiffer}},
  \bibinfo{author}{\bibfnamefont{L.~E.} \bibnamefont{Kidder}},
  \bibnamefont{and} \bibinfo{author}{\bibfnamefont{S.~A.}
  \bibnamefont{Teukolsky}}, \bibinfo{journal}{Phys. Rev.}
  \textbf{\bibinfo{volume}{D82}}, \bibinfo{pages}{124016}
  (\bibinfo{year}{2010}), \eprint{1004.4697}.

\bibitem[{\citenamefont{Buonanno et~al.}(2009)\citenamefont{Buonanno, Iyer,
  Ochsner, Pan, and Sathyaprakash}}]{Buonanno:2009zt}
\bibinfo{author}{\bibfnamefont{A.}~\bibnamefont{Buonanno}},
  \bibinfo{author}{\bibfnamefont{B.~R.} \bibnamefont{Iyer}},
  \bibinfo{author}{\bibfnamefont{E.}~\bibnamefont{Ochsner}},
  \bibinfo{author}{\bibfnamefont{Y.}~\bibnamefont{Pan}}, \bibnamefont{and}
  \bibinfo{author}{\bibfnamefont{B.~S.} \bibnamefont{Sathyaprakash}},
  \bibinfo{journal}{Phys. Rev. D} \textbf{\bibinfo{volume}{80}},
  \bibinfo{pages}{084043} (\bibinfo{year}{2009}).

\bibitem[{\citenamefont{Buonanno et~al.}(2007)\citenamefont{Buonanno, Cook, and
  Pretorius}}]{taylorT3spinning}
\bibinfo{author}{\bibfnamefont{A.}~\bibnamefont{Buonanno}},
  \bibinfo{author}{\bibfnamefont{G.~B.} \bibnamefont{Cook}}, \bibnamefont{and}
  \bibinfo{author}{\bibfnamefont{F.}~\bibnamefont{Pretorius}},
  \bibinfo{journal}{Phys. Rev. D} \textbf{\bibinfo{volume}{75}},
  \bibinfo{pages}{124018} (\bibinfo{year}{2007}).

\bibitem[{\citenamefont{Buonanno et~al.}(2011)\citenamefont{Buonanno, Kidder,
  Mroue, Pfeiffer, and Taracchini}}]{Buonanno:2010yk}
\bibinfo{author}{\bibfnamefont{A.}~\bibnamefont{Buonanno}},
  \bibinfo{author}{\bibfnamefont{L.~E.} \bibnamefont{Kidder}},
  \bibinfo{author}{\bibfnamefont{A.~H.} \bibnamefont{Mroue}},
  \bibinfo{author}{\bibfnamefont{H.~P.} \bibnamefont{Pfeiffer}},
  \bibnamefont{and}
  \bibinfo{author}{\bibfnamefont{A.}~\bibnamefont{Taracchini}},
  \bibinfo{journal}{Phys. Rev.} \textbf{\bibinfo{volume}{D83}},
  \bibinfo{pages}{104034} (\bibinfo{year}{2011}), \eprint{1012.1549}.

\bibitem[{\citenamefont{Husa et~al.}(2008{\natexlab{b}})\citenamefont{Husa,
  Gonzalez, Hannam, Bruegmann, and Sperhake}}]{Husa:2007hp}
\bibinfo{author}{\bibfnamefont{S.}~\bibnamefont{Husa}},
  \bibinfo{author}{\bibfnamefont{J.~A.} \bibnamefont{Gonzalez}},
  \bibinfo{author}{\bibfnamefont{M.}~\bibnamefont{Hannam}},
  \bibinfo{author}{\bibfnamefont{B.}~\bibnamefont{Bruegmann}},
  \bibnamefont{and} \bibinfo{author}{\bibfnamefont{U.}~\bibnamefont{Sperhake}},
  \bibinfo{journal}{Class. Quant. Grav.} \textbf{\bibinfo{volume}{25}},
  \bibinfo{pages}{105006} (\bibinfo{year}{2008}{\natexlab{b}}),
  \eprint{0706.0740}.

\bibitem[{\citenamefont{Loffler et~al.}(2012)}]{Loffler:2011ay}
\bibinfo{author}{\bibfnamefont{F.}~\bibnamefont{Loffler}} \bibnamefont{et~al.},
  \bibinfo{journal}{Class. Quant. Grav.} \textbf{\bibinfo{volume}{29}},
  \bibinfo{pages}{115001} (\bibinfo{year}{2012}), \eprint{1111.3344}.

\bibitem[{\citenamefont{Shibata and Nakamura}(1995)}]{PhysRevD.52.5428}
\bibinfo{author}{\bibfnamefont{M.}~\bibnamefont{Shibata}} \bibnamefont{and}
  \bibinfo{author}{\bibfnamefont{T.}~\bibnamefont{Nakamura}},
  \bibinfo{journal}{Phys. Rev. D} \textbf{\bibinfo{volume}{52}},
  \bibinfo{pages}{5428} (\bibinfo{year}{1995}).

\bibitem[{\citenamefont{Baumgarte and Shapiro}(1998)}]{Baumgarte:1998te}
\bibinfo{author}{\bibfnamefont{T.~W.} \bibnamefont{Baumgarte}}
  \bibnamefont{and} \bibinfo{author}{\bibfnamefont{S.~L.}
  \bibnamefont{Shapiro}}, \bibinfo{journal}{Phys. Rev. D}
  \textbf{\bibinfo{volume}{59}}, \bibinfo{pages}{024007}
  (\bibinfo{year}{1998}).

\bibitem[{\citenamefont{Alcubierre et~al.}(2003)\citenamefont{Alcubierre,
  Br\"ugmann, Diener, Koppitz, Pollney, Seidel, and
  Takahashi}}]{Alcubierre:2002kk}
\bibinfo{author}{\bibfnamefont{M.}~\bibnamefont{Alcubierre}},
  \bibinfo{author}{\bibfnamefont{B.}~\bibnamefont{Br\"ugmann}},
  \bibinfo{author}{\bibfnamefont{P.}~\bibnamefont{Diener}},
  \bibinfo{author}{\bibfnamefont{M.}~\bibnamefont{Koppitz}},
  \bibinfo{author}{\bibfnamefont{D.}~\bibnamefont{Pollney}},
  \bibinfo{author}{\bibfnamefont{E.}~\bibnamefont{Seidel}}, \bibnamefont{and}
  \bibinfo{author}{\bibfnamefont{R.}~\bibnamefont{Takahashi}},
  \bibinfo{journal}{Phys. Rev. D} \textbf{\bibinfo{volume}{67}},
  \bibinfo{pages}{084023} (\bibinfo{year}{2003}).

\bibitem[{\citenamefont{Santamar\'{\i}a
  et~al.}(2010)\citenamefont{Santamar\'{\i}a, Ohme, Ajith, Br\"ugmann, Dorband,
  Hannam, Husa, M\"osta, Pollney, Reisswig et~al.}}]{PhysRevD.82.064016}
\bibinfo{author}{\bibfnamefont{L.}~\bibnamefont{Santamar\'{\i}a}},
  \bibinfo{author}{\bibfnamefont{F.}~\bibnamefont{Ohme}},
  \bibinfo{author}{\bibfnamefont{P.}~\bibnamefont{Ajith}},
  \bibinfo{author}{\bibfnamefont{B.}~\bibnamefont{Br\"ugmann}},
  \bibinfo{author}{\bibfnamefont{N.}~\bibnamefont{Dorband}},
  \bibinfo{author}{\bibfnamefont{M.}~\bibnamefont{Hannam}},
  \bibinfo{author}{\bibfnamefont{S.}~\bibnamefont{Husa}},
  \bibinfo{author}{\bibfnamefont{P.}~\bibnamefont{M\"osta}},
  \bibinfo{author}{\bibfnamefont{D.}~\bibnamefont{Pollney}},
  \bibinfo{author}{\bibfnamefont{C.}~\bibnamefont{Reisswig}},
  \bibnamefont{et~al.}, \bibinfo{journal}{Phys. Rev. D}
  \textbf{\bibinfo{volume}{82}}, \bibinfo{pages}{064016}
  (\bibinfo{year}{2010}).

\bibitem[{\citenamefont{Boyle}(2011)}]{Boyle:2011dy}
\bibinfo{author}{\bibfnamefont{M.}~\bibnamefont{Boyle}},
  \bibinfo{journal}{Phys. Rev.} \textbf{\bibinfo{volume}{D84}},
  \bibinfo{pages}{064013} (\bibinfo{year}{2011}).

\bibitem[{\citenamefont{Ohme et~al.}(2011)\citenamefont{Ohme, Hannam, and
  Husa}}]{Ohme:2011zm}
\bibinfo{author}{\bibfnamefont{F.}~\bibnamefont{Ohme}},
  \bibinfo{author}{\bibfnamefont{M.}~\bibnamefont{Hannam}}, \bibnamefont{and}
  \bibinfo{author}{\bibfnamefont{S.}~\bibnamefont{Husa}},
  \bibinfo{journal}{Phys.Rev.} \textbf{\bibinfo{volume}{D84}},
  \bibinfo{pages}{064029} (\bibinfo{year}{2011}), \eprint{1107.0996}.

\bibitem[{\citenamefont{Buchman et~al.}(2012)\citenamefont{Buchman, Pfeiffer,
  Scheel, and Szilagyi}}]{Buchman:2012dw}
\bibinfo{author}{\bibfnamefont{L.~T.} \bibnamefont{Buchman}},
  \bibinfo{author}{\bibfnamefont{H.~P.} \bibnamefont{Pfeiffer}},
  \bibinfo{author}{\bibfnamefont{M.~A.} \bibnamefont{Scheel}},
  \bibnamefont{and} \bibinfo{author}{\bibfnamefont{B.}~\bibnamefont{Szilagyi}},
  \bibinfo{journal}{Phys.Rev.} \textbf{\bibinfo{volume}{D86}},
  \bibinfo{pages}{084033} (\bibinfo{year}{2012}), \eprint{1206.3015}.

\bibitem[{\citenamefont{Bowen and York}(1980{\natexlab{b}})}]{Bowen1980}
\bibinfo{author}{\bibfnamefont{J.~M.} \bibnamefont{Bowen}} \bibnamefont{and}
  \bibinfo{author}{\bibfnamefont{J.~W.} \bibnamefont{York}},
  \bibinfo{journal}{Phys. Rev. D} \textbf{\bibinfo{volume}{21}},
  \bibinfo{pages}{2047} (\bibinfo{year}{1980}{\natexlab{b}}).

\bibitem[{\citenamefont{Brandt and Br\"{u}gmann}(1997)}]{Brandt1997}
\bibinfo{author}{\bibfnamefont{S.}~\bibnamefont{Brandt}} \bibnamefont{and}
  \bibinfo{author}{\bibfnamefont{B.}~\bibnamefont{Br\"{u}gmann}},
  \bibinfo{journal}{Physical Review Letters} \textbf{\bibinfo{volume}{78}},
  \bibinfo{pages}{3606} (\bibinfo{year}{1997}), ISSN \bibinfo{issn}{1079-7114}.

\bibitem[{\citenamefont{Pollney et~al.}(2011)\citenamefont{Pollney, Reisswig,
  Schnetter, Dorband, and Diener}}]{Pollney:2009yz}
\bibinfo{author}{\bibfnamefont{D.}~\bibnamefont{Pollney}},
  \bibinfo{author}{\bibfnamefont{C.}~\bibnamefont{Reisswig}},
  \bibinfo{author}{\bibfnamefont{E.}~\bibnamefont{Schnetter}},
  \bibinfo{author}{\bibfnamefont{N.}~\bibnamefont{Dorband}}, \bibnamefont{and}
  \bibinfo{author}{\bibfnamefont{P.}~\bibnamefont{Diener}},
  \bibinfo{journal}{Phys. Rev.} \textbf{\bibinfo{volume}{D83}},
  \bibinfo{pages}{044045} (\bibinfo{year}{2011}).

\bibitem[{\citenamefont{Ansorg et~al.}(2004)\citenamefont{Ansorg, Br\"ugmann,
  and Tichy}}]{Ansorg2004}
\bibinfo{author}{\bibfnamefont{M.}~\bibnamefont{Ansorg}},
  \bibinfo{author}{\bibfnamefont{B.}~\bibnamefont{Br\"ugmann}},
  \bibnamefont{and} \bibinfo{author}{\bibfnamefont{W.}~\bibnamefont{Tichy}},
  \bibinfo{journal}{Phys. Rev. D} \textbf{\bibinfo{volume}{70}},
  \bibinfo{pages}{064011} (\bibinfo{year}{2004}).

\bibitem[{\citenamefont{Marronetti et~al.}(2008)\citenamefont{Marronetti,
  Tichy, Br\"ugmann, Gonz\'alez, and Sperhake}}]{Marronetti:2007wz}
\bibinfo{author}{\bibfnamefont{P.}~\bibnamefont{Marronetti}},
  \bibinfo{author}{\bibfnamefont{W.}~\bibnamefont{Tichy}},
  \bibinfo{author}{\bibfnamefont{B.}~\bibnamefont{Br\"ugmann}},
  \bibinfo{author}{\bibfnamefont{J.}~\bibnamefont{Gonz\'alez}},
  \bibnamefont{and} \bibinfo{author}{\bibfnamefont{U.}~\bibnamefont{Sperhake}},
  \bibinfo{journal}{Phys. Rev. D} \textbf{\bibinfo{volume}{77}},
  \bibinfo{pages}{064010} (\bibinfo{year}{2008}).

\bibitem[{\citenamefont{Brown et~al.}(2009)\citenamefont{Brown, Diener,
  Sarbach, Schnetter, and Tiglio}}]{Brown:2008sb}
\bibinfo{author}{\bibfnamefont{D.}~\bibnamefont{Brown}},
  \bibinfo{author}{\bibfnamefont{P.}~\bibnamefont{Diener}},
  \bibinfo{author}{\bibfnamefont{O.}~\bibnamefont{Sarbach}},
  \bibinfo{author}{\bibfnamefont{E.}~\bibnamefont{Schnetter}},
  \bibnamefont{and} \bibinfo{author}{\bibfnamefont{M.}~\bibnamefont{Tiglio}},
  \bibinfo{journal}{Phys. Rev. D} \textbf{\bibinfo{volume}{79}},
  \bibinfo{pages}{044023} (\bibinfo{year}{2009}).

\bibitem[{\citenamefont{Bona et~al.}(1995)\citenamefont{Bona, Masso, Seidel,
  and Stela}}]{Bona:1994dr}
\bibinfo{author}{\bibfnamefont{C.}~\bibnamefont{Bona}},
  \bibinfo{author}{\bibfnamefont{J.}~\bibnamefont{Masso}},
  \bibinfo{author}{\bibfnamefont{E.}~\bibnamefont{Seidel}}, \bibnamefont{and}
  \bibinfo{author}{\bibfnamefont{J.}~\bibnamefont{Stela}},
  \bibinfo{journal}{Phys. Rev. Lett.} \textbf{\bibinfo{volume}{75}},
  \bibinfo{pages}{600} (\bibinfo{year}{1995}), \eprint{gr-qc/9412071}.

\bibitem[{\citenamefont{Thornburg}(2004)}]{Thornburg:2003sf}
\bibinfo{author}{\bibfnamefont{J.}~\bibnamefont{Thornburg}},
  \bibinfo{journal}{Class. Quant. Grav.} \textbf{\bibinfo{volume}{21}},
  \bibinfo{pages}{743} (\bibinfo{year}{2004}), \eprint{gr-qc/0306056}.

\bibitem[{\citenamefont{Dreyer et~al.}(2003)\citenamefont{Dreyer, Krishnan,
  Shoemaker, and Schnetter}}]{Dreyer:2002mx}
\bibinfo{author}{\bibfnamefont{O.}~\bibnamefont{Dreyer}},
  \bibinfo{author}{\bibfnamefont{B.}~\bibnamefont{Krishnan}},
  \bibinfo{author}{\bibfnamefont{D.}~\bibnamefont{Shoemaker}},
  \bibnamefont{and}
  \bibinfo{author}{\bibfnamefont{E.}~\bibnamefont{Schnetter}},
  \bibinfo{journal}{Phys. Rev.} \textbf{\bibinfo{volume}{D67}},
  \bibinfo{pages}{024018} (\bibinfo{year}{2003}), \eprint{gr-qc/0206008}.

\bibitem[{\citenamefont{Reisswig and Pollney}(2011)}]{Reisswig:2010di}
\bibinfo{author}{\bibfnamefont{C.}~\bibnamefont{Reisswig}} \bibnamefont{and}
  \bibinfo{author}{\bibfnamefont{D.}~\bibnamefont{Pollney}},
  \bibinfo{journal}{Class. Quant. Grav.} \textbf{\bibinfo{volume}{28}},
  \bibinfo{pages}{195015} (\bibinfo{year}{2011}), \eprint{1006.1632}.

\bibitem[{\citenamefont{Husa et~al.}(2006)\citenamefont{Husa, Hinder, and
  Lechner}}]{Husa:2004ip}
\bibinfo{author}{\bibfnamefont{S.}~\bibnamefont{Husa}},
  \bibinfo{author}{\bibfnamefont{I.}~\bibnamefont{Hinder}}, \bibnamefont{and}
  \bibinfo{author}{\bibfnamefont{C.}~\bibnamefont{Lechner}},
  \bibinfo{journal}{Comput. Phys. Commun.} \textbf{\bibinfo{volume}{174}},
  \bibinfo{pages}{983} (\bibinfo{year}{2006}), \eprint{gr-qc/0404023}.

\bibitem[{\citenamefont{Thomas and Schnetter}(2010)}]{SimulationFactory}
\bibinfo{author}{\bibfnamefont{M.~W.} \bibnamefont{Thomas}} \bibnamefont{and}
  \bibinfo{author}{\bibfnamefont{E.}~\bibnamefont{Schnetter}},
  \bibinfo{journal}{CoRR} \textbf{\bibinfo{volume}{1008.4571}}
  (\bibinfo{year}{2010}), \eprint{1008.4571}.

\bibitem[{\citenamefont{Hinder and Wardell}(Simulation Tools
  v1.1.0)}]{SimulationTools}
\bibinfo{author}{\bibfnamefont{I.}~\bibnamefont{Hinder}} \bibnamefont{and}
  \bibinfo{author}{\bibfnamefont{B.}~\bibnamefont{Wardell}}
  (\bibinfo{year}{Simulation Tools v1.1.0}),
  \urlprefix\url{http://simulationtools.org}.

\end{thebibliography}

\end{document}